\documentclass[modern, times]{aastex701}

\usepackage{CJK}
\usepackage{longtable}
\usepackage{makecell}
\usepackage{rotating}

\makeatletter
\let\frontmatter@title@above=\relax
\makeatother

\begin{document}
\begin{CJK}{UTF8}{gbsn}  

\title{The bimodal distribution of donor stars in X-ray binaries}

\author[orcid=0000-0003-2017-9151,gname=Jia, sname=Zhang]{Jia Zhang (张嘉)}
\affiliation{Yunnan Observatories, Chinese Academy of Sciences, Kunming 650216, China}
\affiliation{Key Laboratory of the Structure and Evolution of Celestial Objects, Chinese Academy of Sciences, Kunming 650216, China}
\email{yelangdadi@163.com}

\author[gname=Guo-Bao, sname=Zhang]{Guo-Bao Zhang (张国宝)}
\affiliation{Yunnan Observatories, Chinese Academy of Sciences, Kunming 650216, China}
\affiliation{Key Laboratory of the Structure and Evolution of Celestial Objects, Chinese Academy of Sciences, Kunming 650216, China}
\email{zhangguobao@ynao.ac.cn}

\correspondingauthor{Li-Ying Zhu (朱俐颖)} 
\author[gname=Li-Ying, sname=Zhu]{Li-Ying Zhu (朱俐颖)}
\affiliation{Yunnan Observatories, Chinese Academy of Sciences, Kunming 650216, China}
\affiliation{Key Laboratory of the Structure and Evolution of Celestial Objects, Chinese Academy of Sciences, Kunming 650216, China}
\email[show]{zhuly@ynao.ac.cn}

\author[gname=Sheng-Bang, sname=Qian]{Sheng-Bang Qian (钱声帮)}
\affiliation{Department of Astronomy, Key Laboratory of Astroparticle Physics of Yunnan Province, Yunnan University, Kunming 650091, China}
\email{qiansb@ynu.edu.cn}

\author[gname=Xiao, sname=Zhou]{Xiao Zhou (周肖)}
\affiliation{Yunnan Observatories, Chinese Academy of Sciences, Kunming 650216, China}
\affiliation{Key Laboratory of the Structure and Evolution of Celestial Objects, Chinese Academy of Sciences, Kunming 650216, China}
\email{zhouxiaophy@ynao.ac.cn}

\author[gname=Er-Gang, sname=Zhao]{Er-Gang Zhao (赵二刚)}
\affiliation{Yunnan Observatories, Chinese Academy of Sciences, Kunming 650216, China}
\affiliation{Key Laboratory of the Structure and Evolution of Celestial Objects, Chinese Academy of Sciences, Kunming 650216, China}
\email{zergang@ynao.ac.cn}

\begin{abstract}
The classification of X-ray binaries into high- and low-mass types has historically lacked a unified, data-driven quantitative criterion, and large-scale statistical studies of the donor star population have been limited. In this work, we address this gap by compiling data for 3,964 XRBs and deriving a plentiful set of physical parameters (mass, radius, age, and evolutionary stage) for a sub-sample of 288 donor stars using Gaia DR3 spectral data and stellar evolution models. We find a statistically bimodal distribution in the donor star parameters, which is characterized by a valley at approximately 3 $M_{\odot}$ or 11,000 K. We uncover the physical mechanism behind this bimodality: a previously unreported ``parallel tracks'' phenomenon observed in the relationship between the donor's evolutionary stage and its fundamental parameters, such as luminosity and radius. These two tracks represent distinct main-sequence populations, and the valley between them corresponds to the sparsely populated pre- and post-main-sequence evolutionary phases.
\end{abstract}



\section{Introduction} \label{sec:intro}

X-ray Binaries (hereafter XRBs) are binary star systems in which one of the stars is a compact object, such as a white dwarf, neutron star, or a black hole, and the other is usually a main sequence star. These systems are named as such because they release a significant amount of X-ray radiation, which is generated by material from a donor star being pulled onto the compact object and heated to extremely high temperatures \citep{2006csxs.book..623T}. XRBs are an important accretor for studying the physics of extreme environments, as well as for testing theories of gravity in the strong-field regime.

XRBs were often divided into two main types: high-mass XRBs (HMXBs), in which the mass donor star is a high-mass star such as O-B (Be) type stars, and low-mass XRBs (LMXBs), in which the mass donor star is a low-mass star like a red giant or a main sequence star \citep{2006csxs.book..623T}. There is also a type known as intermediate-mass XRBs (IMXBs) as a transitional state between the two previously mentioned types, as proposed by \citet{2003ApJ...597.1036P}.

In this classification, the mass refers to the mass of the donor star, rather than the compact accretor star. Regretfully, there are no unified quantitative criteria for drawing the line between HMXBs and LMXBs. One solar mass was initially designated as the dividing line, but later there were also 1-10 $M_{\odot}$ \citep{2006csxs.book..623T}, 1-5 $M_{\odot}$ \citep{2011hea..book.....L}, or 8 $M_{\odot}$ \citep{2023A&A...671A.149F}. A typical standard dividing line for classification is that high-mass refers to donor stars with masses $>=$ 8 $M_{\odot}$, while low-mass refers to donor stars with masses below this limit. 

The main issue with current XRBs is that there is a lack of samples with reliable physical parameters. The majority of the masses of compact components were not measured, nor the reliable masses of the donor stars \citep{2023A&A...671A.149F, 2023A&A...675A.199A, 2023A&A...677A.134N}. Excitingly, the emergence of large-scale optical sky survey data (e.g., \textit{Gaia}, \textit{LAMOST}, and \textit{TESS}) has provided a new opportunity.


In this study, we collected a large sample of XRBs and used \textit{Gaia} \citep{2023A&A...674A..32B, 2023A&A...674A...1G, 2016A&A...595A...1G, 2023A&A...674A..28F, 2023A&A...674A..26C} spectral data to derive the physical parameters of donor stars for 288 systems. Further, based on these physical parameters, we obtained the parameter distribution and correlations, and offered a new criterion for the classification.

\section{Selection and parameter obtaining of XRBs}

\subsection{The Selection of XRBs by Simbad and three catalogs}

We utilized the Astronomical Data Query Language (ADQL) to retrieve data from the Simbad database \citep{2000A&AS..143....9W} within the Topcat software \citep{2005ASPC..347...29T}. Through this process, we acquired a total of 3884 XRBs or XRBs candidates. Recognizing the potential lag in the Simbad database and its possible omission of the latest research achievements, we augmented our approach by referring to three recent X-ray binary star catalogs \citep{2023A&A...671A.149F, 2023A&A...675A.199A, 2023A&A...677A.134N}. A total of 3642 confirmed XRBs, comprising 1461 HMXBs, 781 LMXBs and 1400 unclassified systems, along with 322 candidate XRBs.

\subsection{Cross-matching XRBs with \textit{Gaia} DR3 for atmospheric parameters}

We cross-matched XRBs with \textit{Gaia} DR3 to obtain three atmospheric parameters of the targets, including surface temperature, surface gravity acceleration, and metallicity.

We used the \textit{SIMBAD} database, where coordinate uncertainties are provided as three parameters: the major axis, minor axis, and position angle of the error ellipse. For this study, we define high-precision coordinates as those with a major axis smaller than 0.4 arcseconds. Among the 3964 collected XRBs, 817 satisfy this criterion and are therefore used for further cross-matching.

The remaining 3147 binaries should not be cross-matched. Although some targets with large coordinate errors can indeed be matched with correct optical counterparts, we have chosen to discard these targets to maximize the accuracy of the cross-matching results.

The XRBs in this paper come from multiple observational sources, so the cross-matching radius with \textit{Gaia} should varies. We cross-matched these 817 targets with Gaia DR3 with a cross-matching radius of 0.5 to 2 arcseconds (see Appendix \ref{radius_of_the_cross-matching} for details on the radius selection). We thoroughly discuss the various possible sources of matching errors and calculate the corresponding false-match rate (see Appendix \ref{mistake_rate_of_cross-matching}). For the 288 XRBs with parameters in this paper, the false-match rate is approximately 1.1\%.

We obtained 514 matched targets with \textit{Gaia} observation. Among the 514 targets with \textit{Gaia} observation, 288 have three atmospheric parameters provided by the GSP-Phot (General Stellar Parametrizer from Photometry) model: temperature \texttt{Teff}, surface gravity \texttt{logg}, and metallicity \texttt{[Fe/H]}.

GSP-Phot is just one of the many models provided by \textit{Gaia}. Other models that can provide atmospheric parameters include GSP-Spec (General Stellar Parametrizer from Spectroscopy), ESP-HS (Extended Stellar Parametrizer for Hot Stars), and ESP-UCD (Extended Stellar Parametrizer for Ultra Cool Dwarfs).

Only GSP-Phot and GSP-Spec can provide all three atmospheric parameters, so we can only choose from these two models. In terms of quantity, GSP-Phot provides far more targets with parameters than GSP-Spec. GSP-Phot provides the atmospheric parameters for 288 XRBs in this paper, while GSP-Spec only provides 8 of them, so we finally chose the GSP-Phot model.

Among the parameter targets we cross-matched, there are 31 and 98 targets from the large Magellanic Cloud (LMC) and small Magellanic Cloud (SMC), respectively. These targets from the Magellanic Clouds (MCs) are an important sample for the analysis in this paper. We examined \textit{Gaia}'s observational capability for MC targets (Appendix~\ref{observational_capabilities_for_MCs}). \textit{Gaia} can observe stars within the MCs down to a lower limit of 1\,$M_{\odot}$, corresponding to a temperature range of approximately 3,900--7,600\,K. This capability is sufficient to cover the XRBs in this paper, which have an average temperature of 28,000\,K and a mass range of 1.6--24\,$M_{\odot}$, with an average mass of 11\,$M_{\odot}$.

\subsection{Obtaining the physical parameters of donor stars in XRBs}\label{subsec:calculating_pars}

To gain a deeper understanding of XRBs, we need to derive the fundamental parameters such as mass, radius, and age for their donor stars. Although \textit{Gaia} provides some parameters (e.g. mass and age) in its database, the quantity is limited. Taking the 288 targets with three atmospheric parameters as an example, \textit{Gaia} only provides mass for 54 targets and age for 37 targets. In order to maximize the available physical parameters, we derived them based on the atmospheric parameters.



This work provides parameters of donor stars in XRBs in bulk, based on spectral observations from \textit{Gaia} and using stellar evolutionary models. The method is basically the isochrone interpolation method, which is a classic method that has been used for a long time. It is fundamentally based on atmospheric parameters and estimates parameters such as mass by comparing them to a library of stellar database. It has been successfully applied and described by \citet{2019ApJS..244...43Z}. Here, we will introduce it briefly.

We already have a stellar parameter database that covers all possible stars, containing complete parameters for all ages, initial masses, and metallicities. Within this comprehensive database, we look for stellar samples with atmospheric parameters closest to our target, usually at least several dozens can be found. Then, the masses and radii of these dozens of stars becomes the estimated values for our target. We take the mass and radius of the sample that has the closest atmospheric parameters as the central value, and all the samples used to estimate the errors.



In this work, we utilized database \textit{MIST} (MESA Isochrones \& Stellar Tracks; \citealp{2016ApJS..222....8D, 2016ApJ...823..102C, 2011ApJS..192....3P, 2013ApJS..208....4P, 2015ApJS..220...15P, 2018ApJS..234...34P}). We did not write our own interpolation program but instead utilized a mature python package called \texttt{Isochrones} \citep{2015ascl.soft03010M}, version 2.1, that can interpolate stellar properties by using the \textit{MIST} database. With the help of the database and \textit{Gaia}'s atmospheric parameters, we obtained the mass, radius, and other physical parameters for 288 XRBs donor stars.


\subsection{Obtaining the X-ray Binary Catalogs}

Based on the Simbad database \citep{2000A&AS..143....9W} and three recent X-ray binary star catalogs \citep{2023A&A...671A.149F, 2023A&A...675A.199A, 2023A&A...677A.134N}, we have compiled a catalog containing 3964 Galactic and extragalactic XRBs (or XRBs candidates), listed in Table \ref{tab:total} in Appendix \ref{catalogs}. By cross-matching this table with \textit{Gaia} DR3, we obtained a catalog of 288 targets with three atmospheric parameters. Leveraging these atmospheric parameters and established stellar evolution databases, we further derived the basic parameters of these binaries' donor stars, which are presented in Table \ref{tab:Gaia_phy} in Appendix \ref{catalogs}.

While determining parameters such as mass and radius, we also obtained another crucial parameter from \textit{MIST} that finely characterizes stellar evolutionary stages. This parameter is EEP (Equivalent Evolutionary Points), which signifies the various phases of stellar evolution. A value of 202 corresponds to the Zero Age Main Sequence (ZAMS), while smaller values indicate pre-main-sequence stages, and 454 represents the Terminal Age Main Sequence (TAMS), with larger values marking post-main-sequence phases. For a detailed explanation of EEP, please refer to Appendix \ref{eep}.

\subsection{Uncertainties of Gaia-Based Donor Star Parameters in X-ray Binaries}\label{subsec:mistake_and_errors}

The primary goal of this work is to investigate the distribution and correlation of XRBs parameters. Therefore, it is essential to first assess the uncertainties and potential systematic error affecting our derived parameters.

To quantify the magnitude of these potential errors, we compared our derived donor masses with previously published values. Of the 288 targets with \textit{Gaia}-based atmospheric parameters in our sample, literature masses were available for 64 systems. 

Of the 64 systems with literature values, 31 masses were derived from spectral classifications (e.g., O6 V for 1FGL J1018.6-5856 and B0-B1 I for SAX J1802.7-2017). We argue that mass derivation via isochrone interpolation using three atmospheric parameters from modern \textit{Gaia} XP spectra is inherently more reliable than estimation from a single, coarse spectral type based on observations made 20-30 years ago.

\begin{figure*}[h]
\gridline{\fig{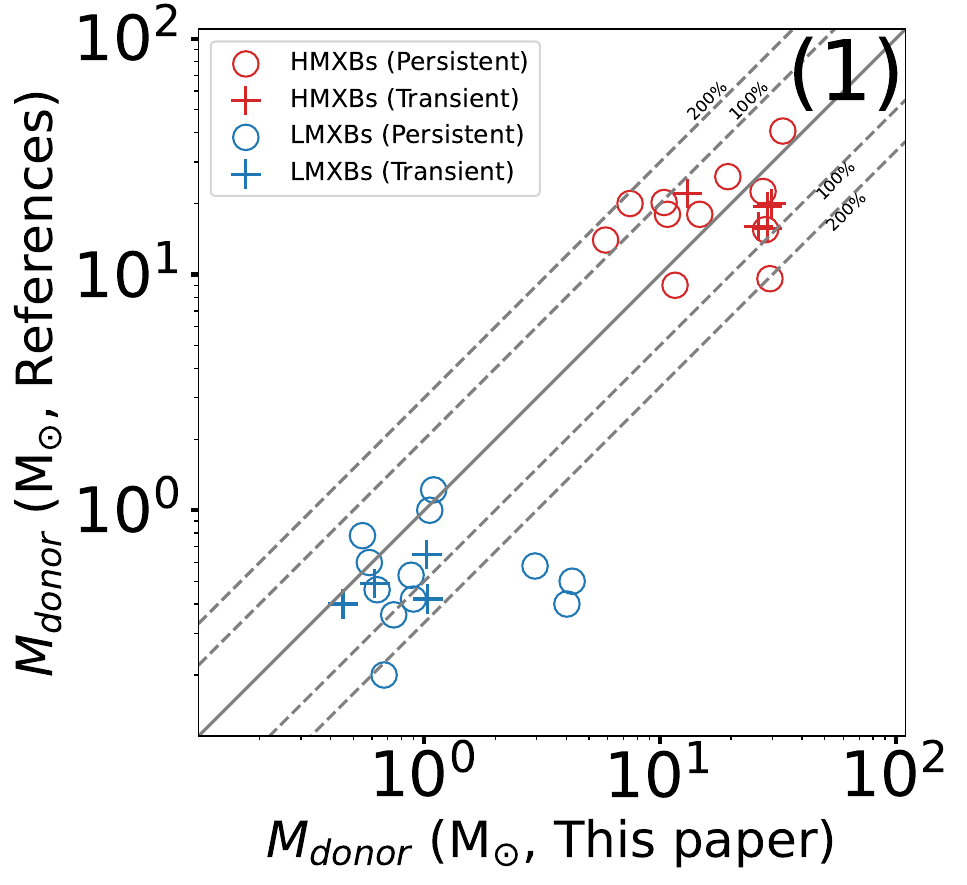}{0.44\textwidth}{}
          \fig{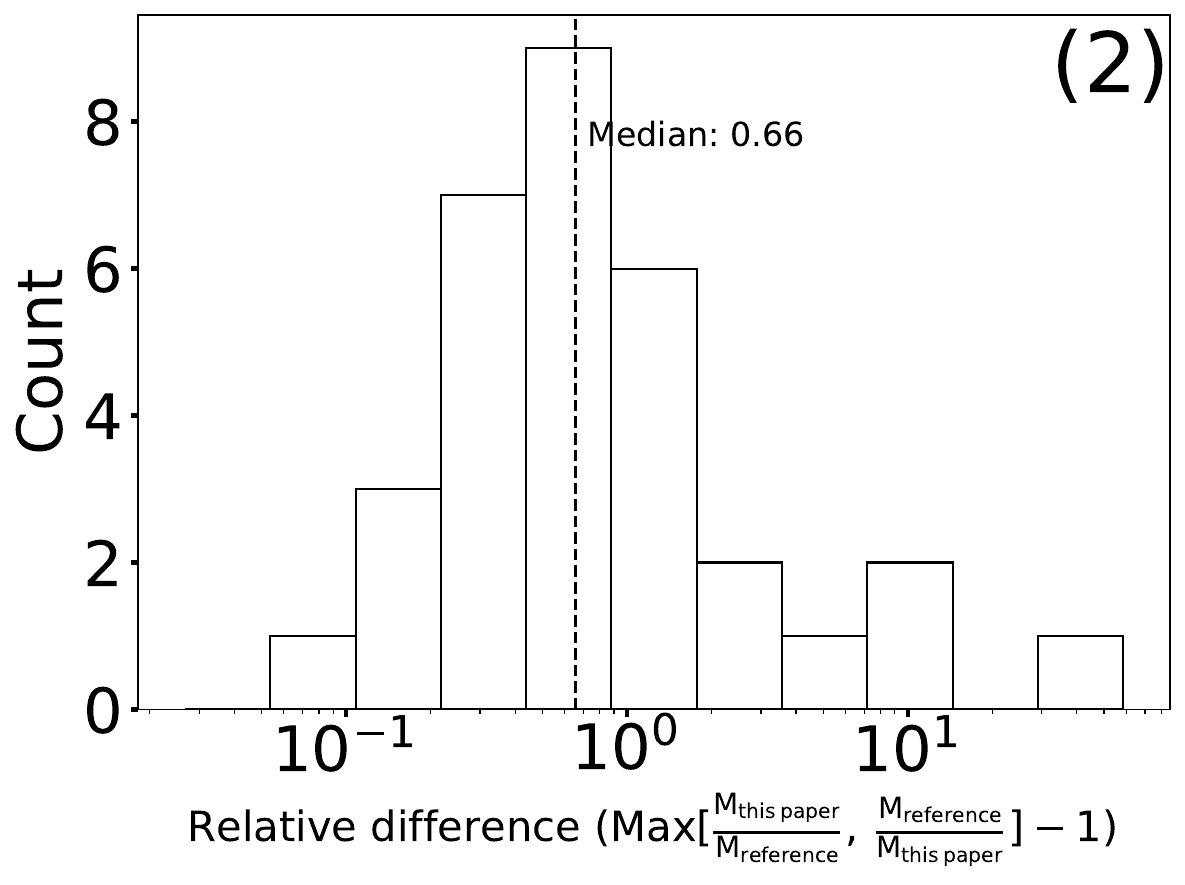}{0.53\textwidth}{}}
\vspace{-0.6cm}
\caption{Comparison of donor star masses between this paper and references. Panel (1): Donor mass derived in this work versus values from the references. Panel (2): Histogram of the relative difference between our mass values and references values. The dashed line indicates the median value. \label{fig:Comparison_mass_Refs}}
\end{figure*}

After removing the 31 spectral-type-based masses, the remaining 33 targets are presented in Figure \ref{fig:Comparison_mass_Refs}. The relative deviations for these 33 systems range from 2.6\% to 5849\%, with a median value of 66\%. One target, V1055 Ori, was excluded from the figure as an extreme outlier; its literature donor mass is 0.0142 M$_{\odot}$, whereas our estimate is 0.84 M$_{\odot}$, a value approximately 60 times larger.

We need to analyze the sources of uncertainty. The final errors are determined by four main factors: 1. The uncertainties in the stellar atmospheric parameters used as input. 2. The reliability of the adopted stellar evolution database. 3. The bias introduced when applying single-star evolutionary models to binary systems. 4. The contribution of accretion disk emission in the optical band, which is particularly severe for LMXBs with luminous accretion disk.

The atmospheric parameters used in this work are taken from the \textit{Gaia} GSP-Phot module. By comparing with the results of APOGEE DR16 \citep{2020AJ....160..120J}, GALAH DR3 \citep{2021MNRAS.506..150B}, \textit{LAMOST} DR4 \citep{2011RAA....11..924W, 2014IAUS..306..340W} and RAVE DR6 \citep{2020AJ....160...83S}, The mean absolute differences are 150-418 K, 0.1-0.4 dex, and 0.24-0.3 dex, and the median absolute differences are 110-169K, 0.06-0.25 dex, 0.20-0.21 dex for $T_{eff}$, \texttt{logg}, and [M/H], respectively. If we want to know the typical error, the median is more worth considering because the mean is easily influenced by individual extreme values and may not represent the overall situation.

These median absolute differences transferred to the stellar mass will cause a typical relative deviation of 3-7\%, 1-7\%, and 4-5\%, respectively. The combination of errors from three targets will result in a typical relative deviation of 2-8\% on mass.

The second factor affecting the final uncertainties is the reliability of the stellar evolution database. We adopted the MIST database, and to assess its reliability, we also employed an alternative evolutionary database \textit{PARSEC} (PAdova and TRieste Stellar Evolution Code; \citealp{2012MNRAS.427..127B, 2014MNRAS.444.2525C, 2015MNRAS.452.1068C, 2014MNRAS.445.4287T, 2017ApJ...835...77M,2019MNRAS.485.5666P,2020MNRAS.498.3283P}), version CMD 3.6. For the same atmospheric parameters of the XRBs in this work, the relative deviations between \textit{PARSEC} and \textit{MIST} are 0.0064\% - 457\% in mass, with a median of 3.2\%. The median relative deviations for radius, luminosity, and age are 1.7\%, 4.3\%, and 24\%, respectively. Considering that the two databases differ in many default parameters and computational methods, the intrinsic uncertainty from the \textit{MIST} database should be smaller than the discrepancies observed between the two models.

The third factor is the applicability of single-star models to binary systems. XRBs are binaries, and their evolutionary paths differ significantly from single stars once mass transfer occurs. In practice, the effect of mass transfer mainly impacts the age, making ages derived from single-star models unreliable. However, parameters such as the current mass and radius remain robust. While it is not feasible to compute large samples of XRBs with full binary models, we can refer to previous studies for guidance. \citet{2020ApJ...895..136C, 2021ApJ...920...76C} derived the parameters of pulsating star of two binary system (KIC 10736223 and OO Dra) using single star and binary star (considering matter accretion) models, respectively. These two pulsating stars have accumulated 2 and 1.7 times their initial mass during the evolution of binary stars, respectively, but the results of the single star model and binary model are very close. The mass deviation derived by the two models is only 2.5\% and 1\%. The deviation of the radius is also very small, at 0.7\% and 0.3\%, respectively. Although accretion significantly increases the mass of a star, its surface atmospheric composition has not changed too much. So the current internal and external structures still conform to the single star model, and their mass and radius can still be accurately determined based on the single star model.

The final and, in our view, most important factor is contamination from the accretion disk in the optical band. The spectra observed by \textit{Gaia} represent the entire binary system. While the contribution of the compact object in the optical can be neglected, the luminosity of the accretion disk may be significant and can even exceed that of the donor star. The flux contribution from the accretion disk can cause the measured donor mass to be overestimated. This effect is particularly pronounced in LMXBs with high accretion rates. For instance, calculations for Scorpius X-1 (V818 Sco) show that its accretion disk luminosity is approximately 4 to 8 times that of the donor star in the optical band \citep{2021MNRAS.508.1389C}. This is consistent with the situation shown in Figure \ref{fig:Comparison_mass_Refs}, where the donor masses of five LMXBs measured in this work are significantly higher than those reported in the literature. We identify at least three LMXBs (V1341 Cyg, V818 Sco, V691 CrA) where our results are significantly biased for this reason. Assuming the literature values represent the true donor masses, 29\% of the LMXBs in our comparison sample have mass errors exceeding 200\%. For HMXBs, stable accretion disk are generally not expected, making disk contamination an unlikely explanation for the systems with large deviations. Figure \ref{fig:Comparison_mass_Refs} shows that transient XRBs (plus markers) exhibit better agreement. This may suggest that for persistent XRBs, stable accretion has a greater influence in the optical band, which is consistent with our expectations.

In summary, the combined impact of the first three factors should not contribute more than about 15\% median uncertainty in mass. The last factor can account for the large deviations observed in some LMXBs with bright accretion disks, but it is less likely to explain the substantial discrepancies in HMXBs. Further considering that parameters reported in the literature also carry uncertainties, the typical relative error in the donor masses derived in this work should be less than 66\%. Acknowledging these limitations and potential biases is essential for conducting a robust statistical analysis and for assessing the reliability of the conclusions.

\section{The observed parameter distribution and occurrence rate} \label{sec:distribution}

\subsection{The observed distribution of the donor stars}

\begin{figure*}[h]
\gridline{\fig{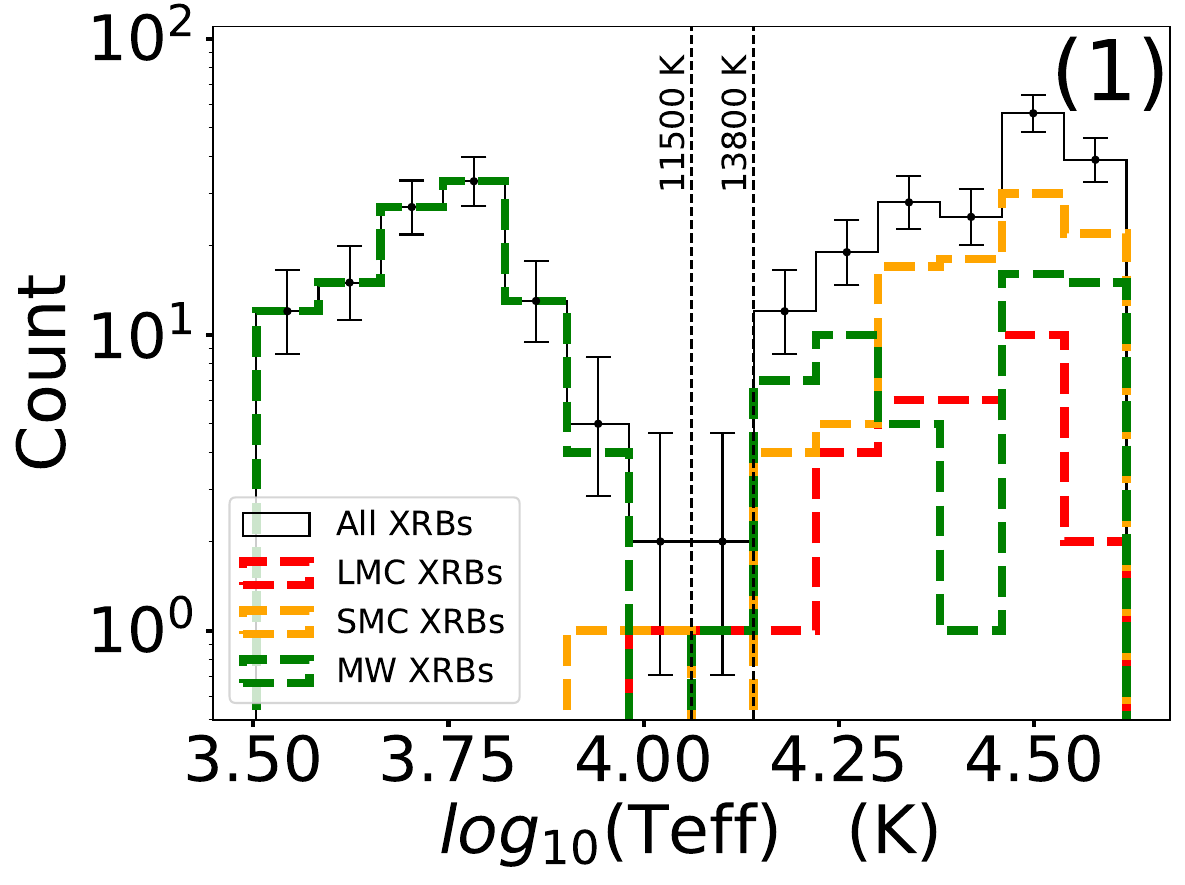}{0.33\textwidth}{}
          \fig{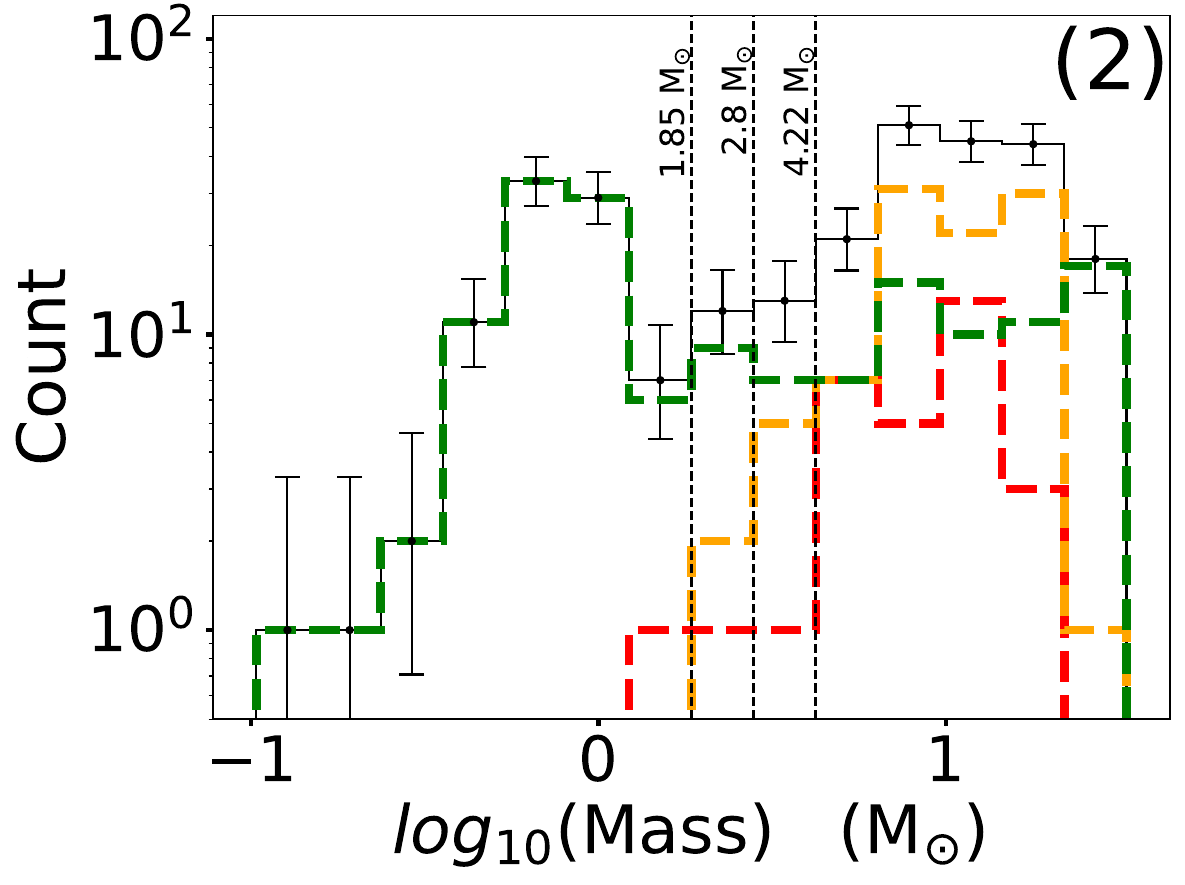}{0.33\textwidth}{}
          \fig{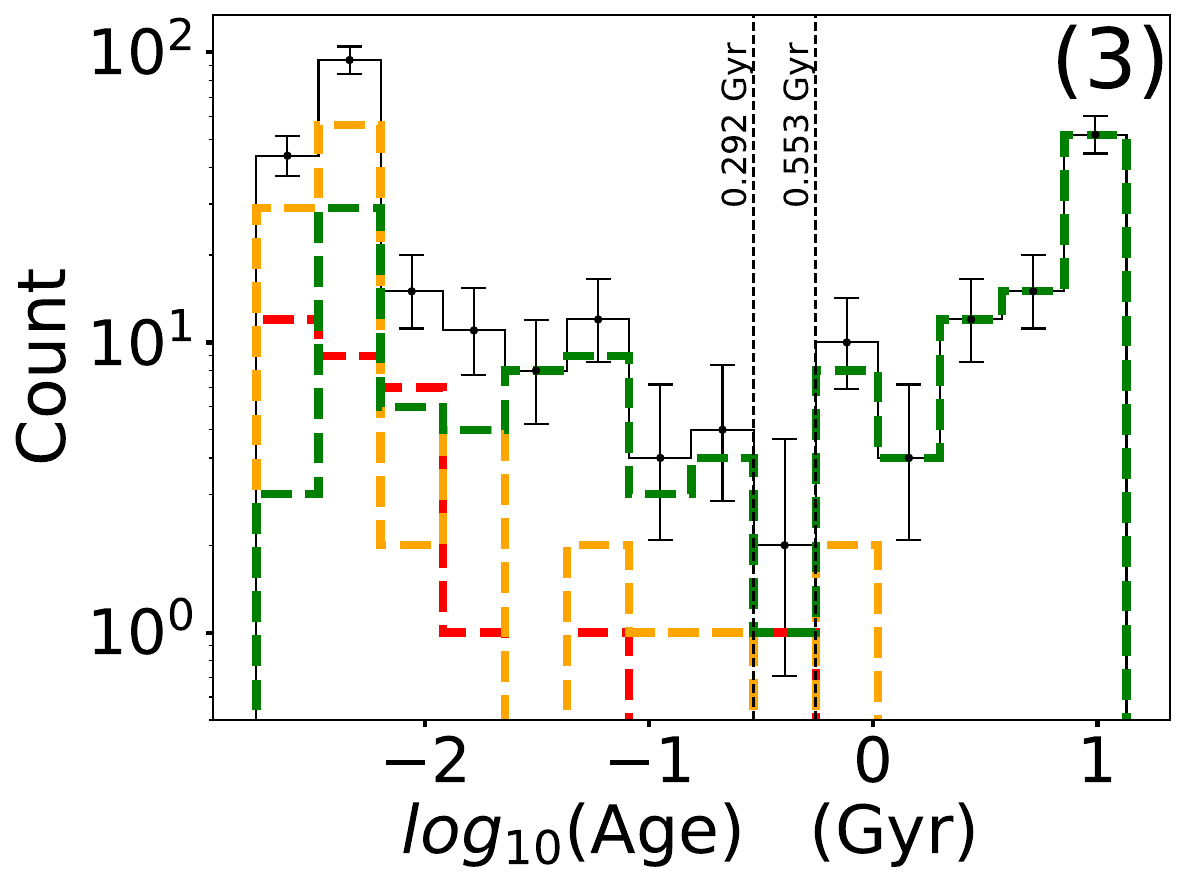}{0.33\textwidth}{}
              }
\vspace{-0.8cm}
\gridline{\fig{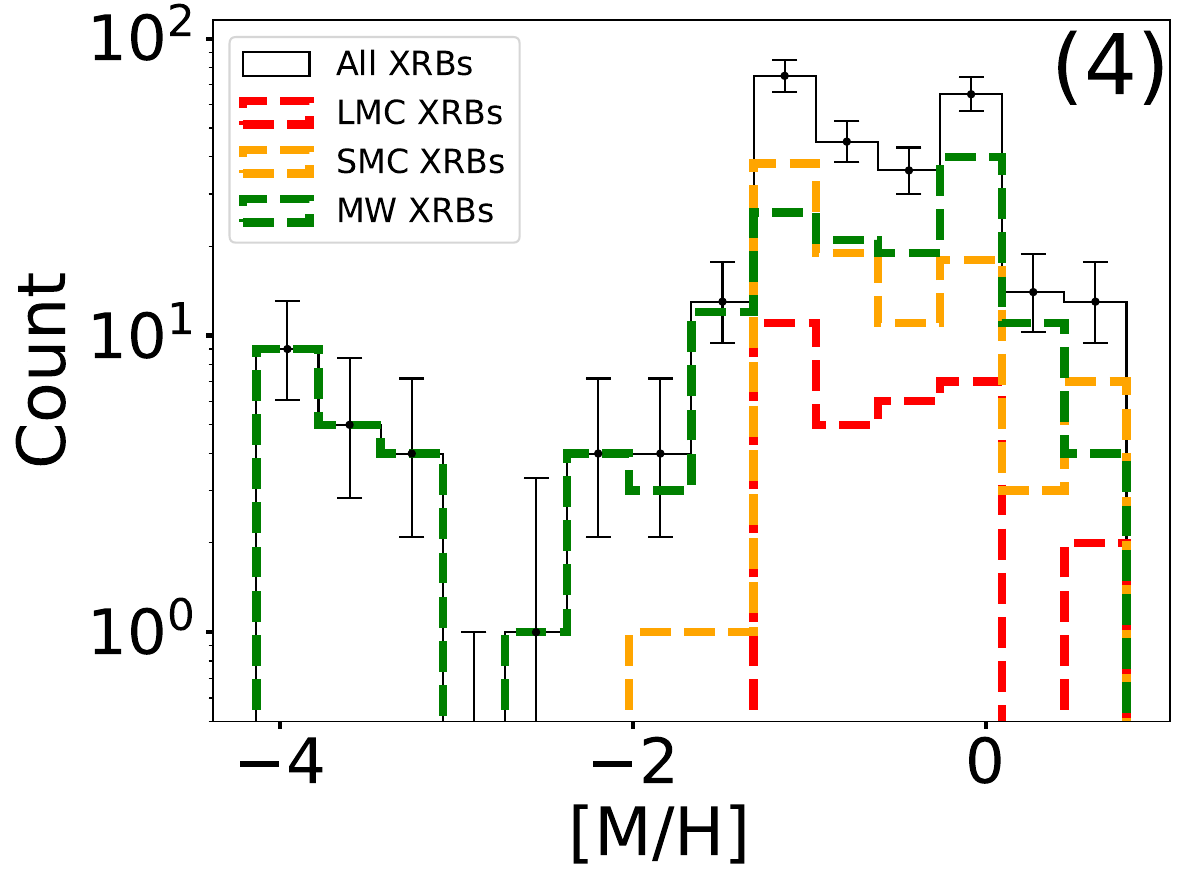}{0.33\textwidth}{}
          \fig{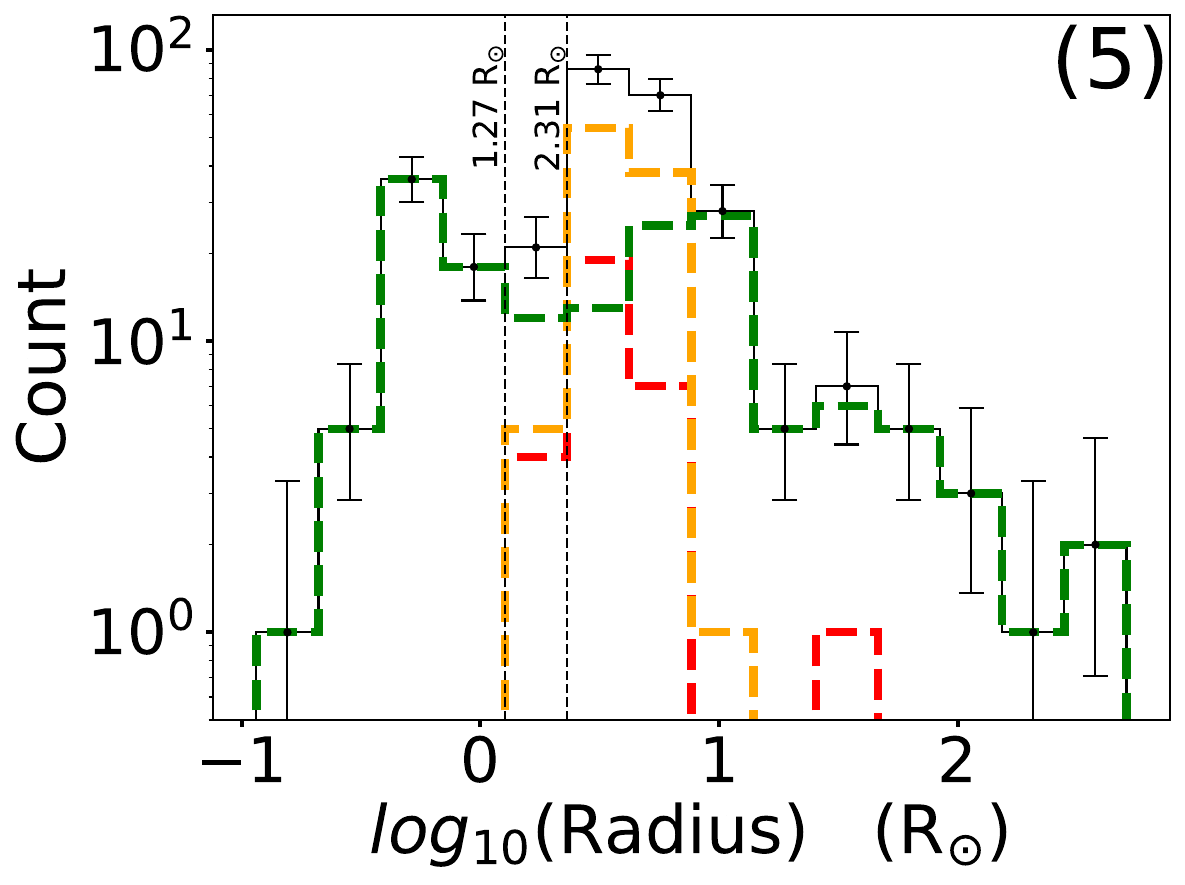}{0.33\textwidth}{}
          \fig{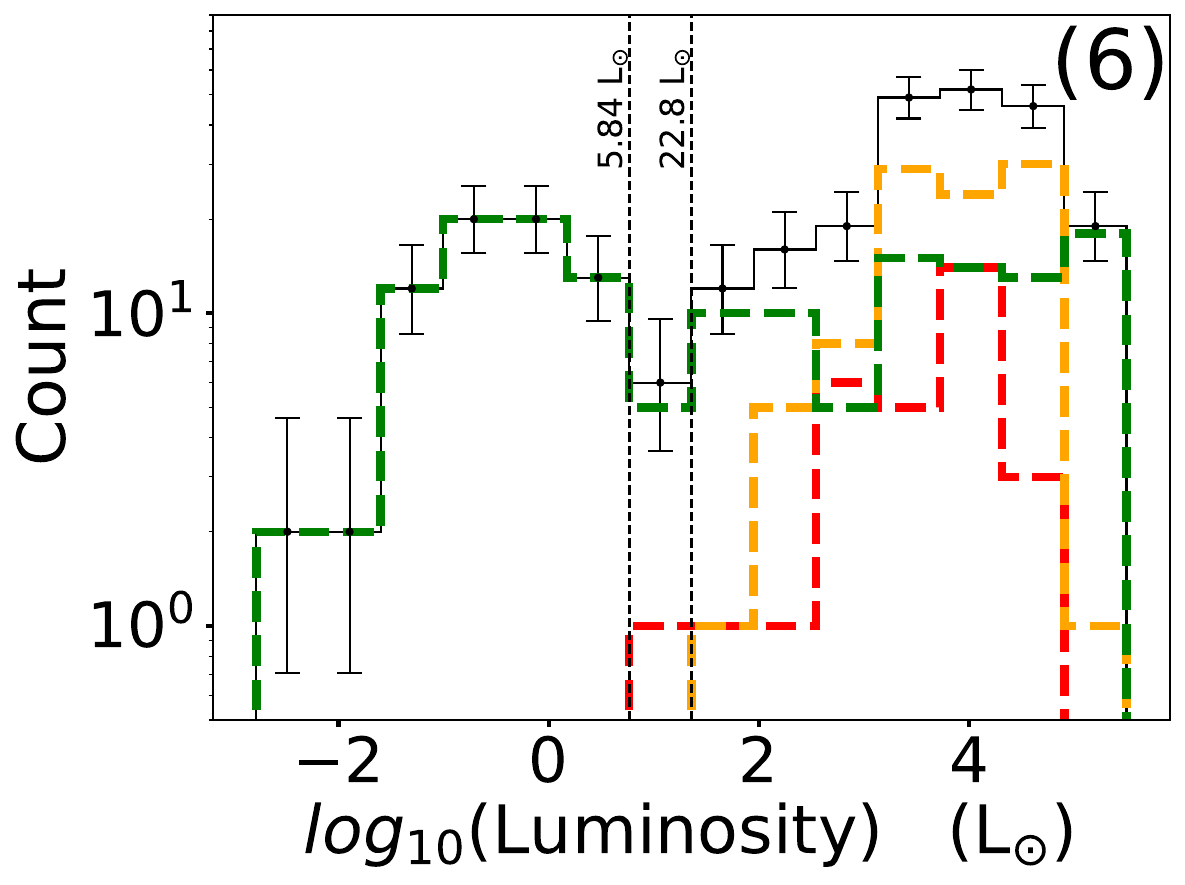}{0.33\textwidth}{}
              } 
\vspace{-0.8cm}
\gridline{\fig{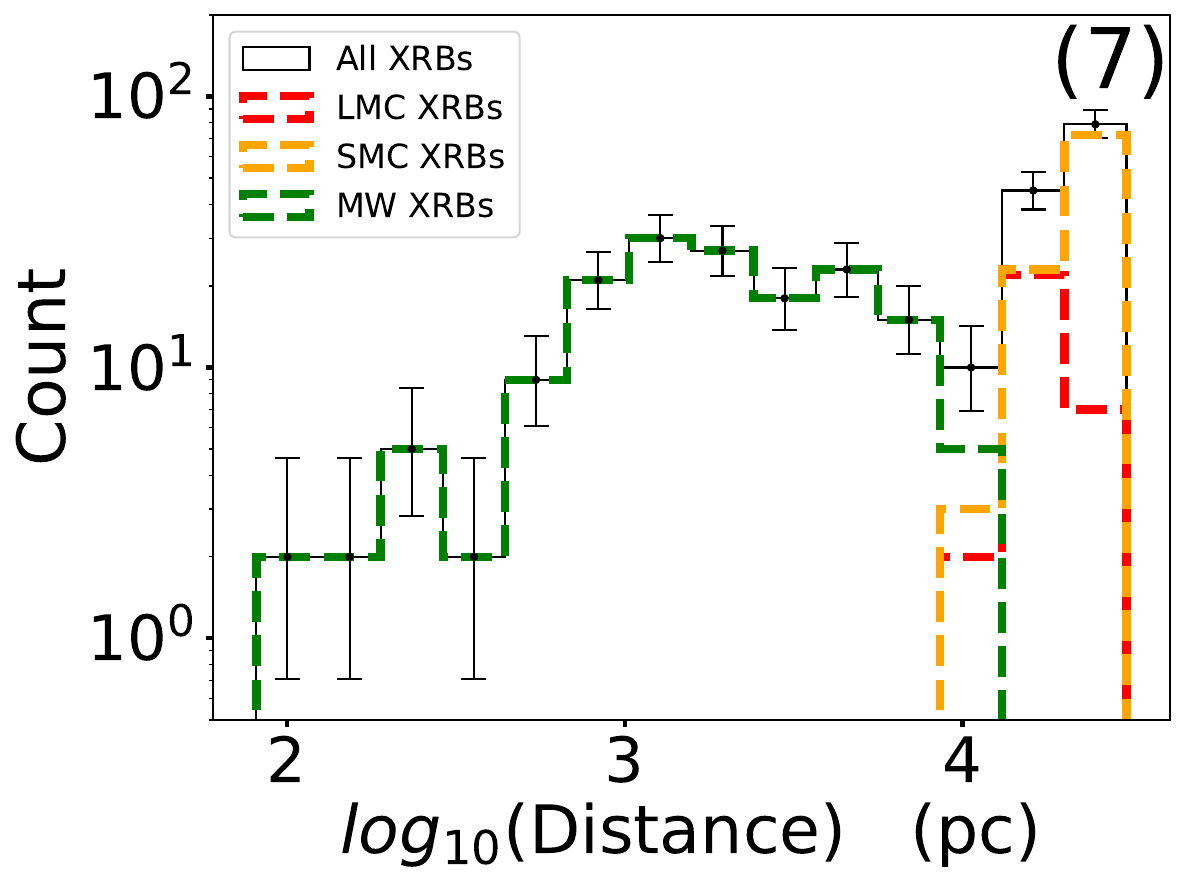}{0.33\textwidth}{}
          \fig{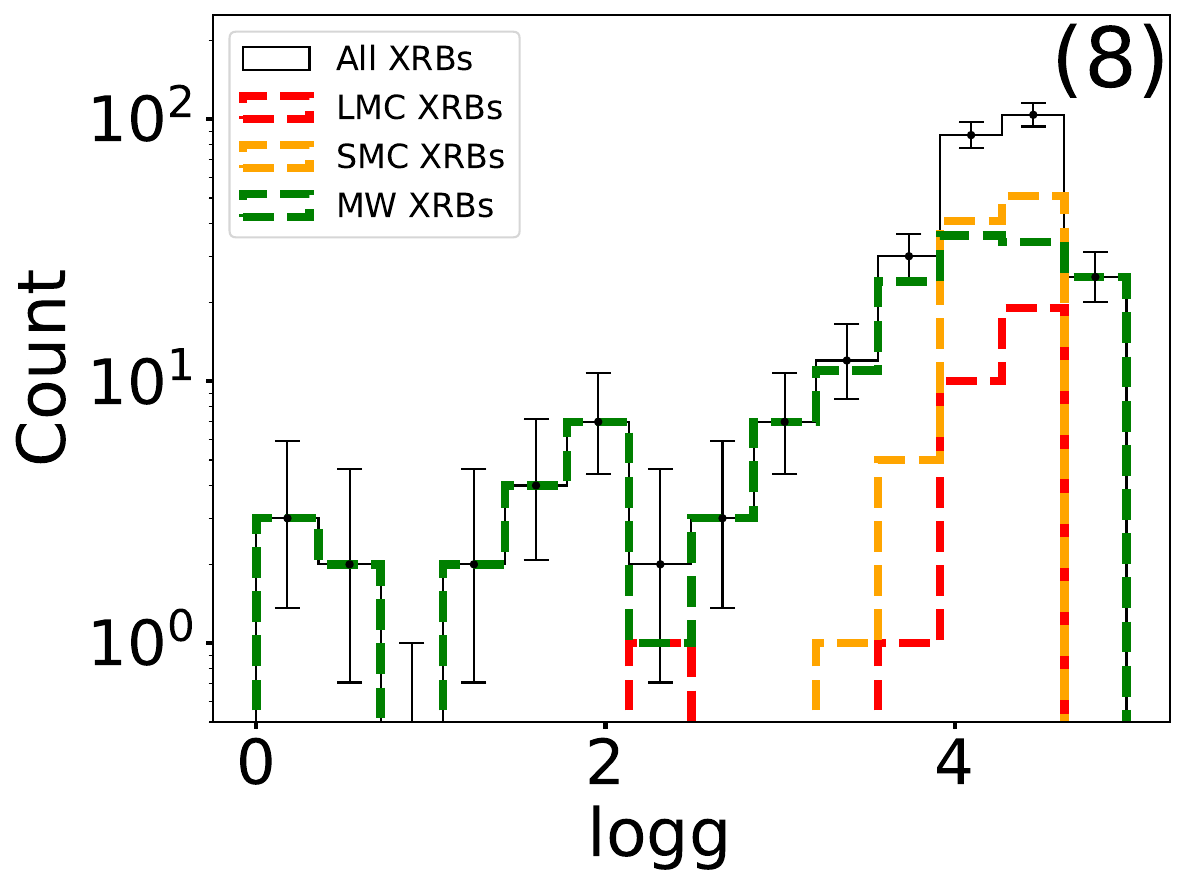}{0.33\textwidth}{}
          \fig{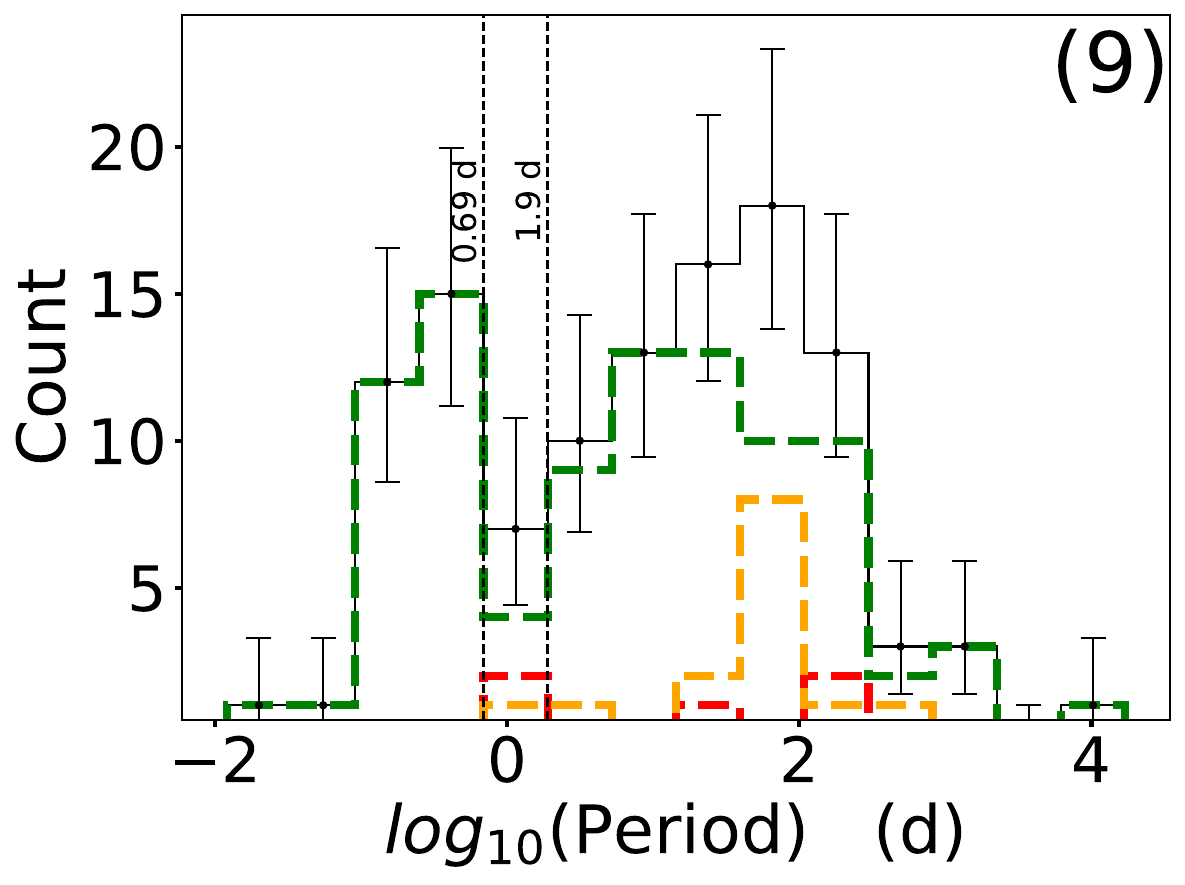}{0.33\textwidth}{}
             }
\vspace{-0.6cm}
\caption{The observed distribution of the donor stars in XRBs. Different colors represent different locations. Green represents MW, and red and orange represents LMC and SMC respectively. Black represent all XRBs. \label{fig:dist_observed}}
\end{figure*}

Panels 1-3 of Figure \ref{fig:dist_observed} present the logarithmic distribution of temperature, mass, and age, respectively. A noticeable dip can be observed in the middle region, particularly evident in the temperature panel 1, demarcated by two vertical dashed lines. The plots reveal that XRBs from the LMC and SMC dominate the high-temperature, high-mass, and young age end of the distribution, while those from the MW populate the low-temperature, low-mass, and older end.

To statistically validate the bimodal distributions of donor mass and temperature, we applied a two-component Gaussian Mixture Model (GMM). This model assumes the observed data are a superposition of two distinct Gaussian populations. We used the Expectation-Maximization (EM) algorithm to fit the model parameters. To assess the goodness-of-fit, we employed a bootstrap-corrected Kolmogorov-Smirnov (K-S) test. This robust approach mitigates the issue of inflated p-values that arises when model parameters are derived directly from the data. For both stellar mass and temperature, the maximum vertical distances between the empirical and theoretical cumulative distribution functions (CDFs) were small. The bootstrap p-values, estimated from 1000 resamples, were significantly greater than 0.05 (approximately 0.95 for mass and 0.68 for temperature, respectively). These results provide statistical support for the bimodal distributions of stellar mass and temperature. See Appendix \ref{GMM_and_KStest} for more detailed fitting parameters and explanations.

To further support the existence of a dip in our observed distributions, we conducted Hartigan's dip test for unimodality. Using the Python package \text{diptest}, we obtained a Dip statistic of 0.048 for mass and 0.066 for temperature. The corresponding bootstrap p-values were $4.6\times10^{-5}$ for mass and 0 for temperature. Both of these values are extremely small, leading us to reject the null hypothesis of a unimodal distribution. This result strongly supports the existence of a dip and confirms the bimodal nature of our distributions.

In addition to the quantitative analysis of the bimodal distribution above, we also need to investigate whether the observed bimodal structure holds up against the large parameter errors, or even incorrect values, discussed previously. To do this, we applied the distribution of deviations found in our 33-star comparison sample (Figure~\ref{fig:Comparison_mass_Refs}) to the entire XRB sample. Specifically, the mass of each XRB was randomly perturbed following that deviation distribution, and this process was repeated 5,000 times to generate 5,000 new mass distributions. The results of this Monte Carlo test, presented in Figure~\ref{fig:distribution_random} in Appendix~\ref{distribution_random}, demonstrate that the bimodal distribution remains robust.

It is well-established that the number and luminosity distribution of HMXBs in external galaxies correlate with the host galaxy's star formation rate, whereas low-mass systems correlate with the host galaxy's mass \citep{2003MNRAS.339..793G, 2004MNRAS.349..146G, 2006ARA&A..44..323F, 10.1093/mnras/staa1063}. Therefore, the observed differences in XRBs populations between the MCs and the MW could be attributed to the varying stellar formation rates and galaxy masses. Furthermore, the younger age of the MCs may contribute to the paucity of old LMXBs within these galaxies.

Panels 4-6 of Figure \ref{fig:dist_observed} show the distributions of metallicity, radius, and luminosity, which display similar bimodal structures to those in panels 1-3. Panels 7-9 depict the distributions of distance, log \( g \), and orbital period. The origins of the locations from the MW and MCs are clearly visible in the distance distribution.



\subsection{The distribution of occurrence rate}

The observed number distribution of XRBs, as presented earlier, is not corrected for selection effects and therefore does not represent the true intrinsic distribution for the full XRB populations in the MW and MCs. The sample in this study is compiled from 117 literature sources (41 for XRB types, 7 for coordinates, and 110 for various parameters), which introduces complex and intractable selection biases. These biases arise from numerous factors, including differing telescope sensitivities, sky coverage areas, coordinate accuracies, limiting magnitudes, and the fraction of XRBs in active states, in addition to potential human selection tendencies. The sheer complexity of these combined effects makes a direct correction to the number distribution unfeasible.

Please allow us to illustrate our intention with an analogy. Suppose we want to study the age distribution of men within a general population. After surveying, we find a peak in the absolute number of men in the 30--40 year-old age range. If we were to conclude that men have a tendency to be 30--40 years old, this conclusion would be totally misleading, because we have not considered the age distribution of the entire population. If, instead, we analyze the proportion of men relative to the total population as a function of age, we would find that the proportion remains near 0.5 at all ages, with no significant peaks or valleys. This would indicate that men do not have an intrinsic preference for any particular age group. The peak in the number distribution of men exists just because the underlying population is largest in that age range, not because of any special character of men themselves.

To apply this analogy to the present study: XRBs are equivalent to the men, parameters like temperature and mass are equivalent to age, and the entire population of stars represents the total population. While we cannot obtain the true number distribution of XRBs, we can determine the proportion/ratio of XRBs to the background stellar population. This proportion/ratio is not only physically meaningful but also can circumvent many selection effects. The age distributions of different human populations can vary drastically (e.g., a kindergarten versus a nursing home), but the proportion of men within them will not differ significantly. Similarly, Observations are more likely to miss faint XRBs compared to bright ones. However, among a sample of equally faint stars, the observed proportion of XRBs should be unaffected by this bias and should approximate the true value.

To implement this, the number of XRBs in each parameter bin should be divided by the number of background stars in the corresponding bin, yielding the occurrence rate. We selected a total of 186,126 random background stars with atmospheric parameters from \textit{Gaia}~DR3. Subsequently, we determined other parameters, such as mass, for 900 of these targets (300 each from the MW, LMC, and SMC). Before the number dividing, the background random star samples for the MW, LMC, and SMC were first normalized to match the total number of XRBs in our sample from each respective region: 159 for the MW, 31 for the LMC, and 98 for the SMC.

\begin{figure*}[h]
\gridline{\fig{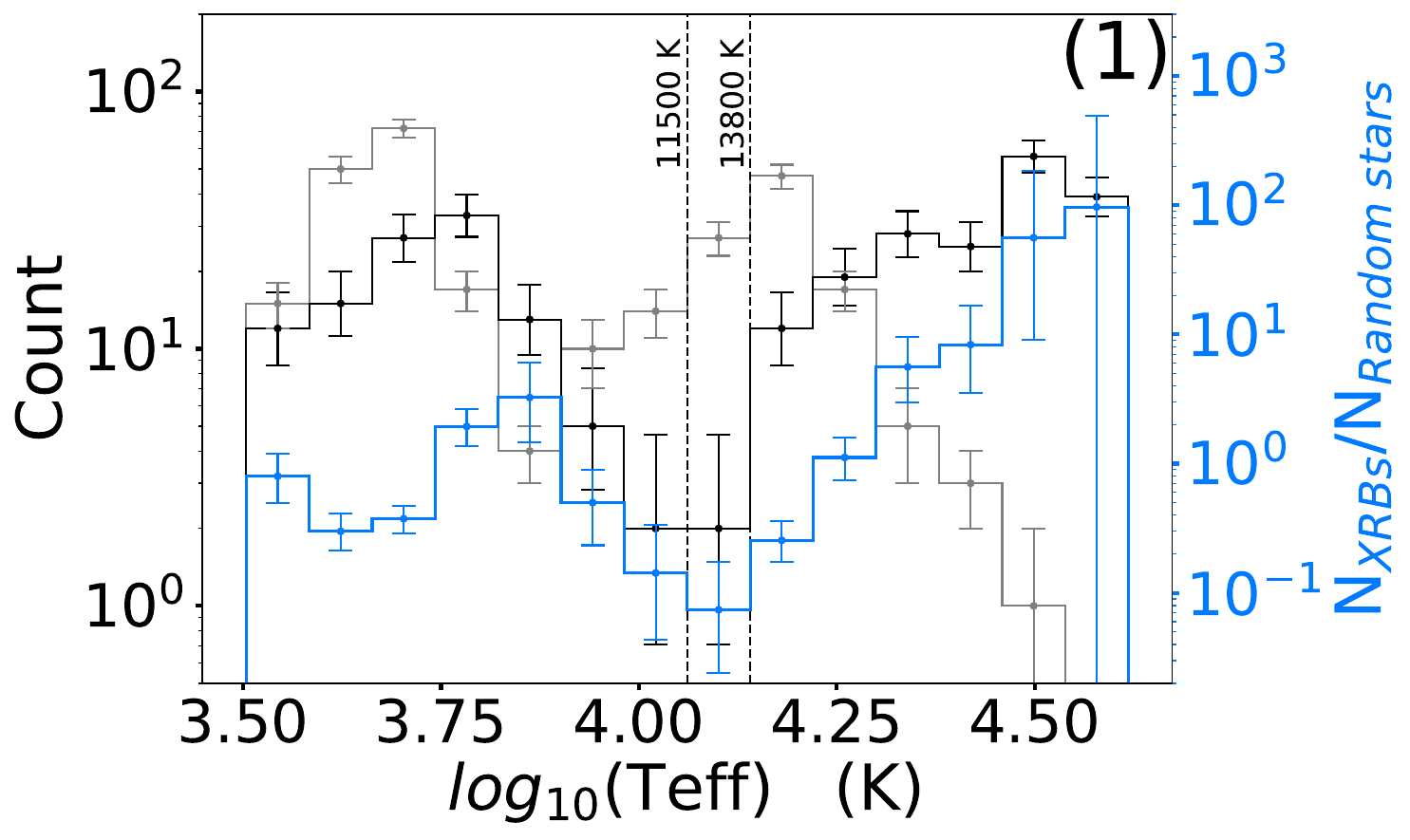}{0.495\textwidth}{}
          \fig{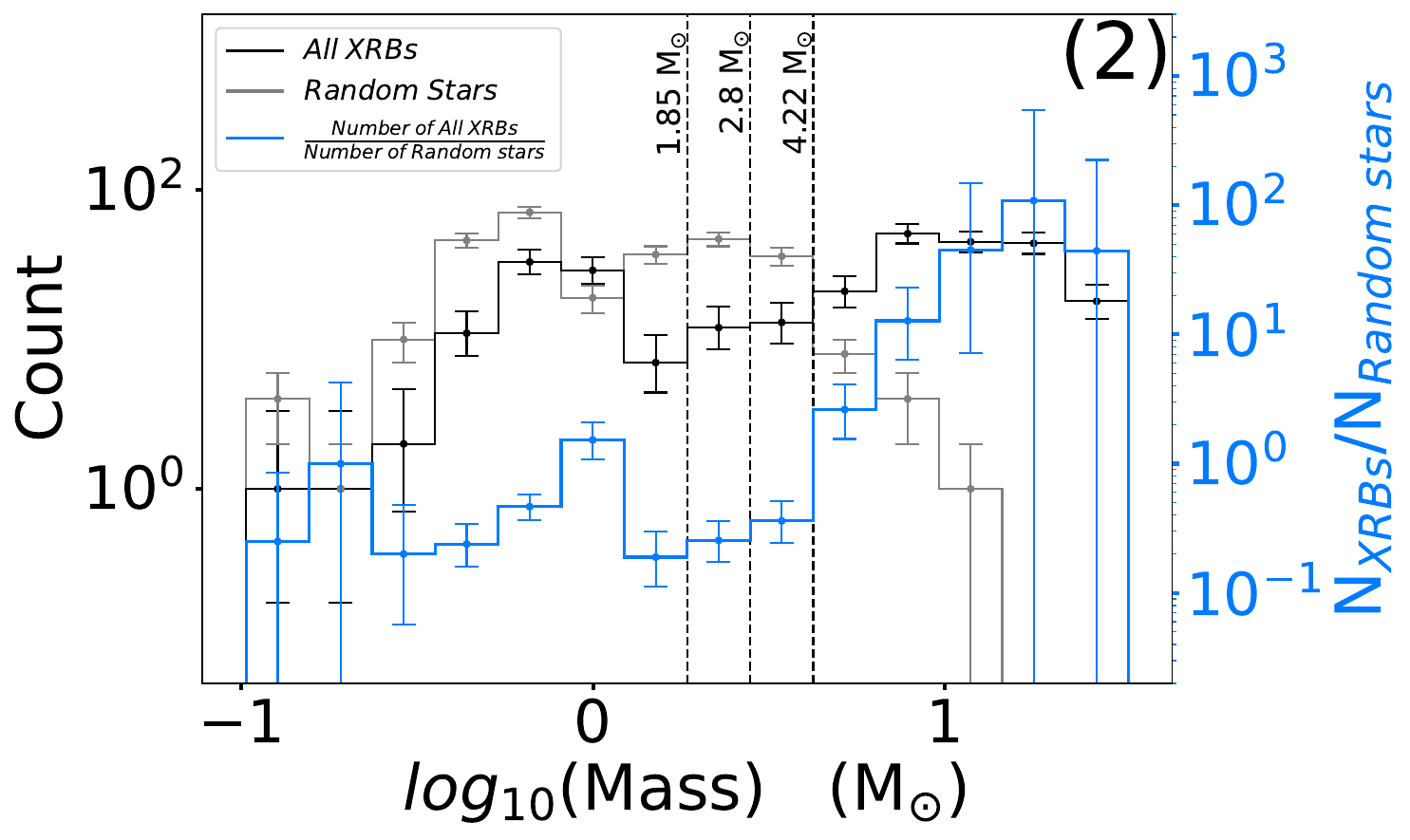}{0.495\textwidth}{}
              }
\vspace{-0.8cm}
\gridline{\fig{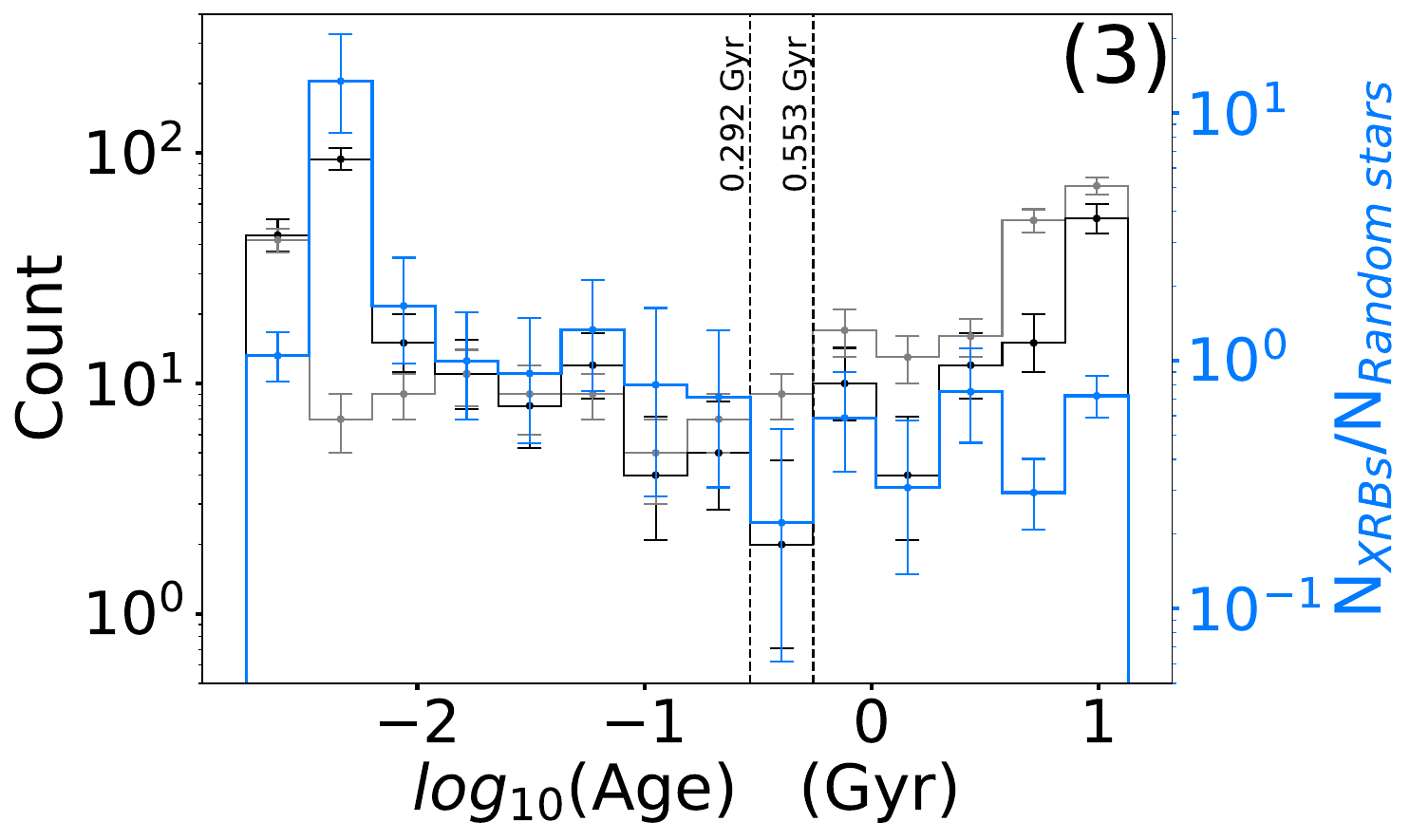}{0.495\textwidth}{}
          \fig{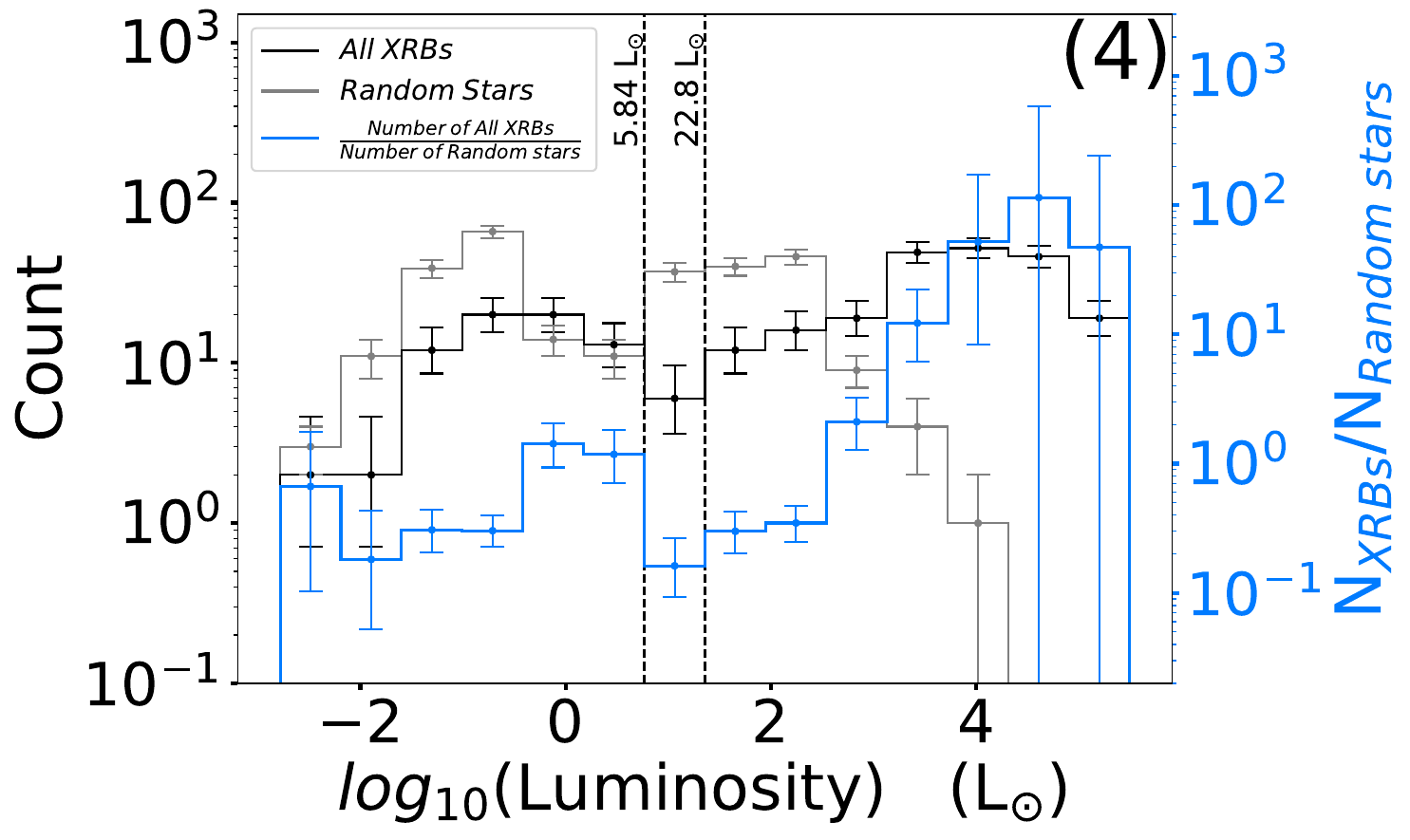}{0.495\textwidth}{}
              }
\vspace{-0.6cm}
\caption{The occurrence ratio distribution of the donor stars in XRBs. The black lines represent the observed distribution of XRBs, while the gray lines represent the distribution of random background stars from the MW and the MCs. The blue lines are the ratio of the black lines to the gray lines, representing the occurrence ratio of XRBs after removing the influence of background stars distribution. \label{fig:dist_intrinsic}}
\end{figure*}

Figure \ref{fig:dist_intrinsic} is the distribution of occurrence rate of XRBs by donor parameters, considering the background stars (random stars around the XRBs) distribution. The black lines represent all XRBs, and the gray lines represent the random background stars. The blue lines, which are the ratio of the black lines to the gray lines, represent the occurrence rate of XRBs. The error bars on the blue line are derived from the error propagation of both the black and gray lines.

In the temperature plot, the blue line (representing the occurrence rate) exhibits a pronounced valley in the middle region. The height difference between the peaks on either side of this valley is more than tenfold, necessitating the use of a logarithmic scale for representation. While the peaks of the black line (the observed number) are relatively similar in height, the peaks in the blue line show a substantial disparity, which is due to differences in the underlying distribution of background stars.

In the distribution plots for mass, age, and luminosity, the valley is less pronounced than in the temperature distribution, but it remains distinct and cannot be ignored. The mass distribution plot also shows a prominent peak for donor masses in the range of 15--22\,$M_{\odot}$.

Given that most of the HMXBs in our sample are located in the MCs, while nearly all LMXBs originate from the MW, this distribution closely reflects the known stellar population differences between the two galaxies. One might therefore be concerned that the observed bimodality—whether in the raw number distribution or in the occurrence rate after correcting for background stars—could merely arise from the distinct stellar populations, rather than representing an intrinsic feature of the XRBs themselves. To address this concern, we examined the MW sample independently, as shown in Figure \ref{fig:CDF_MW} in Appendix \ref{GMM_and_KStest}. The results demonstrate that the MW XRBs alone also exhibit a bimodal distribution.

The occurrence rate is not uniform; XRBs are much more frequent with donor stars of high temperatures, high masses, younger ages, and high luminosities. If we take the mass range of 1.23--1.85\,$M_{\odot}$ as the dividing line, the occurrence rate of HMXBs is over 50 times that of LMXBs. While we do not know the mass distribution of the progenitor stars that form XRBs, our results indicate that XRBs are intrinsically more likely to feature high-mass donor stars. This conclusion does not contradict the observation of numerous LMXBs, as these systems are found almost exclusively in the Milky Way, a galaxy composed of a vast population of low-mass stars.

Similar to our testing on the observed donor mass distribution, we used a Monte Carlo method to check if the occurrence rate distribution is robust against huge parameter errors. The results, shown in Panel~2 of Figure~\ref{fig:distribution_random} in Appendix~\ref{distribution_random}, demonstrate that the simulated distributions closely match the original data.

The distribution of occurrence rate gives us stronger confidence that the valley is an intrinsic feature of the XRBs population itself, and not an artifact of observational selection effects. In the following section, we will provide a physical explanation for this valley.

\section{The relationship among parameters and the parallel tracks}

\subsection{The relationship among parameters of donor stars}

\begin{figure*}[h]
\gridline{\fig{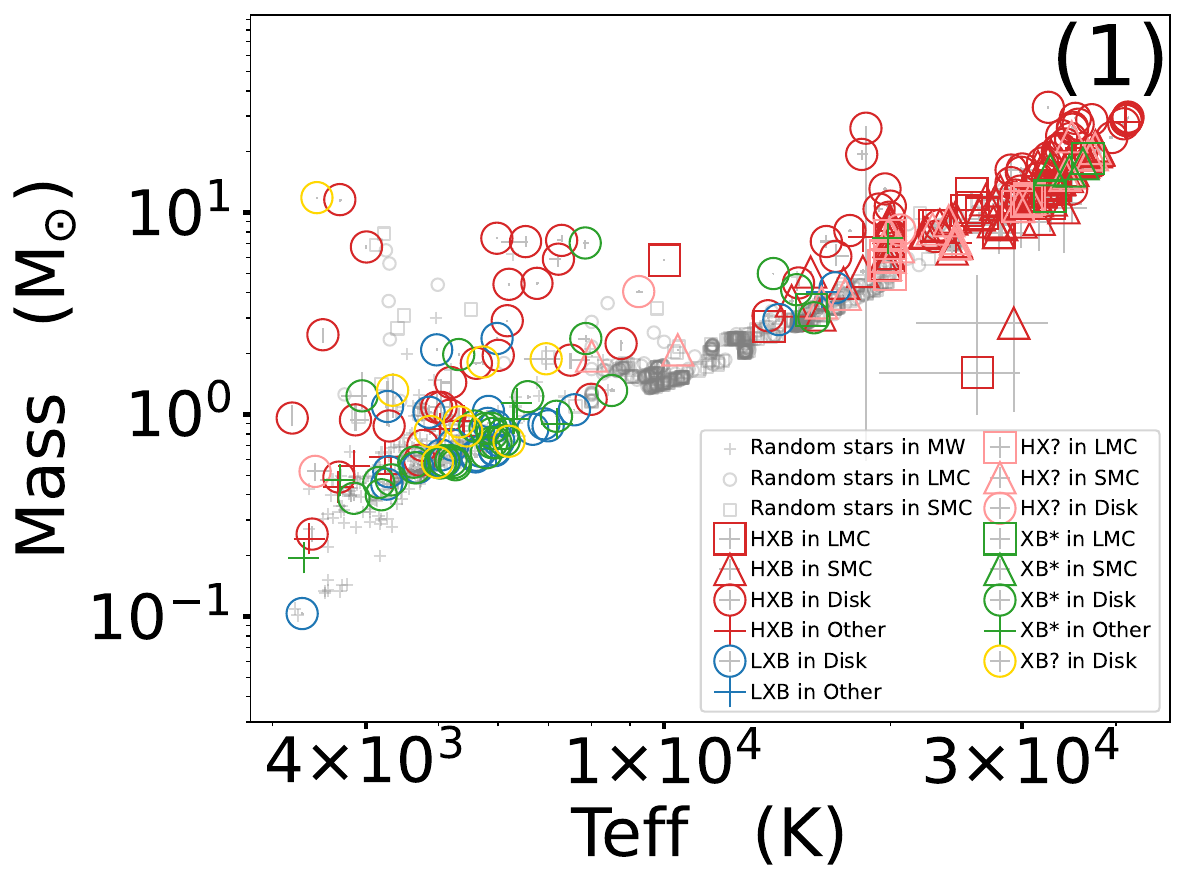}{0.33\textwidth}{}
          \fig{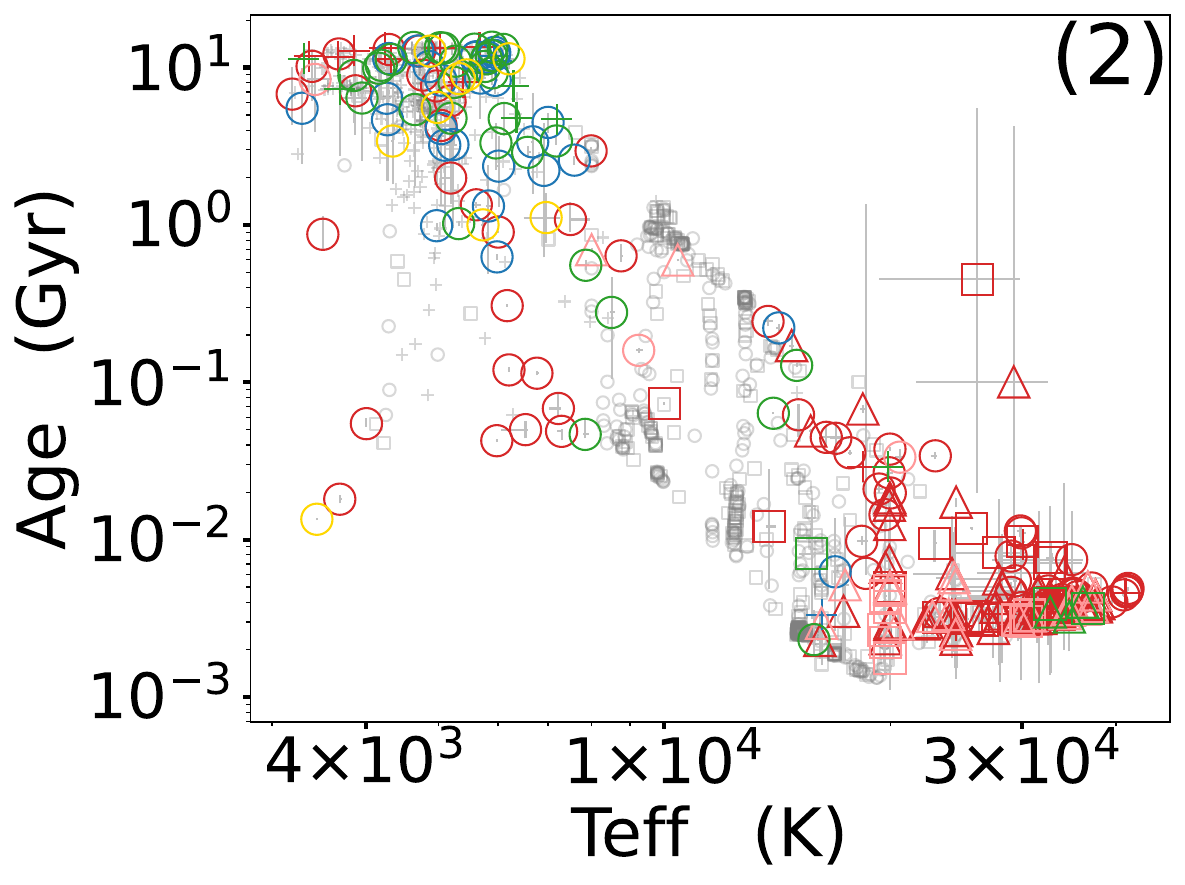}{0.33\textwidth}{}
          \fig{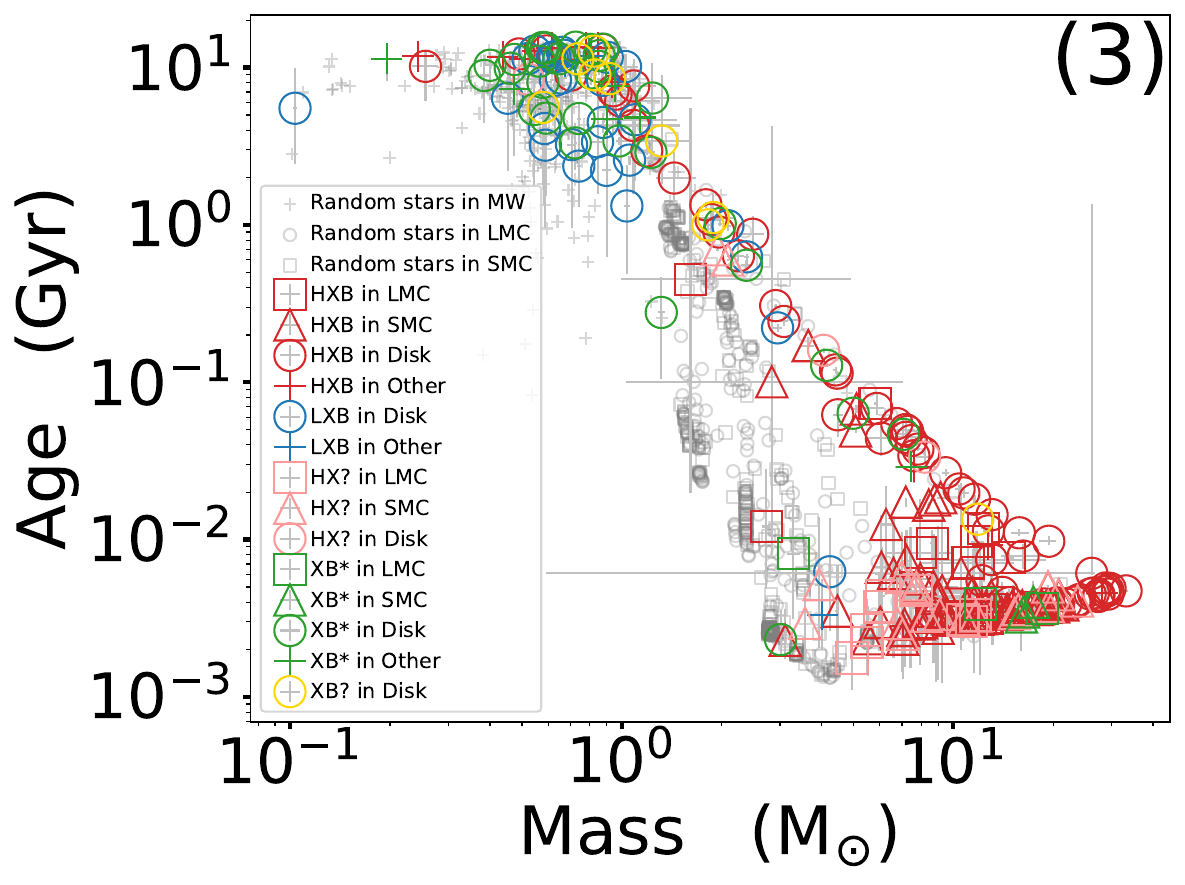}{0.33\textwidth}{}
              }
\vspace{-0.8cm}
\gridline{\fig{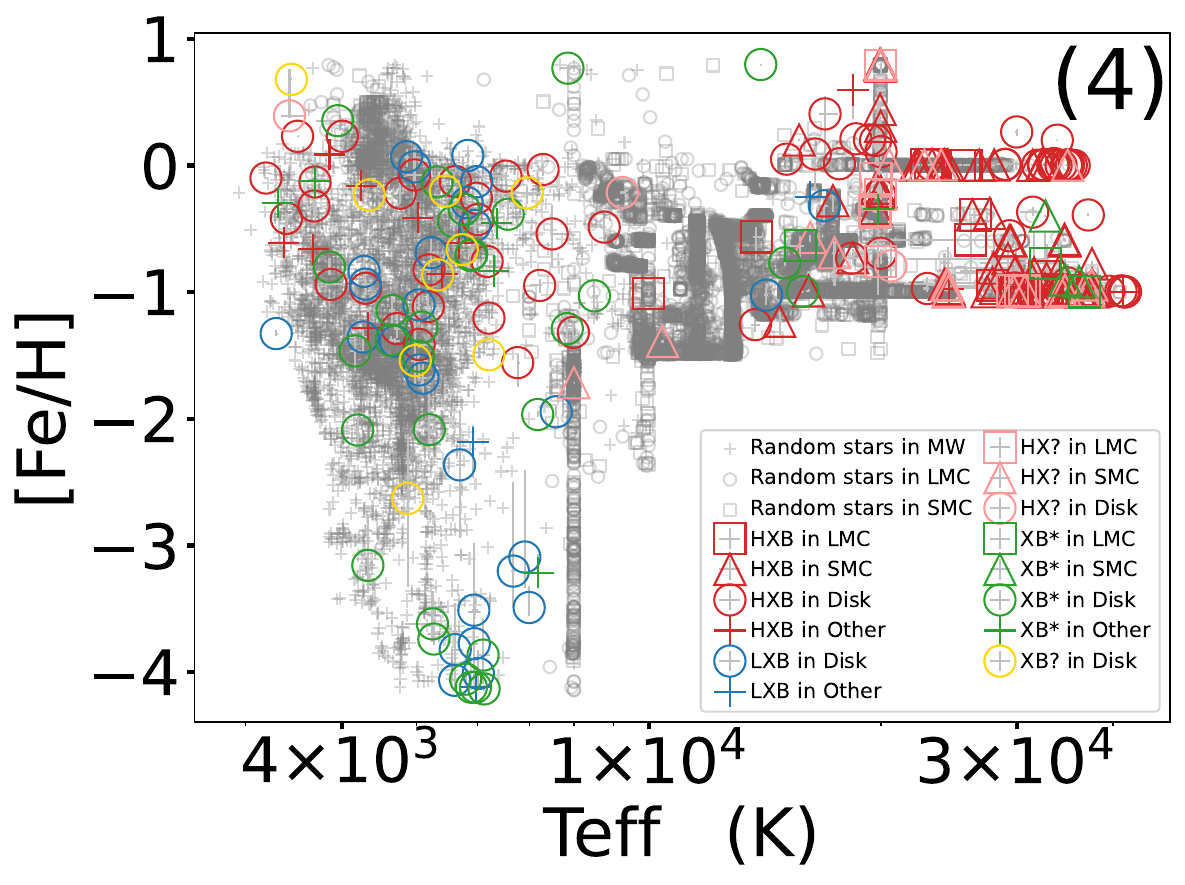}{0.33\textwidth}{}
          \fig{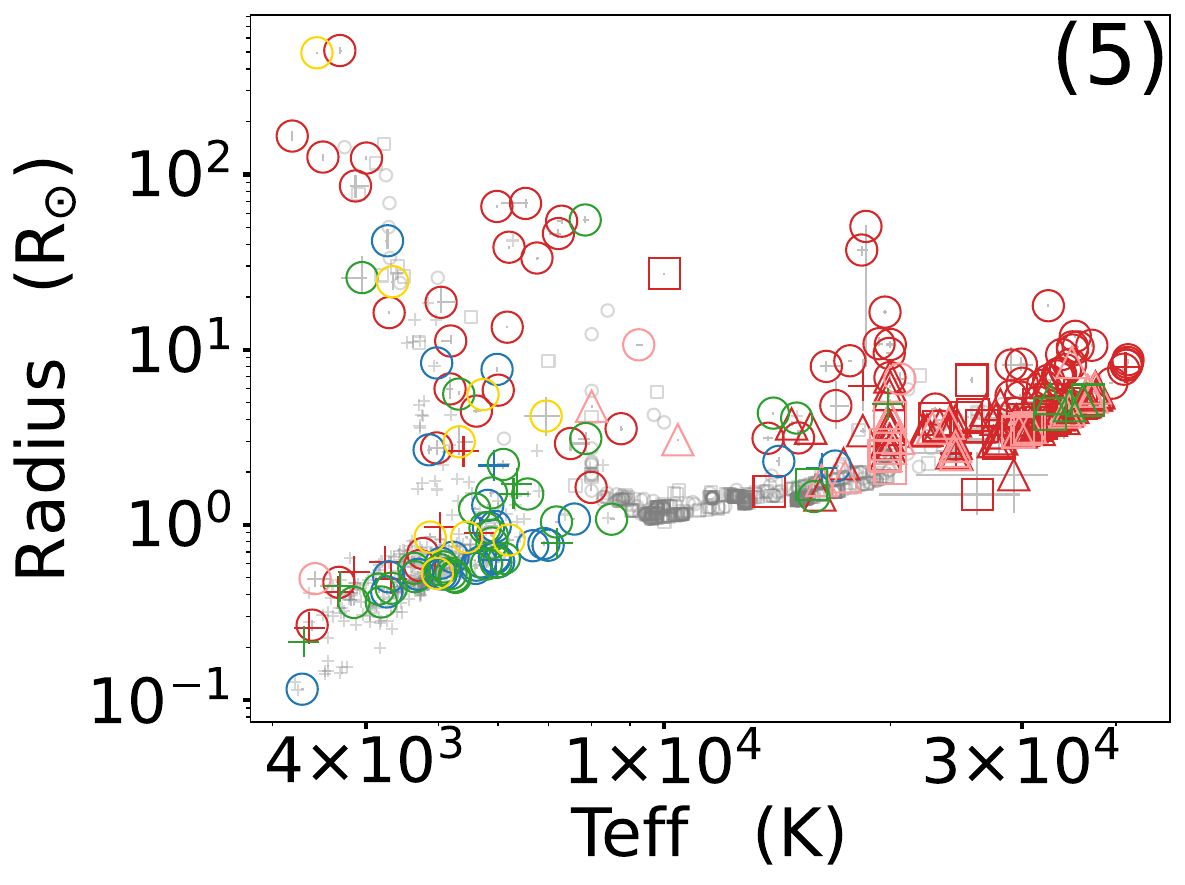}{0.33\textwidth}{}
          \fig{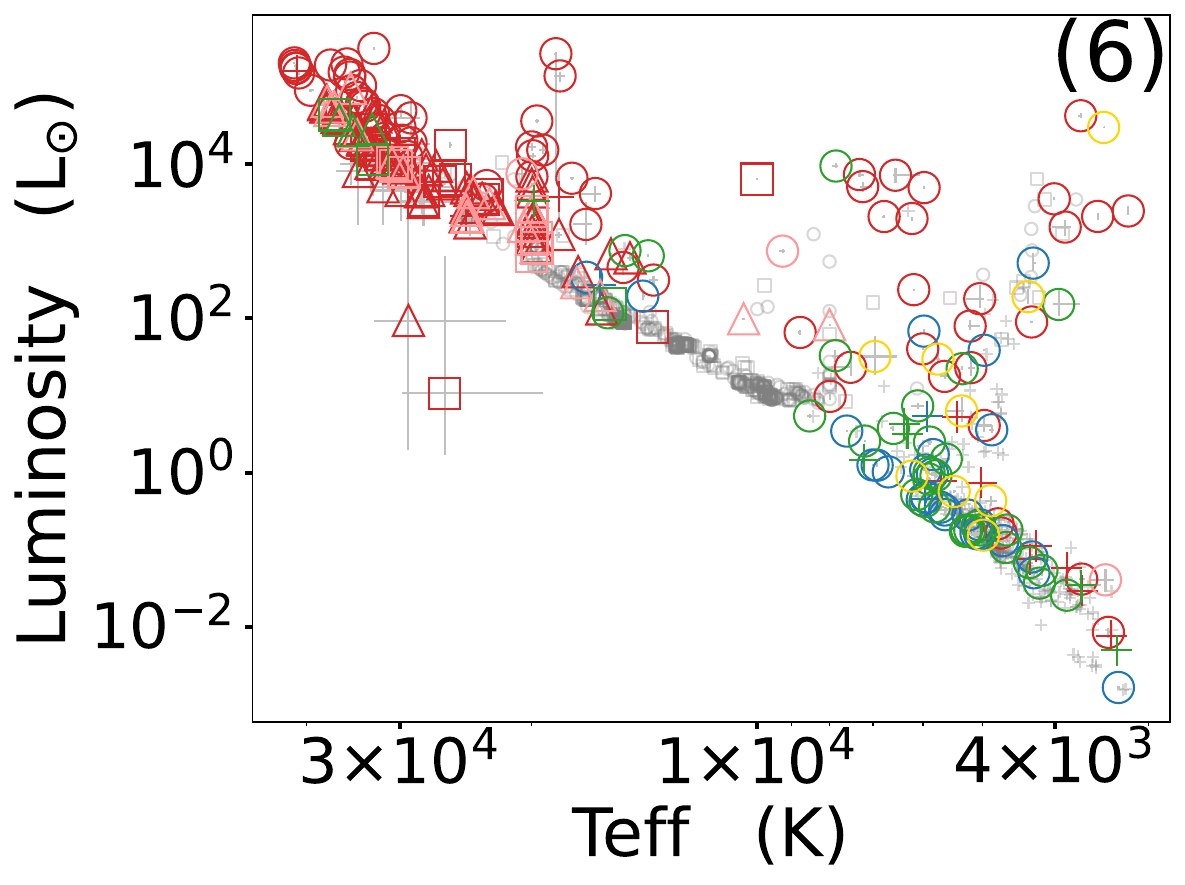}{0.33\textwidth}{}
              }
\vspace{-0.8cm}
\gridline{\fig{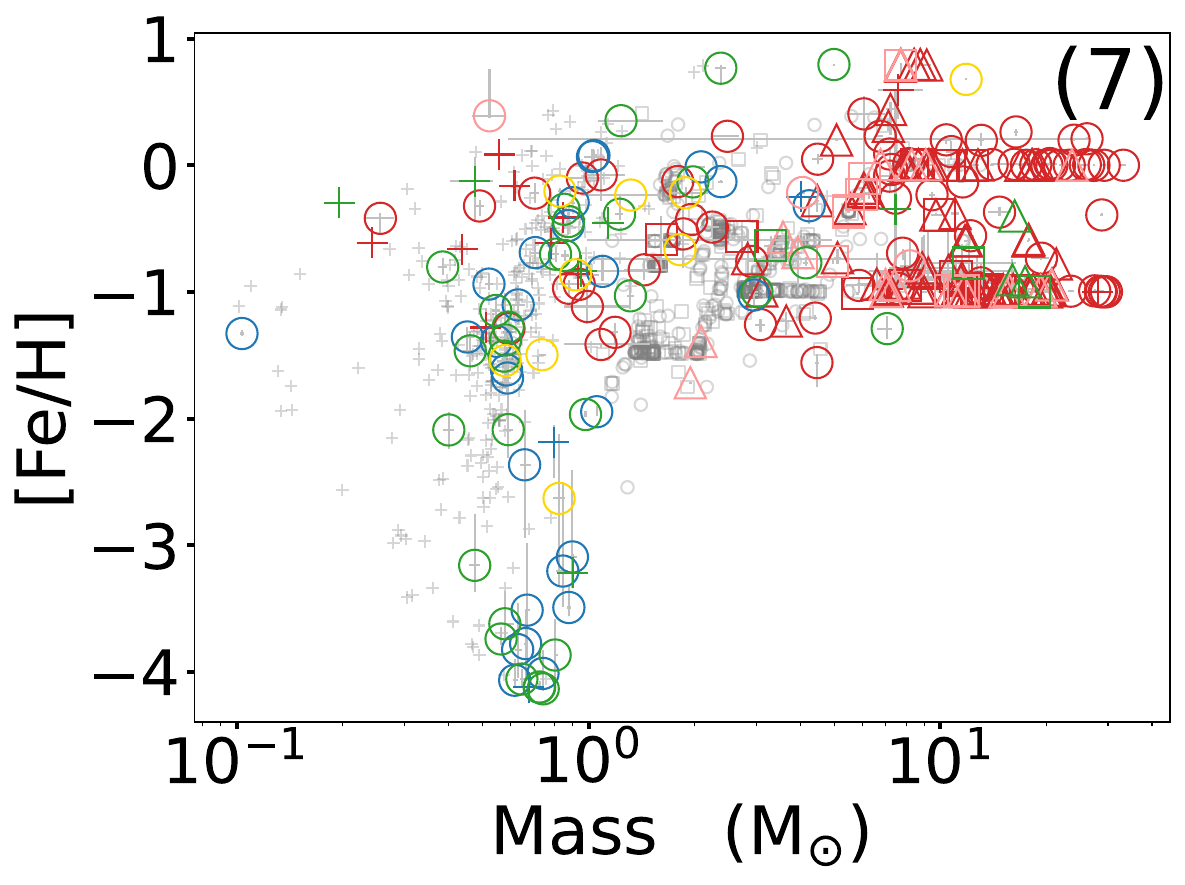}{0.33\textwidth}{}
          \fig{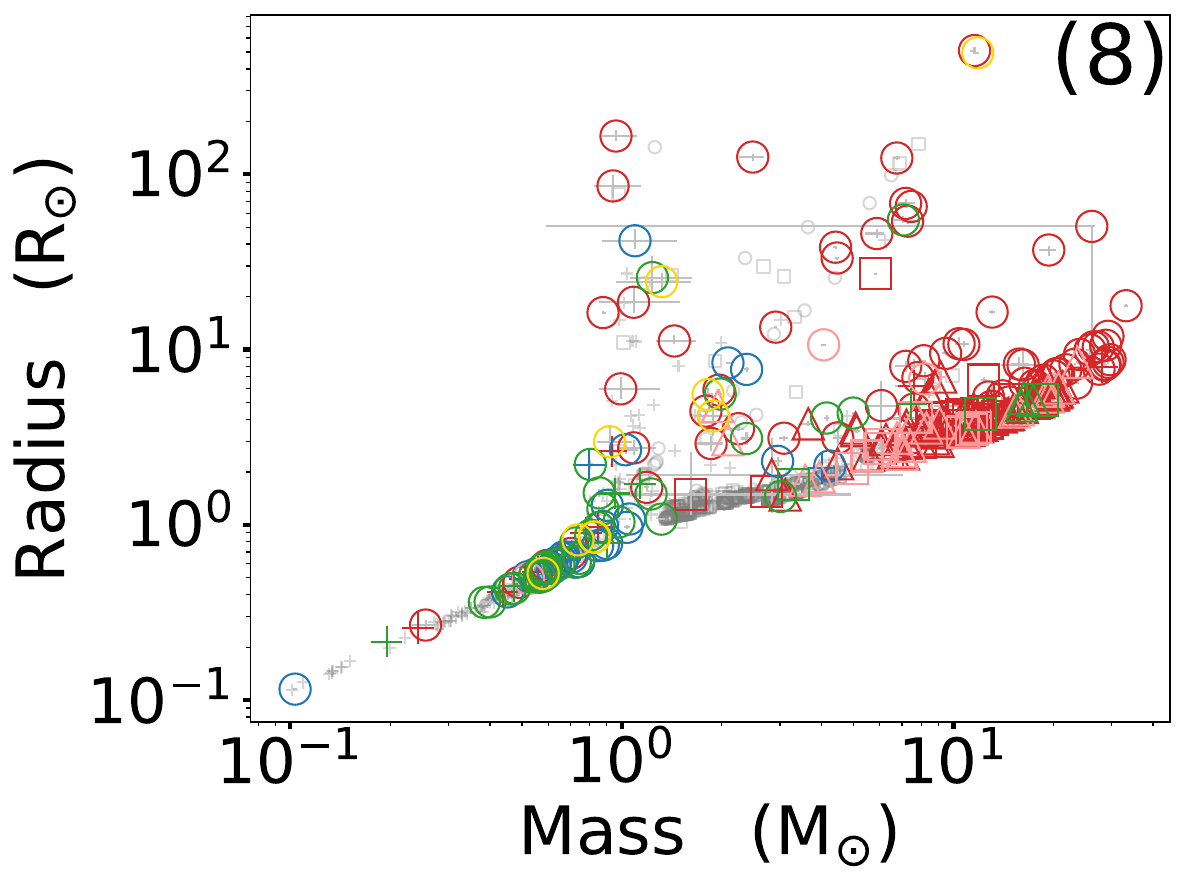}{0.33\textwidth}{}
          \fig{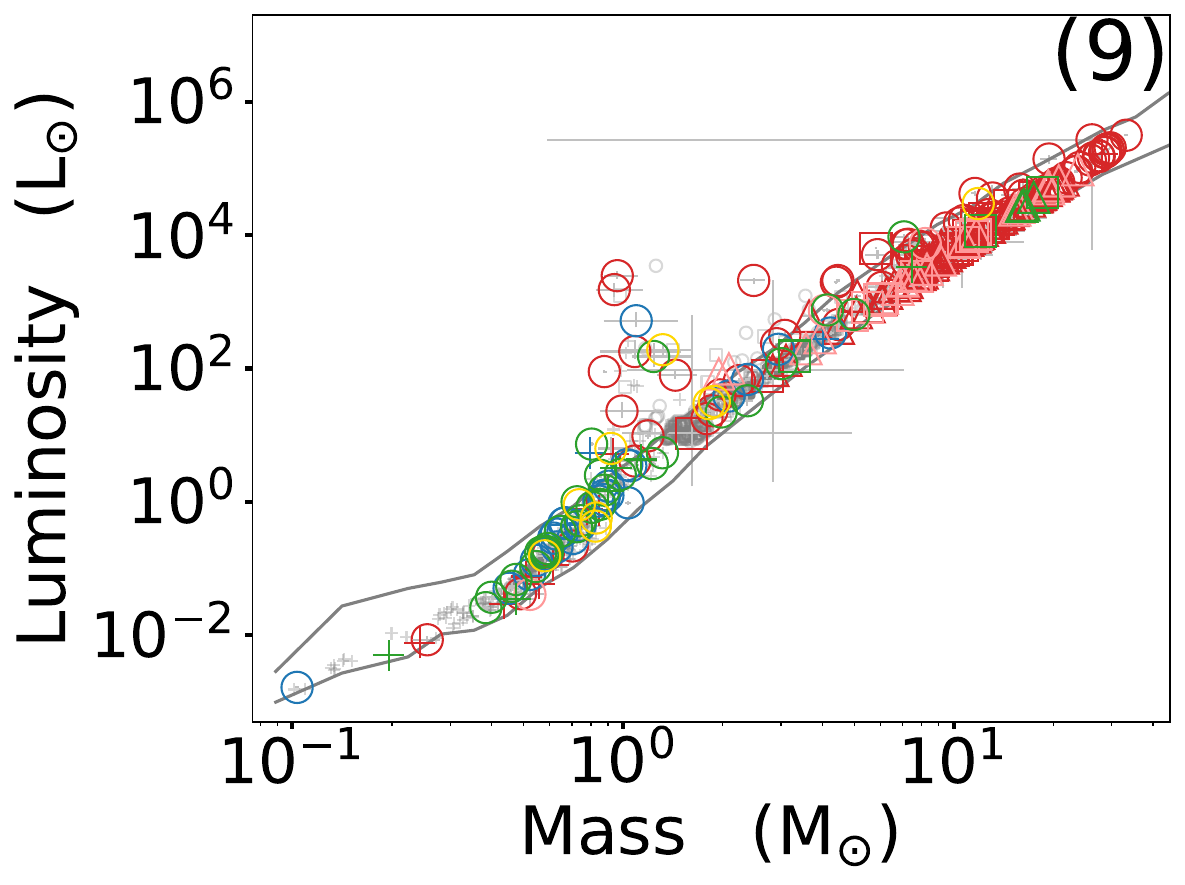}{0.33\textwidth}{}
             }
\vspace{-0.6cm}
\caption{Parameters relationship for the donor stars in XRBs. Different colors represent different types of XRBs from literature. Different colors represent different types of XRBs, that are originate from the original literature, as described in Section \ref{catalogs}. Different shapes represent different locations, LMC and SMC refer to the Large and Small Magellanic Clouds, respectively. Disk and Other denote the Galactic disk direction (-15 $<$ b $<$ 15) and other directions (halo), respectively. Two gray lines in panel 9 enclose the region of main-sequence stars. \label{fig:relation}}
\end{figure*}

The panels 1-3 in Figure \ref{fig:relation} depict the relationships between temperature, mass, and age. It can be observed that there is a strong correlation between the parameters, which is essentially the main sequence relation. Since most donor stars are main sequence stars, the correlation between the parameters is strong. Panel 6 is the Hertzsprung-Russell diagram, from which the main sequence and scattered post-main-sequence stars can be clearly seen.

The different colors of points in Figure \ref{fig:relation} represent different types of XRBs. These types come from the original literature. It can be seen that blue points (LMXBs) are indeed more inclined to appear on the side of low-temperature and low-mass, and the red points (HMXBs) are more inclined to the other side. Of course, this is because they are primarily main sequence stars.

It can be observed that many red points also appear on the low-temperature, low-mass side, and a few blue points are on the high-temperature, high-mass side. Since the colors representing HMXBs and LMXBs classifications are from the original literature, rather than based on the classification in this paper, some population mixing is inevitable. If we use the mass of 3 $M_{\odot}$ as the boundary line to distinguish HMXBs from LMXBs (see the valley position shown in Section \ref{sec:distribution}), the agreement rate between this classification and those from previous studies is about 86\%.

The panels 4-6 in Figure \ref{fig:relation} show the relationship between temperature and other three parameters (metallicity, radius, and luminosity), and in panel 7-9, temperature is substituted with stellar mass.

In Panel 4, the lower right region shows no data points. This blank area occurs because the atmospheric parameter measurement technique fails in that parameter range. Nevertheless, we argue that the blank region is expected to contain very few stars, insufficient to alter the observed distributions. See Appendix \ref{absence_on_highT_low_metal} for more detailed explanation.

In panel 9, we can observe a tight linear relationship between mass and luminosity (in logarithm). This relationship is the mass-luminosity relation for main-sequence stars, and the scattered points above the linear relationship are post-main-sequence stars. The two gray lines in panel 9 enclose the region where main-sequence stars are located, noting that post-main-sequence stars may also appear in this region.

The gray points in Figure \ref{fig:relation} represent random stars used only for comparison with XRBs. It can be observed that the gray points from the MW cover roughly the same region as the local XRBs. However, the gray points from the MCs noticeably deviate from their local XRBs, primarily filling the intermediate blank regions.

In panel 4, the gray points are abundant due to their direct source from \textit{Gaia}. Noticeable straight lines features are present, attributed to discontinuities in the spectral fitting templates. Fundamentally, this arises from the challenge of obtaining atmospheric parameters in regions of high temperature and low metallicity. Hence, XRBs in high-temperature regions exhibit larger parameter uncertainties compared to those in low-temperature regions.

\begin{figure*}[h]
\gridline{\fig{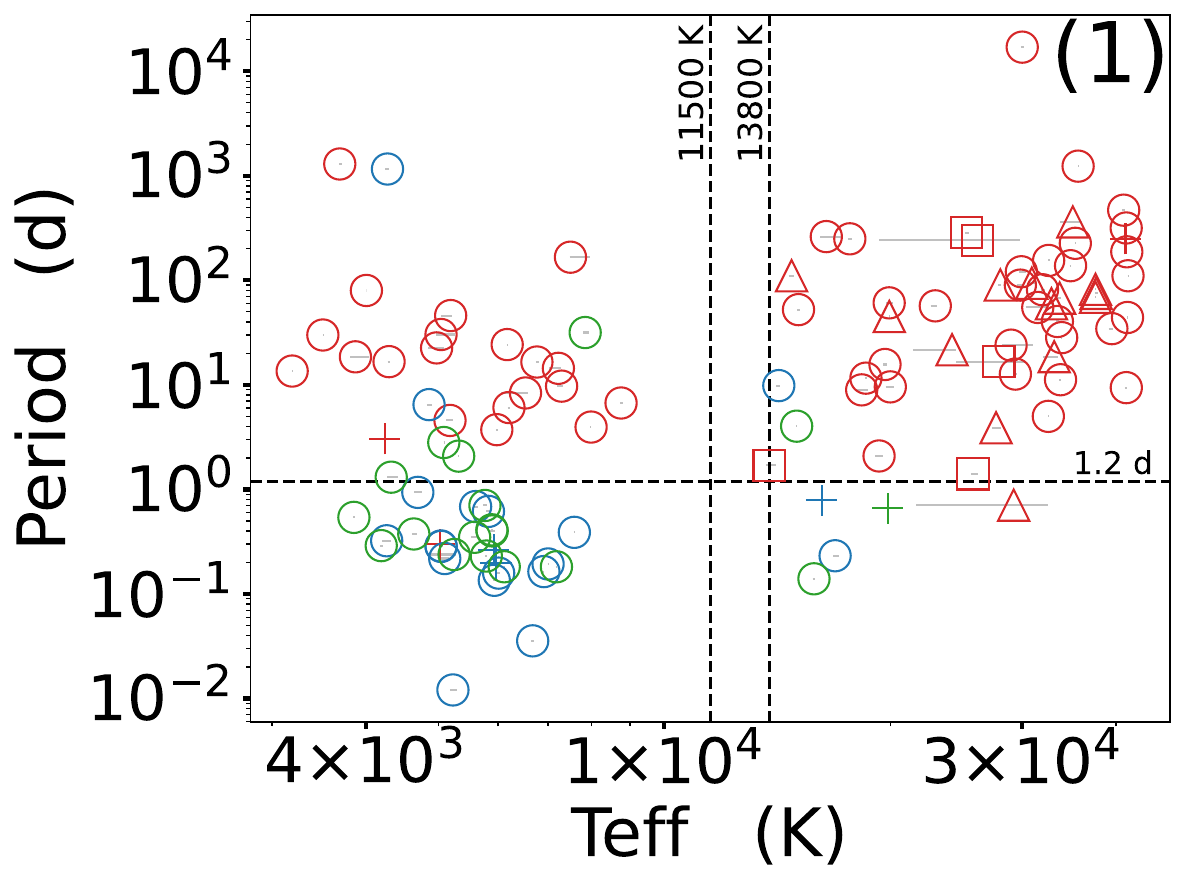}{0.33\textwidth}{}
          \fig{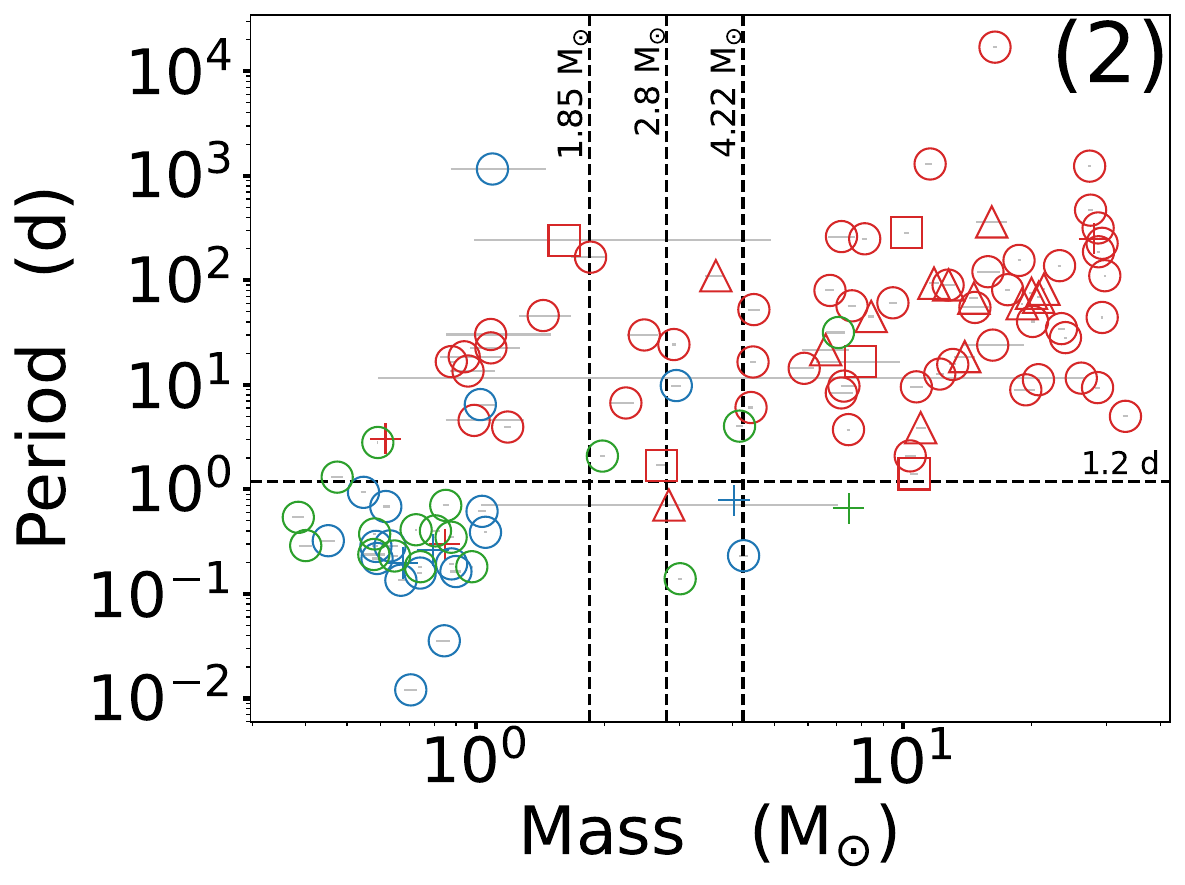}{0.33\textwidth}{}
          \fig{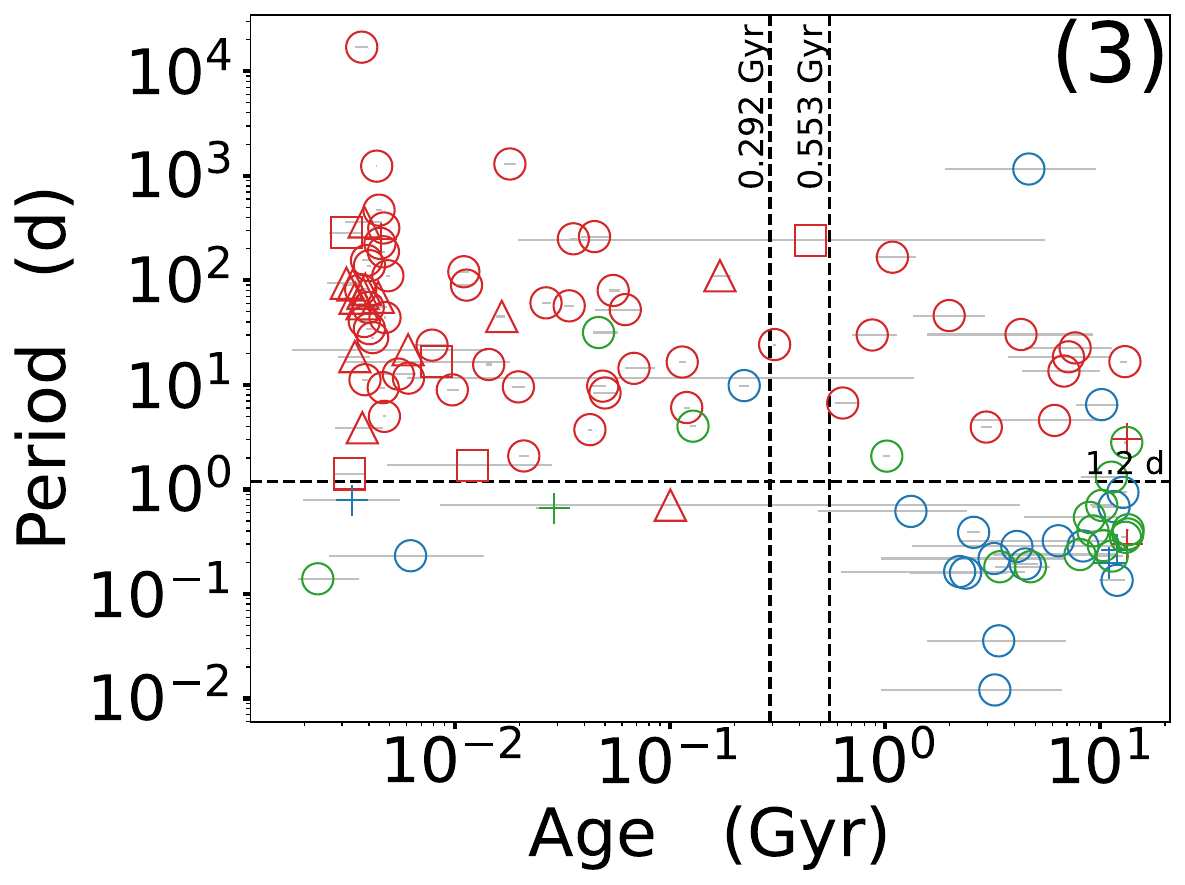}{0.33\textwidth}{}
             }
\vspace{-0.8cm}
\gridline{\fig{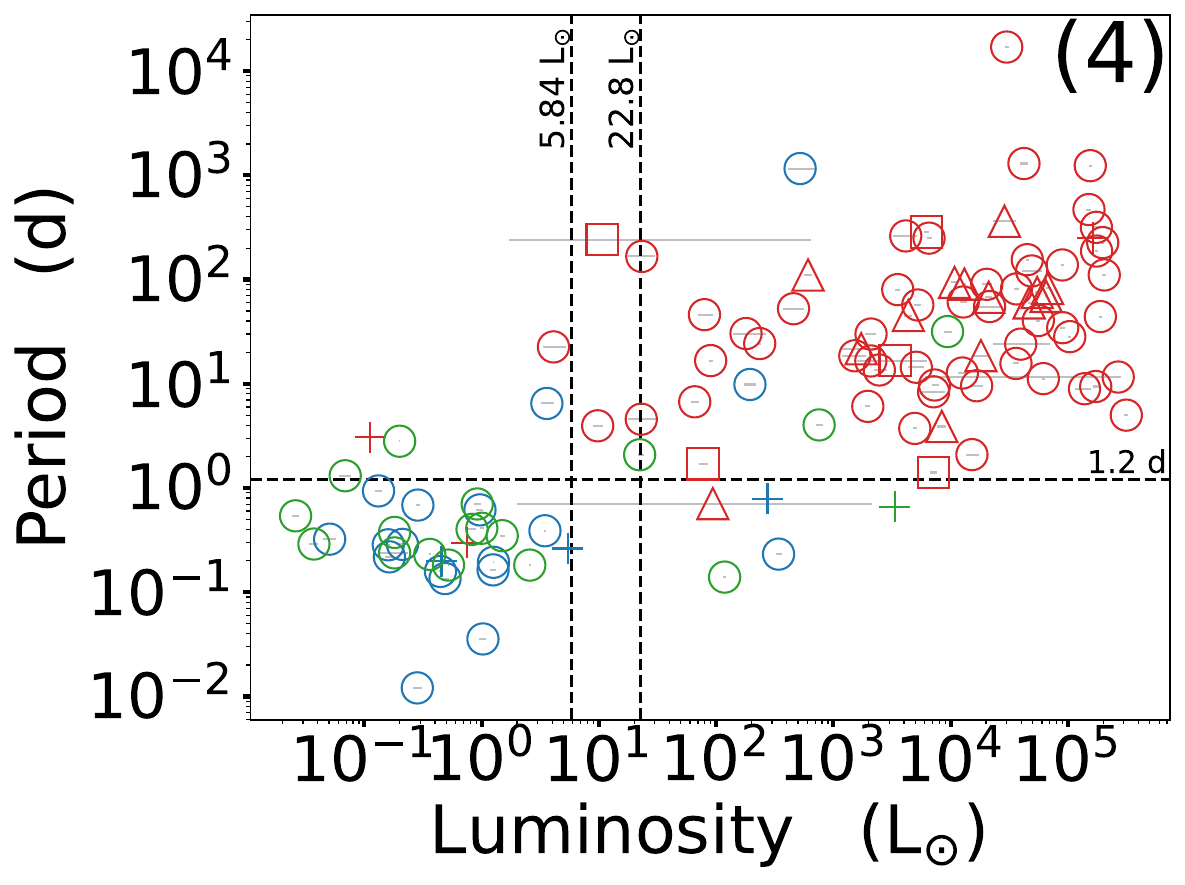}{0.33\textwidth}{}
          \fig{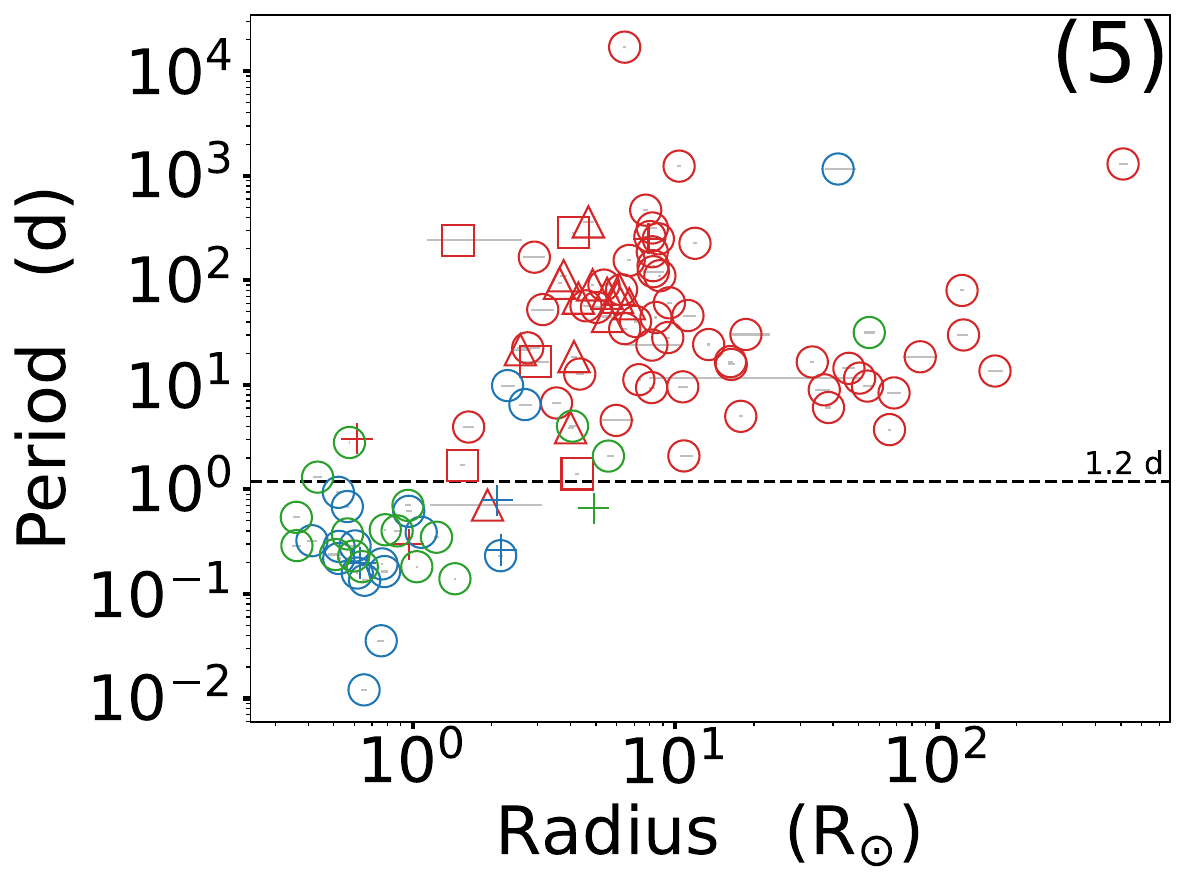}{0.33\textwidth}{}
          \fig{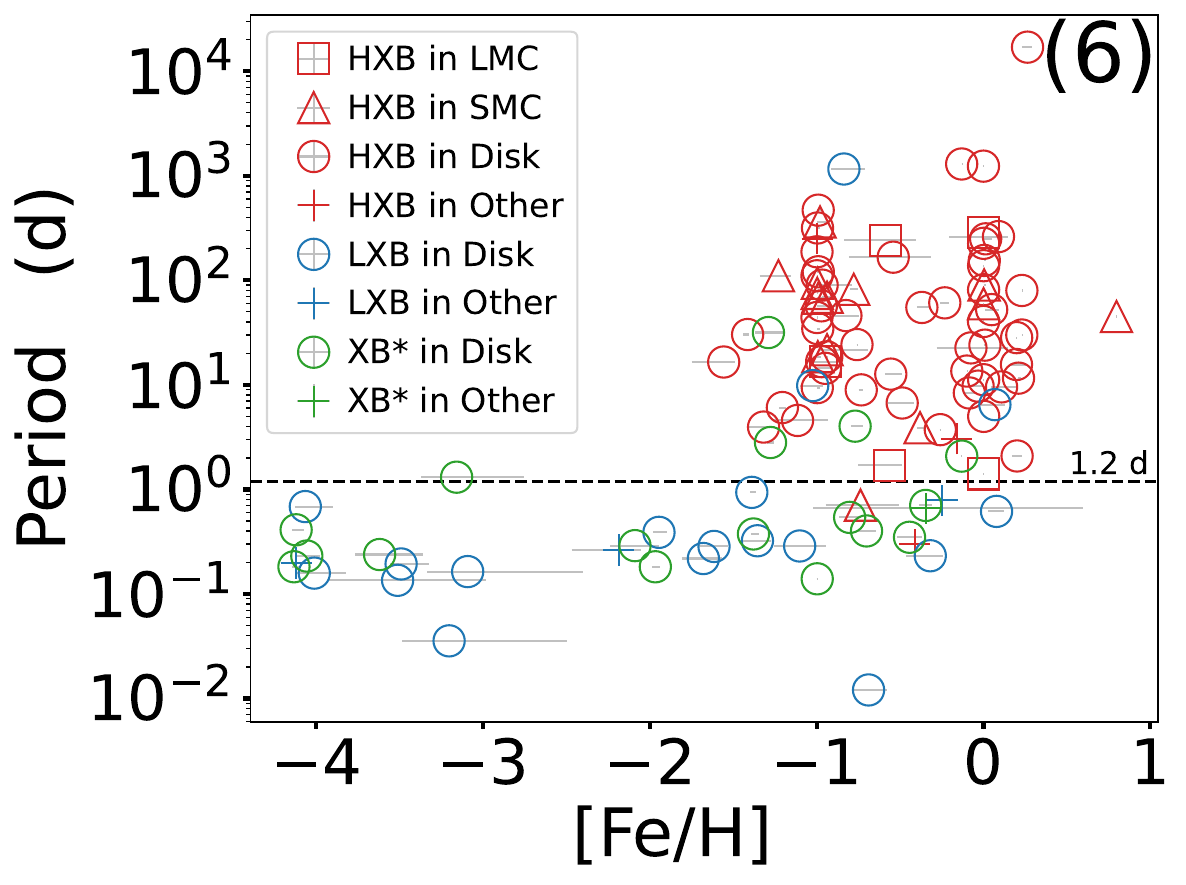}{0.33\textwidth}{}
             }
\vspace{-0.6cm}
\caption{Same as Fig. \ref{fig:relation} but for period and other parameters. \label{fig:relation2}}
\end{figure*}

Figure \ref{fig:relation2} illustrates the relationship between binary period and other parameters. The period parameters are sourced from \citet{2023A&A...671A.149F}, \citet{2023A&A...675A.199A}, and \citet{2023A&A...677A.134N}. It can be observed that there is a clear positive correlation between period and mass, radius, and luminosity. Interestingly, the periods correspond well to the classifications provided in previous literature, with 1.2 days serving as an effective dividing line to separate the two types. We also marked the valley positions derived from Figure \ref{fig:dist_intrinsic}.

\subsection{Parallel tracks in the main sequence region}\label{Parallel_tracks}

\begin{figure*}[h]
\gridline{\fig{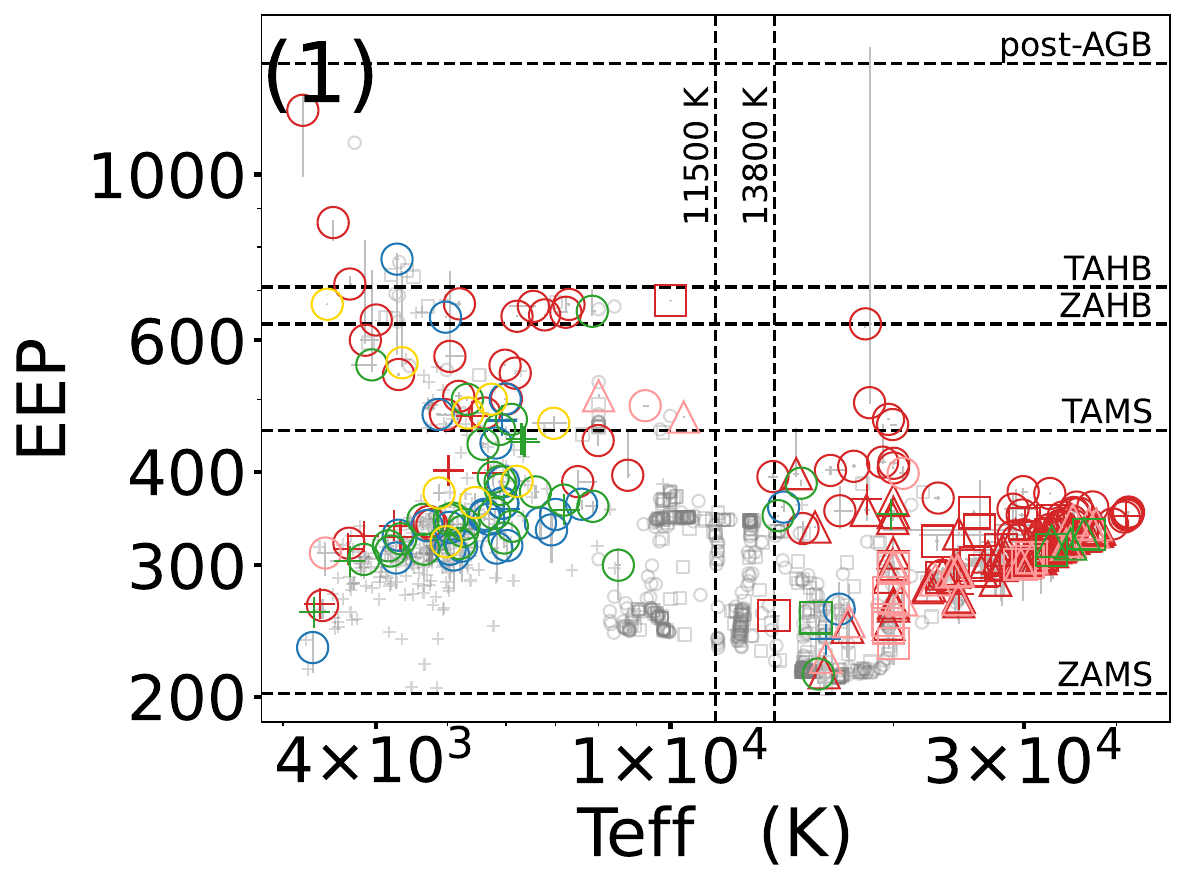}{0.495\textwidth}{}
          \fig{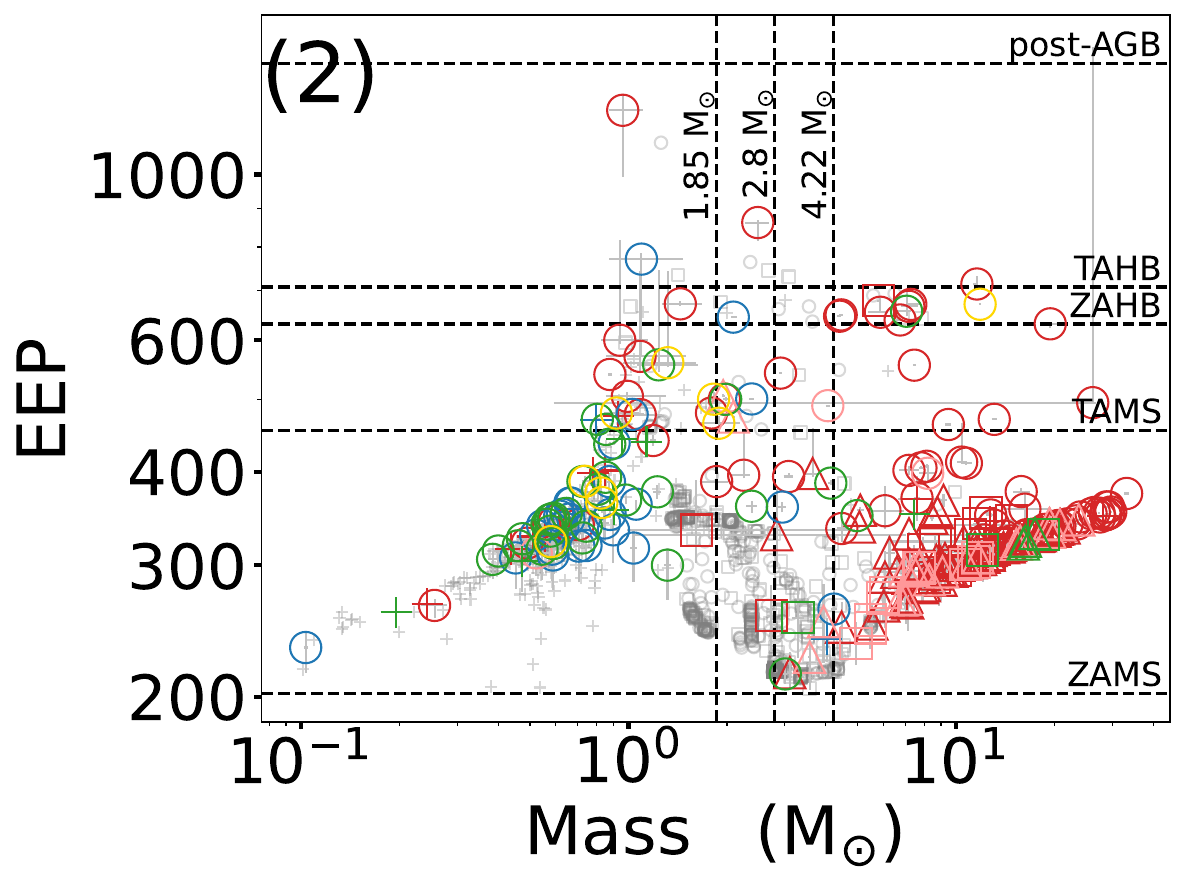}{0.495\textwidth}{}
             }
\vspace{-0.8cm}
\gridline{
          \fig{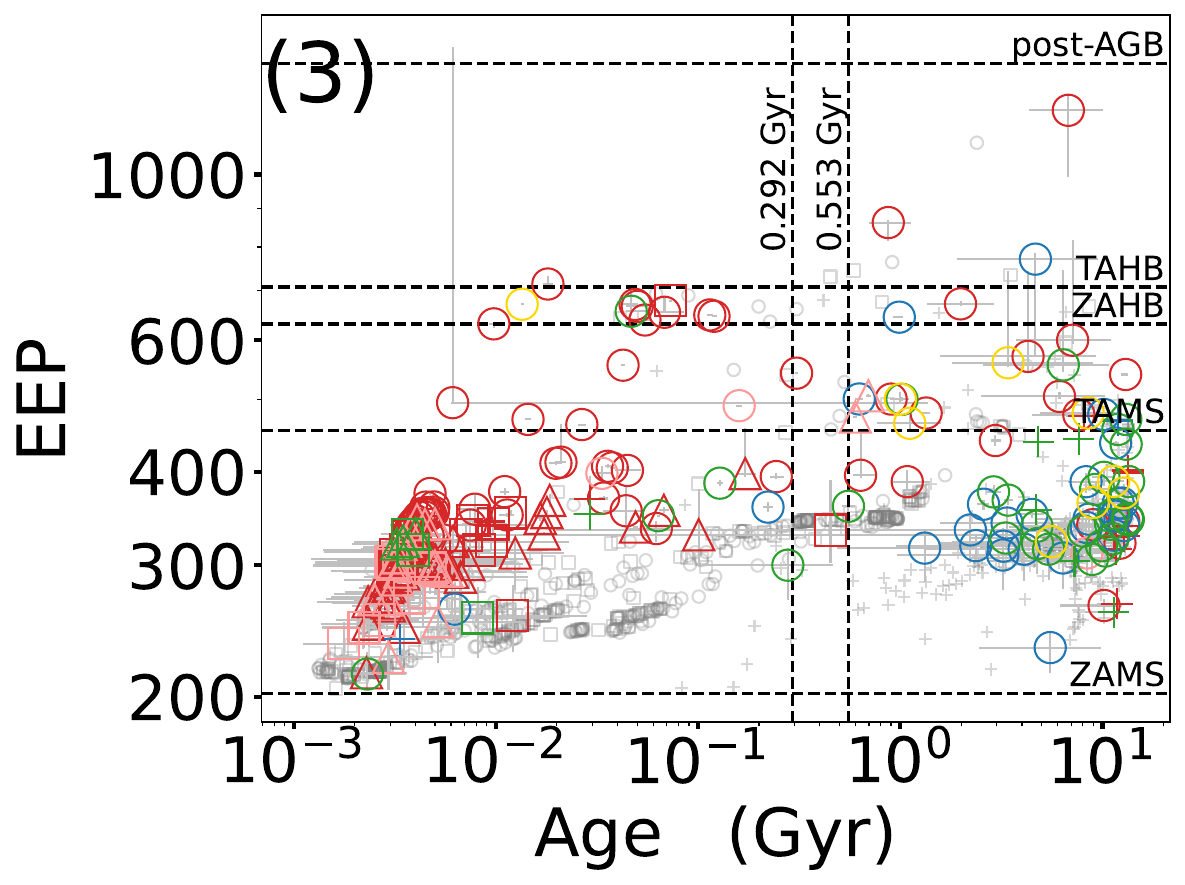}{0.495\textwidth}{}
          \fig{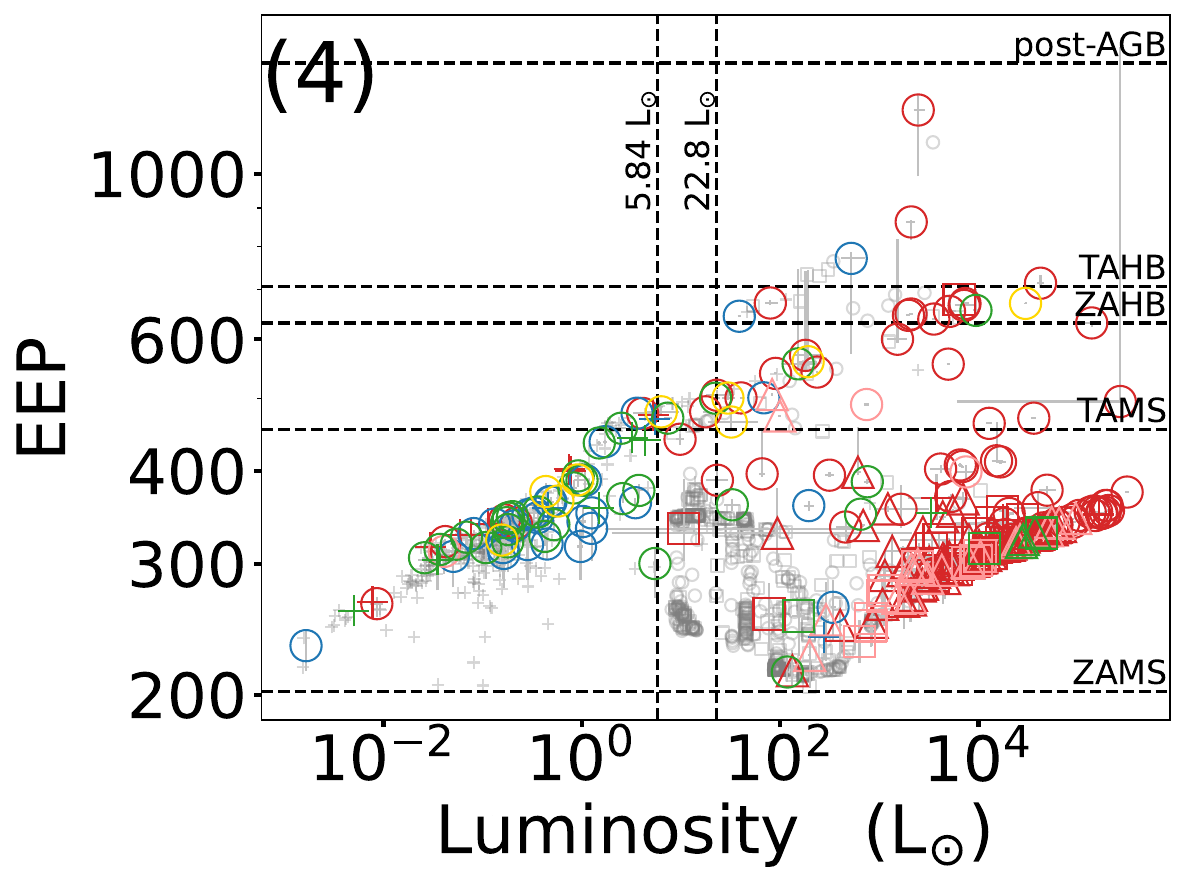}{0.495\textwidth}{}
             }
\vspace{-0.8cm}
\gridline{
          \fig{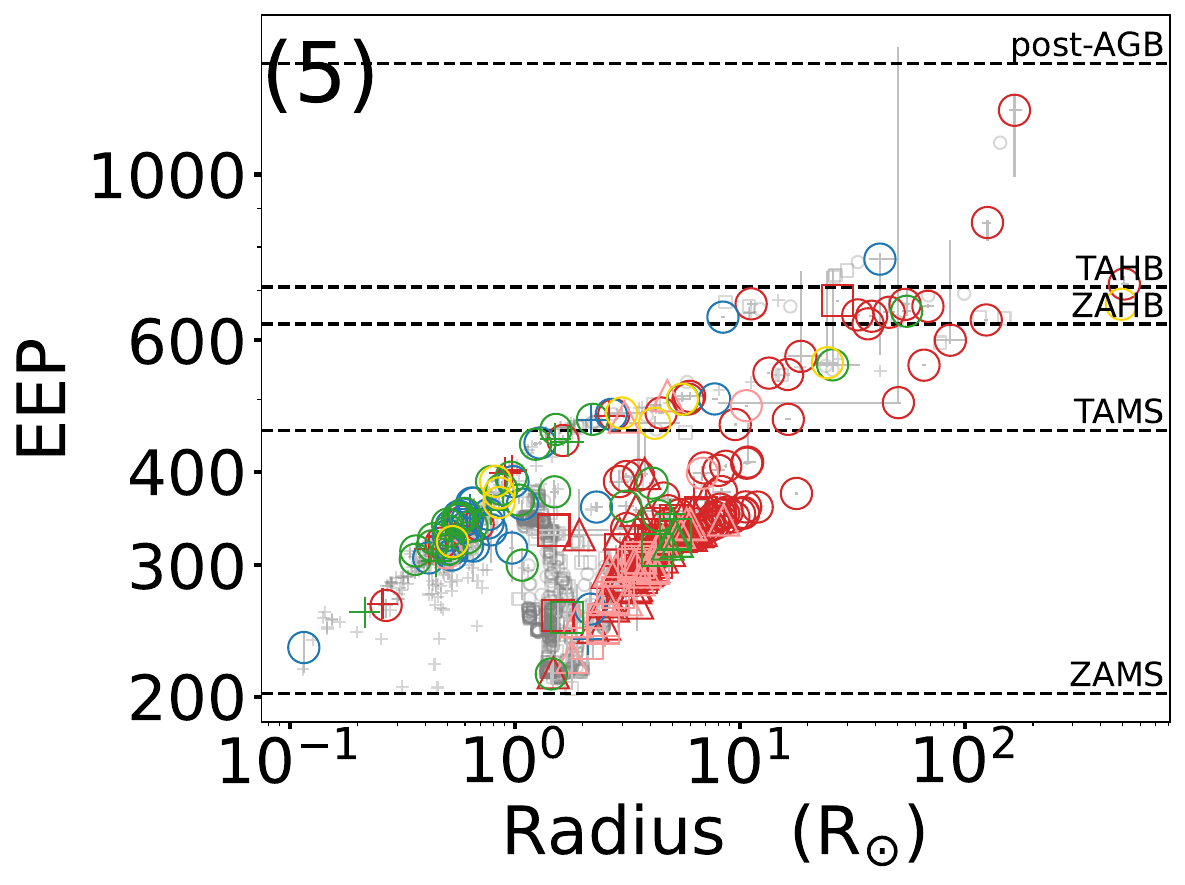}{0.495\textwidth}{}
          \fig{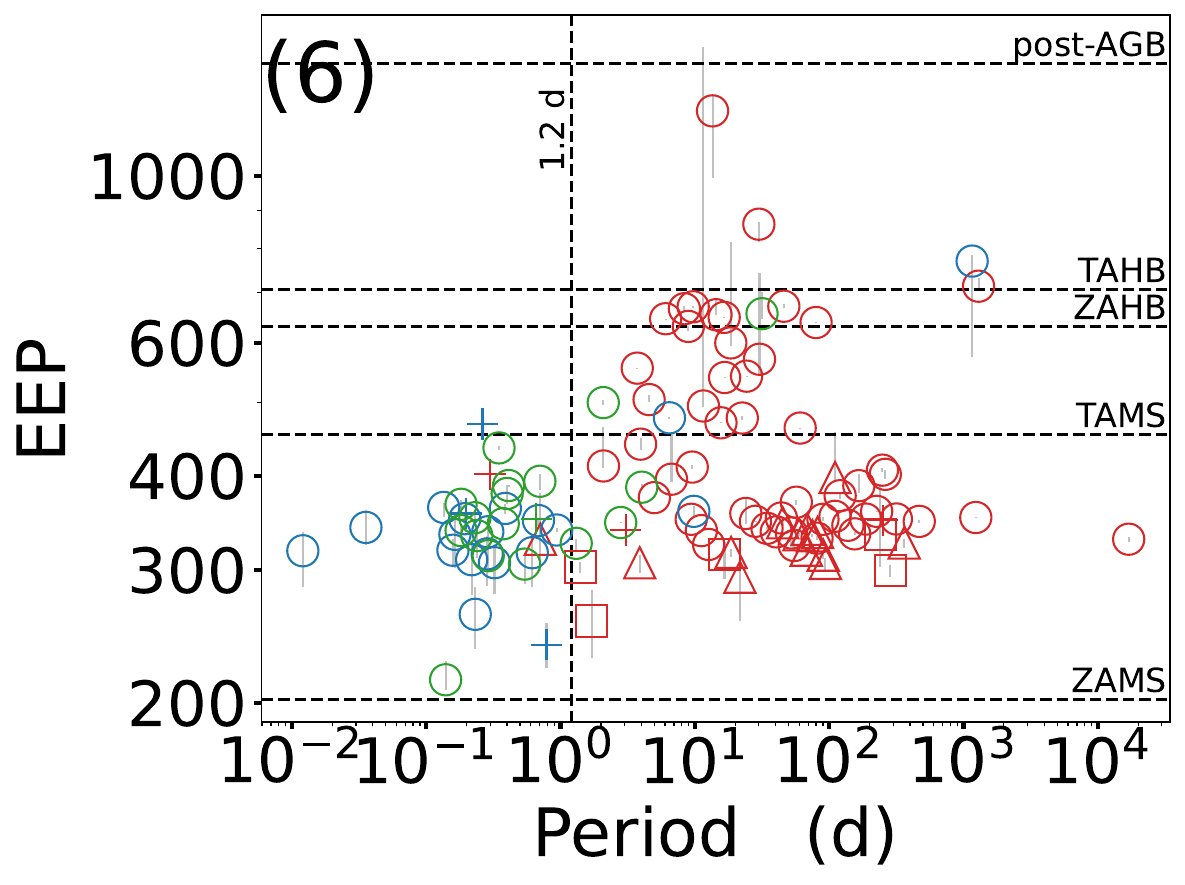}{0.495\textwidth}{}
             }
\vspace{-0.6cm}
\caption{Same as Fig. \ref{fig:relation} but for EEP and other parameters. On the horizontal lines, ZAMS, TAMS, ZAHB, TAHB, and post-AGB represent zero-age main sequence (202), terminal-age main sequence (454), zero-age horizontal branch (631), terminal-age horizontal branch (707), and post-asymptotic giant branch (1409), respectively. \label{fig:relation3}}
\end{figure*}

Figure \ref{fig:relation3} show the relation between EEPs and other donor parameters. We overlaid the previously obtained valley positions and marked several critical EEPs values on the plots.

In the panels 1, 2, 4, and 5, two parallel tracks in the main sequence region can be seen. The most distinct parallel tracks appear in the luminosity and radius plots (panel 4 and 5). These two tracks generally fall on either side of the valley positions. As shown in above figures, the random stars (gray points) as background were also shown in Figure \ref{fig:relation3}. Random stars from the MW generally overlap with local XRBs, but those from the MCs show clear deviation from local XRBs and exhibit irregular patterns.

The two parallel tracks phenomenon naturally explains the earlier observed valley in distribution. Around the valley, the evolutionary stages are either post-main sequence or pre-main sequence. The number of non-main sequence stars is evidently much lower, resulting in a reduced quantity near the dividing lines. These parallel tracks align well with, and so strongly supports, the above valley in distribution.

\subsection{Locations of XRBs} \label{subsec:Location}

\begin{figure*}[h]
\gridline{\fig{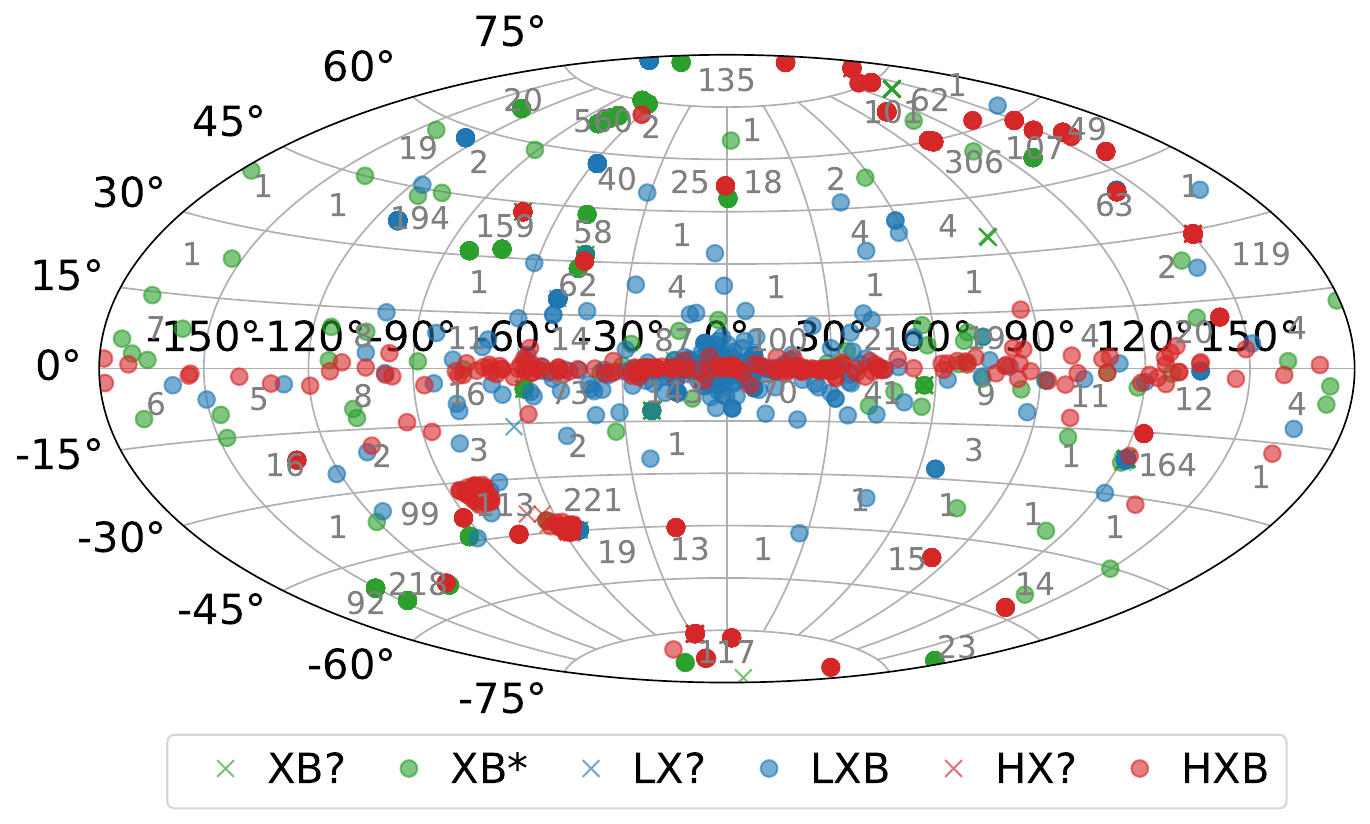}{0.5\textwidth}{(1)}
          \fig{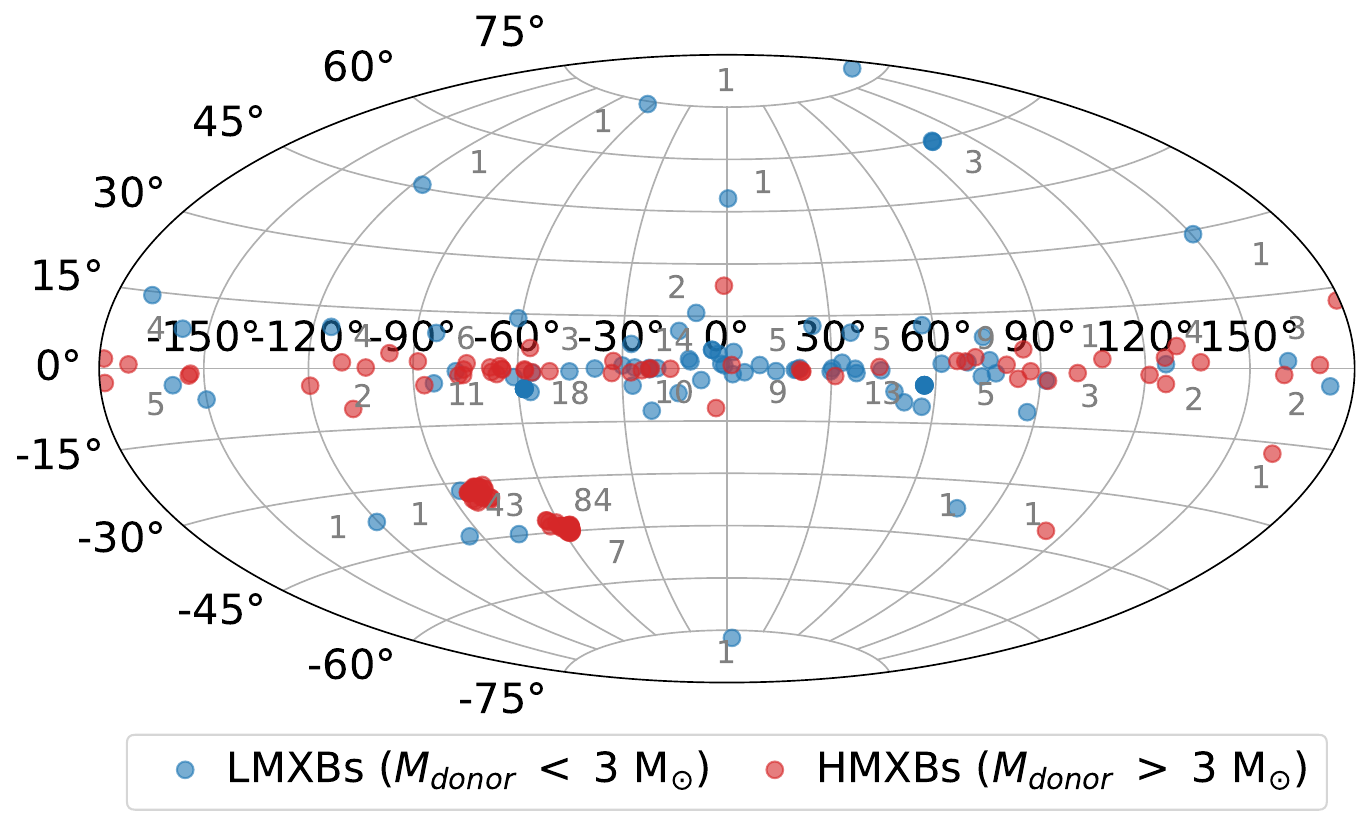}{0.5\textwidth}{(2)}
         }
\caption{Panel 1: Galactic Coordinates distribution of all 3964 XRBs, The definition of colors and shapes is labeled in the figure. Panel 2: Galactic Coordinates distribution of 288 XRBs with donor masses, different colors represent types based on the new criterion of 3 $M_{\odot}$. The number of targets within each grid by latitude and longitude lines were marked. \label{fig:coord}}
\end{figure*}

The panel 1 of figure \ref{fig:coord} shows the Galactic Coordinates distribution of all 3964 XRBs, including both high-precision and low-precision coordinates. Different types are distinguished by different symbols and colors, and these types are from the original literature. The panel 2 of figure \ref{fig:coord} is the position distribution of 288 XRBs with donor mass. The red and blue points represent the donor mass above and below 3 $M_{\odot}$. They are HMXBs and LMXBs based on the criteria of 3 $M_{\odot}$ derived from the distribution study above.

To begin with, it should be noted that in panel 1, it may seem that the majority of the points are concentrated in the MW disk and the MCs, but this is not the true case. Many points that appear as a single point actually represent tens to hundreds of targets that have very similar coordinates because they belong to the same galaxy. Therefore, the points appear close together on the all-sky map and look like a single point. This is why it is necessary to mark the number of targets in each grid.

In fact, the number of XRBs in directions outside the MW disk is much greater than in the direction of the MW disk. The number of targets in the direction of the Galactic disk (where l is within +/- 15 degrees) is 693, while the number in the direction of the LMC and SMC is 83 and 183 (without filter by distance), and the number in other directions is 3005. In this paper, there are at least 3002 XRBs from outside the MW, accounting for 3002/3964=76\% of the total.

For the extragalactic XRBs collected in the paper, we list the works who observed and obtained the most targets. \citet{2008ApJ...689..983H} presented 1,132 low-mass XRBs (LMXBs) from 24 early-type galaxies. The closest galaxy is NGC 3115, located 9 Mpc away, while the farthest is IC 4296, located 50.8 Mpc away. The criterion used to identify LMXBs is an X-ray luminosity of $L_{x} > 10^{37}$ erg/s. \citet{2011A&A...533A..33Z} presented 185 LMXBs, 12 of which are from the MW and the rest from seven other galaxies. \citet{2016ApJS..222...15T} presented 187 XRBs from NGC 3115 with the assuming distance of 9.7 Mpc. \citet{2013A&A...555A..65H} presented 86 binaries, all from the central region of M31. \citet{2012MNRAS.419.2095M} provided 1,026 high-mass XRBs (HMXBs) from 29 nearby galaxies, with the closest galaxy, NGC 5474, located 6.8 Mpc away, and the farthest, CARTWHEEL, located 122.8 Mpc away. \citet{2015AJ....150...94B} provided 132 binaries from three galaxies, NGC 55, NGC 2403 and NGC 4214.

From Figure \ref{fig:coord}, it can be seen that in the two MCs, almost all (252/253 = 99.6\%) the XRBs are HMXBs, which is significantly different from that of the MW. This aligns with previous research that the X-ray luminosity of LMXBs is positively correlated with the stellar mass, while the X-ray luminosity of HMXBs is positively correlated with the Star Formation Rate \citep{2019ApJS..243....3L, 2021ApJ...907...17L}. Because the MCs are low stellar mass galaxies, we do not expect many LMXBs. On the other hand, the age of the MCs is lower than that of the MW, making it more challenging to form the old LMXBs.

Except for the MCs, the reliable spectral observations of XRBs in more distant galaxies is quite difficult to obtain, and therefore, their atmospheric parameters and further fundamental parameters cannot be determined. The 288 XRBs with parameters provided in this paper all come from the MW and the two MCs, with 159 from the MW, and 31 and 98 from the MCs (filtered by distance larger than 10,000 pc), respectively. Given the significant differences in types between the MW and the MCs, these patterns are challenging to generalize to other distant galaxies.

Although \textit{Gaia} provides distances, the values are not reliable for targets outside the MW. We know that the distance to the LMC and SMC is 50,000 and 60,000 pc, while the distances given by \textit{Gaia} DR3 are 10,000-40,000 pc (see the panel 7 in Figure \ref{fig:dist_observed}), significantly smaller than the true values.

\section{Conclusions and discussions}

\subsection{The valley on the distribution and the explanation}

Neither the directly observed number distribution of XRBs, nor the ratio of XRBs to background stars (which we term the ``occurrence rate''), exhibits a flat distribution. XRBs from the MW and MCs display a bimodal structure with a valley around 3\,$M_{\odot}$ or 11,000\,K. This valley persists even when considering only XRBs in the MW. As shown in Figure~\ref{fig:dist_observed}, the MW XRBs (gray dashed lines) exhibit an increase on the high-temperature, high-mass side, despite the scarcity of random stars from the MW in those regions.

The most surprising observation stems from the strong correlation between the stellar evolutionary stage (EEP) and other parameters. Figure \ref{fig:relation3} displays two parallel and distinctly separated strip-like tracks in the main sequence region, which we term the ``parallel tracks" phenomenon. 

The ``parallel tracks" phenomenon naturally explains the valley in the distribution of mass and temperature, or in other words, the low number in the intermediate region. This is because the intermediate region corresponds to either post-main-sequence for low-mass, low-temperature donor stars or pre-main-sequence for high-mass, high-temperature donor stars, resulting in lower abundance compared to the main sequence.

Combined with the differences in mass ratios and accretion modes between the two types of XRBs, we suggest that categorizing XRBs into two main types is appropriate, making the transitional classification of intermediate-mass X-ray Binaries (IMXBs) less essential. The valley position at 3 $M_{\odot}$, or the range from 1 to 4 $M_{\odot}$, can serve as a criterion for distinguishing HMXBs from LMXBs.

Our proposed classification aligns well with the previously established categories of HMXBs and LMXBs in the literature. Among the 288 binaries with atmospheric parameters, 237 had prior classifications (HMXBs or LMXBs), with 203 of these classifications consistent with our results based on the 3 $M_{\odot}$ threshold, yielding a consistency rate of 85.6\%.

\subsection{The post-main-sequence XRBs}

With the help of EEP, we can easily screen out post-main-sequence XRBs. Among the 288 XRBs, 43 are identified as post-main-sequence XRBs, accounting for 43/288=14.9\%, higher than the proportion of post-main-sequence stars from random stars derived in this work (MW 9\%, LMC 6\%, and SMC 4\%). Among them there are 27 sub-giants, 0 red giants, 12 central helium-burning stars, 2 thermally pulsating asymptotic giant branch (TPAGB) stars, and 2 post asymptotic giant branch (post-AGB) stars.

\subsection{The differences on mass ratio and accretion modes between HMXBs and LMXBs}

\begin{figure*}[h]
\gridline{\fig{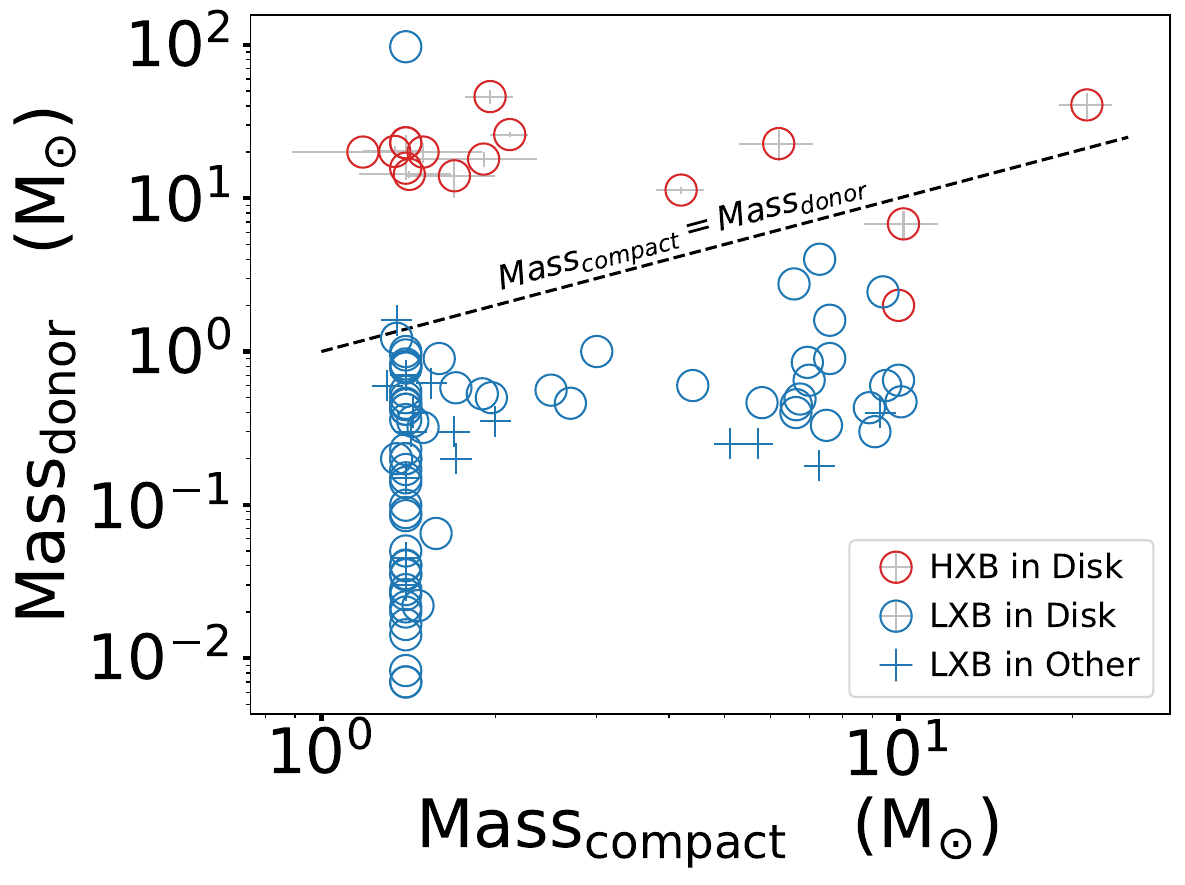}{0.5\textwidth}{}
         }
\vspace{-0.6cm}
\caption{The mass donor star and compact star in XRBs. The masses are all from \citet{2023A&A...671A.149F}, \citet{2023A&A...675A.199A} and \citet{2023A&A...677A.134N}. The color and symbol are the same as Figure \ref{fig:relation} and \ref{fig:relation2}. \label{fig:Mx_Mo}}
\end{figure*}

Figure \ref{fig:Mx_Mo} depicts the mass comparison between the donor star and the compact star for 100 XRBs.

It's important to note that the data here is entirely sourced from \citet{2023A&A...671A.149F}, \citet{2023A&A...675A.199A}, and \citet{2023A&A...677A.134N}, and not from the this paper. Here, we only focus on the mass ratio of XRBs. As in previous literature, the determination of the masses is often done by first determining the mass ratio and then the mass of one component given the mass of the other one. Therefore, to more accurately represent the mass ratio, we utilize the masses of the two stars as provided in the literature.

It can be observed that for HMXBs (red dots), the mass of their donor stars is mostly higher than that of the compact stars, indicating that the mass ratios of the donor stars to the compact stars are mostly greater than 1. Conversely, for LMXBs (blue dots), the mass ratio are mostly less than 1.

Figure \ref{fig:Mx_Mo} shows that the donor stars of HMXBs generally have larger masses than the compact stars, indicating that it is unlikely for them to accrete through Roche-lobe overflow (RLOF), as such accretion mode would not be sustainable in this configuration. When a higher mass donor star fill its Roche lobe and transfers material to the lower mass companion star, the radius of its Roche lobe decreases, causing the donor star to overflow its Roche lobe much more, thus accelerating the mass transfer. This timescale is dynamic and likely on the order of centuries. Our calculations indicate that stable mass transfer is only possible when the mass ratio between the donor star and the compact star is less than 0.788 (see Appendix \ref{q0788} for the calculation and explanation).

Based on this, we conclude that the majority of accretion modes in HMXBs are unlikely to be RLOF (Roche Lobe Overflow) and are more likely to be through stellar wind accretion.

For LMXBs, if their stellar winds are sufficiently strong, wind accretion can also be an effective accretion mode. However, stellar winds of stars with low temperature ($<$ 11,000 K) or low masses ($<$ 3 $M_{\odot}$) are unlikely to form effective wind accretion, so we tend to believe that LMXBs should primarily undergo RLOF. This is also consistent with long-standing views \citep{1985SSRv...40..167W, 1993PASP..105..961C}.

\subsection{A suggestion on searching HMXBs candidates by optical observations}

The analysis of the relationship between the EEP and other stellar parameters has revealed a promising feature. As shown in Figure~\ref{fig:relation3}, high-luminosity and large-radius XRBs occupy a distinct, concentrated strip-like region in parameter space. The most striking characteristic of this region is that it is almost entirely devoid of normal, random field stars from either the MW or the MCs. This unique feature presents a potential opportunity for a new, highly efficient search strategy for HMXBs candidates.

This strategy could leverage large-scale optical surveys as the primary filter, and aid in the discovery of new XRBs candidates which typically begins with X-ray detection. By specifically searching for stars within that parameter space, the optically-selected stars can be then cross-matched with the wide-field X-ray surveys, such as Einstein Probe (EP) or the eROSITA. A confirmed match between an optical candidate and a known X-ray target would serve as evidence of its XRBs nature. This approach would be exceptionally efficient due to the large amount of survey data release.

Regardless of whether this region is exclusively populated by HMXBs or simply enriched with them, the search strategy still valid, and the stars located there warrant detailed investigation. By leveraging the predictive power of stellar evolutionary parameters from huge optical data, we expect a promising discovery channel for peculiar systems.

\newpage
\appendix
\renewcommand\thefigure{Appendix\arabic{figure}}    
\setcounter{figure}{0}

\section{The radius of the cross-matching \label{radius_of_the_cross-matching}}

The key challenge in cross-matching is selecting an appropriate cross-matching radius. According to the method proposed by \citet{2023A&A...671A.149F}, to find unambiguous counterparts, the cross-matching radius should be dynamically selected based on the precision of both the catalog as a whole and the individual input coordinates. The cross-matching radius should be chosen based on the larger of the uncertainties in the two. According to \citet{2023A&A...671A.149F}, for X-ray space telescopes, the telescope with the best accuracy is Chandra, with a corresponding cross-matching radius of 3 arcseconds, which is about twice the worst astrometric performance of the telescope. Additionally, for telescopes with coordinate precision less than 1 arcsecond, an artificial increase of 0.5 arcsecond in coordinate uncertainty is required. This is because different telescopes provide excessively precise coordinates for the same target, making them unable to match each other.

In this study, the binaries that require cross-matching mostly come from telescopes with coordinate precision worse than 1 arcsecond, namely Swift, XMM-Newton, and Chandra, with cross-matching radii ranging from 3 to 8 arcseconds. The astrometric accuracy of the sources in \textit{Gaia} catalog is 20 mas. The coordinate errors of the XRBs requiring cross-matching in this study are all less than 0.4 arcseconds.

Based on the above situation, the cross-matching radius should be chosen between 3 and 8 arcseconds. However, this paper used a more stringent 0.5 to 2 arcseconds. If there is no overall offset between the catalogs, a smaller radius will inevitably reduce the number of matched targets, but it will also increase the reliability of the matched targets. A larger radius will increase the number of matched targets, but it will also increase the probability of false matches. We found that 0.5 to 2 arcseconds is a relatively balanced choice. There are two reasons why we can use a very small cross-matching radius: first, the coordinate accuracy of our targets are very high (less than 0.4 arcseconds), and second, the coordinate accuracy of the \textit{Gaia} targets is very high (0.02 arcseconds).

In this paper, we analyzed a total of 288 targets with atmospheric parameters, and their coordinates were obtained from seven sources. 279 targets' coordinates were provided by five sources (\textit{Gaia} DR1, \textit{Gaia} DR2, \textit{Gaia} EDR3, 2MASS, and Hubble), all of which have astrometric accuracy of less than 0.1 arcseconds. Accordingly, we chose to use a cross-match radius of 0.5 arcseconds.

Eight targets' coordinates were provided by Chandra \citep{2019ApJ...887...20A}, and we used a cross-match radius of 2 arcseconds due to its astrometric accuracy of 1.1-1.6 arcseconds\footnote{\url{https://cxc.harvard.edu/cal/ASPECT/celmon/\#\:\~:text=Chandra\%20absolute\%20astrometric\%20accuracy,Performance\%20varies\%20slightly\%20between\%20detectors.}}.

One target's coordinates were provided by Spitzer \citep{2009AJ....138.1003B}, and we used a cross-match radius of 1 arcsecond due to its astrometric accuracy of 0.5 arcseconds\footnote{\url{https://irsa.ipac.caltech.edu/data/SPITZER/docs/irac/iracinstrumenthandbook/22/}}.

The coordinates provided by different telescopes may have an overall offset between each other. Ideally, we should first align the different coordinate frames before performing the cross-matching. For the targets in this paper, we did not perform this alignment. The reason is that the offsets of each telescope to \textit{Gaia} DR3 (coordinate reference) are smaller than their respective astrometric accuracies. The two sources with the poorest astrometric accuracy have offsets from \textit{Gaia} of 0.57-0.94 arcseconds (Chandra, \citealp{2024ApJ...966..217V}) and 0.02$\pm$0.02 arcseconds (Spitzer, \citealp{2023ApJ...958...33F}), both of which are much smaller than their respective astrometric accuracies. Therefore, we can use an enlarged cross-match radius to cover the impact of these offsets.

\section{The false-match rate of the cross-matching \label{mistake_rate_of_cross-matching}}

\subsection{The causes of the mismatches}

First, we will describe the method of false-match rate (fraction of false matching resulting from the cross-matching) of the cross-matching in this paper. We will consider three scenarios that could lead to erroneous matching:

\begin{enumerate}
    \item Incorrect input coordinates: In this paper, we refer to "incorrect input coordinates" as instances where the given coordinates for a target are significantly inaccurate, deviating from the true position by more than 0.5-2 arcseconds. One common example of this is when the coordinates of a nearby object are erroneously assigned to the target, resulting in errors as large as 20 arcseconds. In such cases, any matched \textit{Gaia} targets are incorrect. If a \textit{Gaia} target is matched, it is not the true match. If no \textit{Gaia} target is matched, it will not enter the parameter table, thus not affecting subsequent analysis.

The number of erroneously matched targets equals the number of incorrect input coordinate multiplied by the probability of matching a \textit{Gaia} target. The probability of erroneous cross-matching results equals the number of erroneously matched targets divided by the total number of cross-matching results.

It's difficult to estimate how many of the 817 targets with high-precision coordinates have incorrect coordinates. However, fortunately, the probability of matching a \textit{Gaia} target from an erroneous coordinate is very low (shown in the first item of next Section \ref{sec:cal_rate_missmatching}), making the error rate negligible. We will determine the false-match rate specifically later on.

    \item Correct input coordinates with multiple cross-matching results: Even with correct input coordinates, erroneous matching results may occur if multiple \textit{Gaia} targets are matched within the cross-matching radius. If multiple \textit{Gaia} targets are found, we can only choose one as the corresponding counterpart, which may lead to a false selection. Given \textit{Gaia}'s high coordinate precision (0.02 arcseconds),we consider the errors caused by \textit{Gaia} coordinates themselves to be negligible.

The false-match rate in this scenario equals the probability of multiple matching results occurring multiplied by the probability of making a false selection.

It's relatively easy to calculate the proportion of multiple matching results, which are listed in Table \ref{tab:statistic}. Although the probability of false selection is difficult to estimate, it should be much less than 0.5. This is because in cases of multiple matching results, we always choose the matched target with the smallest coordinate deviation. Probability-wise, two stars closer in coordinates are more likely to be the same star.

    \item Correct input coordinates with a single cross-matching result: Even with correct input coordinates and a single matching result, it's not guaranteed to be the correct counterpart. A seemingly single matching result may actually be a manifestation of multiple matching results due to insufficient observational limits. Due to \textit{Gaia}'s observational limitations, it may not have observed the true counterpart, but another target may coincidentally appear at the same position, leading to the erroneous selection of this target as the counterpart.

The false-match rate in this scenario equals the probability of two stars coincidentally appearing at the same position. We believe this probability is equivalent to the probability of matching one star from a random position within the same density region and the same cross-matching radius.
\end{enumerate}

To obtain the probabilities mentioned above, we compiled the statistics of cross-matched XRBs in Table \ref{tab:statistic}.

Additionally, to determine the probability of randomly generated coordinates matching \textit{Gaia} targets, we presented the statistics of matches for random targets. Random coordinates were generated by randomly selecting coordinates within a range of 1-3 arcminutes around each high-precision X-ray binary. This ensured that the stellar density around each random coordinate equal to that of X-ray binary. This allowed us to demonstrate the true probability of matching a \textit{Gaia} target when input coordinates are incorrect.

To ensure reliable probability statistics by avoiding sparse numbers, we generated 100 random stars around each X-ray binary, totaling 81700 random stars.

To demonstrate the false-match rate for different cross-match radii (0.5-2 arcseconds), we calculated the results for 0.5 and 2 arcseconds separately.

To examine the situations in the dense stellar fields of the LMC and SMC, we separately listed the statistics for the two MCs.

To review the cross-matching results of targets with parameters (as the main conclusions are based on the analysis results of parameter targets), we also included the cross-matching results of parameter targets.

\centerwidetable 
\begin{deluxetable*}{l|l|l}
\tablecaption{statistic on Cross-match between X-ray binary catalog and \textit{Gaia} DR3 with comparison of random stars}  \label{tab:statistic}
\tablehead{
\colhead{    } & \colhead{X-ray Binary Catalog}  &  \colhead{Random stars} }
\startdata
All targets & 3964 & ---  \\
High-precision coordinate targets & 817 & 817*100=81700 \\
\hline
\multicolumn{3}{c}{0.5 arcseconds radius of cross-matching} \\
\hline
Probability of Single Matches & 513/817=0.63 & 1290/81700=0.016 \\
Probability of Multiple Matches & 1/817=0.001 & 8/81700=0.0001 \\
Proportion of Multiple Matches in all Matches & 1/(513+1)=0.002 & 8/(1290+8)=0.006 \\
Probability of Single Matches with parameters & 288/817=0.35 & 233/81700=0.003 \\
Probability of Multiple Matches with parameters & 0/817=0 & 0/81700=0 \\
\hline
Probability of Single Matches in LMC   & 34/41=0.83 &  85/4100=0.02 \\
Probability of Multiple Matches in LMC & 1/41=0.02  &   0/4100=0 \\
Proportion of Multiple Matches in all Matches in LMC & 1/(34+1)=0.03 & 0/(85+0)=0 \\
Probability of Single Matches with parameters in LMC   & 34/41=0.83 & 20/4100=0.005 \\
Probability of Multiple Matches with parameters in LMC & 0/41=0     & 0/4100=0 \\
\hline
Probability of Single Matches in SMC   & 104/108=0.96 &  337/10800=0.03 \\
Probability of Multiple Matches in SMC & 0/108=0      &   4/10800=0.0004 \\
Proportion of Multiple Matches in all Matches in SMC & 0/(104+0)=0 & 4/(337+4)=0.012 \\
Probability of Single Matches with parameters in SMC   & 99/108=0.92 & 76/10800=0.007 \\
Probability of Multiple Matches with parameters in SMC & 0/108=0     & 0/10800=0 \\
\hline
\multicolumn{3}{c}{2 arcseconds radius of cross-matching} \\
\hline
Probability of Single Matches & 565/817=0.69 &  14475/81700=0.18 \\
Probability of Multiple Matches & 43/817=0.05 & 2883/81700=0.035 \\
Proportion of Multiple Matches in all Matches & 43/(565+43)=0.07 & 2883/(14475+2883)=0.17 \\
Probability of Single Matches with parameters & 288/817=0.35 & 3154/81700=0.04 \\
Probability of Multiple Matches with parameters & 9/817=0.01 & 690/81700=0.008 \\
\hline
Probability of Single Matches in LMC     & 37/41=0.90 &  1121/4100=0.27 \\
Probability of Multiple Matches in LMC   & 1/41=0.02  &  165/4100=0.04 \\
Proportion of Multiple Matches in all Matches in LMC   & 1/(37+1)=0.03 & 165/(1121+165)=0.13 \\
Probability of Single Matches with parameters in LMC     & 35/41=0.85 & 229/4100=0.06 \\
Probability of Multiple Matches with parameters in LMC   & 0/41=0     & 52/4100=0.013 \\
\hline
Probability of Single Matches in SMC     & 100/108=0.93 &  3659/10800=0.34 \\
Probability of Multiple Matches in SMC   & 6/108=0.06   &  676/10800=0.06 \\
Proportion of Multiple Matches in all Matches in SMC   & 6/(100+6)=0.06 & 676/(3659+676)=0.16 \\
Probability of Single Matches with parameters in SMC     & 96/108=0.89 & 834/10800=0.08 \\
Probability of Multiple Matches with parameters in SMC   & 4/108=0.04     & 196/10800=0.018 \\
\enddata
\end{deluxetable*}

\subsection{The calculation of the false-match rates of cross-matching} \label{sec:cal_rate_missmatching}

Now we calculate the false-match rates separately for the three scenarios, and then combine the three probabilities together to obtain the total false-match rate. Since the cross-match radii in this paper are not uniform and the majority (279 out of 288, or 97\%) of the targets have a cross-match radius of 0.5 arcseconds, the following section focuses on the false-match rate results for 0.5 arcseconds radius. The results for other radii and the combined results for different radii are presented later.

\begin{enumerate}
    \item For the first scenario of incorrect input coordinates, we need to determine the number of targets with erroneous coordinates, the probability of matching a \textit{Gaia} target from erroneous coordinates, and the final number of matched targets.

Although it's challenging to estimate the number of targets with erroneous coordinates, let's make an exaggerated guess, assuming half of the targets have incorrect coordinates, so the number is 408.5 (half of 817). The probability of matching a \textit{Gaia} target from erroneous coordinates (i.e. random stars), as seen from the Table \ref{tab:statistic}, is 0.016 + 0.0001 = 0.016 (0.5 arcseconds, the probability of single matches and multiple matches for random stars). The final number of matched targets is 513 + 1 = 514. Therefore, the false-match rate for this scenario is 408.5 * 0.016 / 514 = 0.013 (0.5 arcseconds). For targets with atmospheric parameters, this probability is 408.5 * 0.003 / 288 = 0.004 (0.5 arcseconds).

Similarly, the false-match rates for LMC and SMC are 0.012 and 0.016 (0.5 arcseconds) , respectively.

Note that these probabilities are based on the assumption that half of the input coordinates are incorrect, so the actual error rates are much lower than these values.

    \item For the second scenario of correct input coordinates but multiple cross-match results, we need to determine the probability of multiple match results occurring and the probability of incorrect selection. From the Table \ref{tab:statistic}, the probability of multiple match results occurring is 0.002 (0.5 arcseconds), and let's assume the probability of incorrect selection is 0.5. Therefore, the false-match rate for this scenario is 0.002 * 0.5 = 0.001. For targets with parameters, this probability is 0 because there are no multiple match results.

Similarly, the false-match rates for LMC and SMC are 0.015 and 0 (0.5 arcseconds), respectively.

    \item For the third scenario of correct input coordinates with a single cross-match result, we only need to know the probability of matching a \textit{Gaia} target with erroneous coordinates (random stars), which is 0.016 + 0.0001 = 0.016 (0.5 arcseconds). This is the false-match rate for this scenario. The false-match rate for targets with parameters is 0.003.

Similarly, the false-match rates for LMC and SMC are 0.02 and 0.03 (0.5 arcseconds), respectively.
\end{enumerate}

Combining the three false-match rates mentioned above, the total false-match rate (0.5 arcseconds) is 0.013 + 0.001 + 0.016 = 0.03, or 3\%.

For targets with atmospheric parameters, the total false-match rate (0.5 arcseconds) of cross-matching results is 0.004 + 0 + 0.003 = 0.007, or 0.7\%. 

All the false-match rates are listed in Table \ref{tab:mistake_rate},  for each block there are 8 rates corresponding to different locations, temperatures, and all or parameters only. 

For the two blocks above (false-match rates for 0.5 and 2 arcseconds separately), each false-match rate is the sum of three numbers, which correspond to the false-match rates in the three scenarios described above. Their sum is the total false-match rate.

The bottom block (combined false-match rate from different radii) shows the final false-match rate after combining different cross-matching radii. These results are calculated based on the two blocks above, with the integers in the equations representing the number of targets with different cross-matching radii. For example, 491 represents the number of targets with a 0.5 arcseconds radius, and 23 represents the number of 2 arcseconds. It should be noted that, because there are very few targets with a 1 arcsecond cross-matching radius (only one target with parameter), we included the 1 arcsecond targets in the 2 arcsecond count. This slightly increases the final false-match rate, but still not enough to affect the subsequent statistical analysis.

The final false-match rate of all the XRBs is $(491\times0.0295+23\times0.3906)/514=0.0457$, or 4.6\%. This number represents the probability of erroneous targets in Table \ref{tab:total}.

The final false-match rate of the XRBs with parameters is $(279\times0.0069+09\times0.1269)/288=0.0107$, or 1.1\%. This number represents the false-match probability of targets in Table \ref{tab:Gaia_phy}, which forms the basis of the statistical analysis in this paper.

From Table \ref{tab:mistake_rate}, it can be seen that for 0.5 arcseconds radius, the cross-matching false-match rates are all below 5\%. In contrast, the false-match rates for 2 arcseconds are all above 10\%, with the highest reaching 63\%. Therefore, a smaller cross radius can reduce the false-match rate.
\begin{deluxetable*}{l|l}
\tablecaption{false-match rate of cross-matching}  \label{tab:mistake_rate}
\tablehead{
\colhead{XRBs}  &  \colhead{false-match rate} }
\startdata
\multicolumn{2}{c}{false-match rate of cross-matching by 0.5 arcsecond radius} \\
\hline
ALL XRBs                 &               0.0126 + 0.0010 + 0.0159 = 0.0295 \\
ALL XRBs with parameters &               0.0040 + 0.0000 + 0.0029 = 0.0069 \\
ALL XRBs with temperature $>$  12000 K & 0.0006 + 0.0000 + 0.0003 = 0.0009 \\
ALL XRBs with temperature $<$  12000 K & 0.0098 + 0.0000 + 0.0026 = 0.0124 \\
LMC XRBs                 &               0.0121 + 0.0143 + 0.0207 = 0.0472 \\
LMC XRBs with parameters &               0.0029 + 0.0000 + 0.0049 = 0.0078 \\
SMC XRBs                 &               0.0164 + 0.0000 + 0.0316 = 0.0480 \\
SMC XRBs with parameters &               0.0038 + 0.0000 + 0.0070 = 0.0109 \\
\hline
\multicolumn{2}{c}{false-match rate of cross-matching by 2.0 arcsecond radius}\\
\hline
ALL XRBs                 &              0.1427 + 0.0354 + 0.2125 = 0.3906 \\
ALL XRBs with parameters &              0.0647 + 0.0152 + 0.0471 = 0.1269 \\
ALL XRBs with temperature $>$ 12000 K & 0.0095 + 0.0112 + 0.0041 = 0.0248 \\
ALL XRBs with temperature $<$ 12000 K & 0.1609 + 0.0229 + 0.0429 = 0.2268 \\
LMC XRBs                 &              0.1692 + 0.0132 + 0.3137 = 0.4960 \\
LMC XRBs with parameters &              0.0401 + 0.0000 + 0.0685 = 0.1087 \\
SMC XRBs                 &              0.2045 + 0.0283 + 0.4014 = 0.6342 \\
SMC XRBs with parameters &              0.0515 + 0.0200 + 0.0954 = 0.1669 \\
\hline
\multicolumn{2}{c}{Combined false-match rate from different radii}\\
\multicolumn{2}{c}{(Most 0.5 arcseconds and a small portion of 1-2 arcseconds.)}\\
\hline
ALL XRBs                                 & $(491\times0.0295+23\times0.3906)/514=0.0457$ \\
ALL XRBs with parameters                 & $(279\times0.0069+09\times0.1269)/288=0.0107$ \\
ALL XRBs with temperature   $>$ 12000 K  & $(173\times0.0009+08\times0.0248)/181=0.0020$ \\
ALL XRBs with temperature   $<$ 12000 K  & $(106\times0.0124+01\times0.2268)/107=0.0144$ \\
LMC XRBs                                 & $(034\times0.0472+01\times0.4960)/035=0.0600$ \\
LMC XRBs with parameters                 & $(033\times0.0078+01\times0.1087)/034=0.0108$ \\
SMC XRBs                                 & $(096\times0.0480+08\times0.6342)/104=0.0931$ \\
SMC XRBs with parameters                 & $(091\times0.0109+08\times0.1669)/099=0.0235$ \\
\hline
\enddata
\end{deluxetable*}

Parameterized targets are usually brighter compared to non-parameterized ones. As shown in Table \ref{tab:mistake_rate}, the false-match rates for XRBs with parameters are significantly lower compared to those for all XRBs, indicating a significant reduction of false-match rates for bright sources. This is consistent with the statistical conclusions by \citet{2009ApJ...697.1695A} for the SMC.

The false-match rates in dense regions (LMC and SMC) are significantly higher than the overall false-match rate, which is consistent with our expectations, and is also the reason why we need to calculate the false-match rates in dense regions separately.

Further dividing parameterized targets into high-temperature and low-temperature reveals that the false-match rate for high-temperature stars is relatively lower, which is because high-temperature stars are usually brighter. This result is also consistent with the statistical findings by \citet{2009ApJ...697.1695A}. The temperature threshold of 11,000 K is chosen because it is the boundary used in this study to classify XRBs into high and low-temperature categories.

From the calculations above for the three scenarios, it's clear that multiple cross-match results are not the dominant factor causing errors; rather, it's the probability of a random position matching a \textit{Gaia} target. Ideally, the probability of a random coordinate matching a target should be very low, preferably less than 5\%. This requires a sufficiently small cross-matching radius. Additionally, the error in input coordinates should align with the cross-matching radius. In this paper, the coordinate errors for the 817 targets are all less than 0.4 arcseconds, which fit the 0.5-2 arcseconds cross-matching radius.


\section{The observational capabilities of \textit{Gaia} for MCs \label{observational_capabilities_for_MCs}}

To fully understand the observational capabilities of \textit{Gaia} for MCs, we selected more than 100,000 MCs targets in the \textit{Gaia} database. The selection criteria are that the coordinates are located in the MCs regions and the distance is greater than 20,000 pc. It should be noted that the distances of MCs given by \textit{Gaia} are not correct. Most of the distances given by \textit{Gaia} are less than 30,000 pc, but the actual distances of the two MCs are 50,000 to 60,000 pc. Nevertheless, selecting targets over 20,000 pc can well exclude targets in the Milky Way (hereafter MW) and obtain high-purity MCs targets.

After checking the selected MCs targets, we found that GSP-Phot can observe the temperature range of 3,318-41,000 K and the magnitude of 10-19. The temperature range between 18.5 and 19 magnitudes, which is close to the limiting magnitude, is 3,923-15,000 K. This shows that close to the limiting magnitude of 19, the observational capability of GSP-Phot can basically cover the low temperature part. We performed a small-scale determination of the masses of the MCs targets (the method is the same as XRBs in this paper as described in section \ref{subsec:calculating_pars}). The lower limit of the mass is 1 solar mass, corresponding to a temperature of 3,900-7,600 K. Therefore, for the MCs, GSP-Phot can observe stars with a mass greater than 1 solar mass. Stars with smaller masses or lower temperatures cannot obtain atmospheric parameters.

For the MCs XRBs in this paper, the temperature range is 8,000-37,920 K, with an average value is 28,000 K, and the mass range is 1.6-24, with an average value of 11. Therefore, the observational capability of GSP-Phot is enough to cover the XRBs in MCs.

\section{The X-ray binary catalogs} \label{catalogs}

Table \ref{tab:total} is the full X-ray binary catalogs with the basic information which are star name, coordinates and their errors from the Simbad database, the X-ray binary type, and the reference bibcode of the coordinate and the type.

\begin{deluxetable*}{llllllll}
\tablecaption{Catalog of X-ray binary systems} \label{tab:total}
\tabletypesize{\scriptsize}
\tablehead{
\colhead{Name} & \colhead{RA} & \colhead{DEC} & \colhead{Major axes} & \colhead{Minor axes} & \colhead{Type} & \colhead{Reference} & \colhead{Reference} \\[-10pt]
\colhead{from Simbad} & \colhead{} & \colhead{} & \colhead{of the error ellipses} & \colhead{of the error ellipses} & \colhead{} & \colhead{of coordinate} & \colhead{of Type} \\[-4pt]
\colhead{} & \colhead{(deg)} & \colhead{(deg)} & \colhead{(sec)} & \colhead{(sec)} & \colhead{} & \colhead{(Bibcode)} & \colhead{(Bibcode)}
}
\startdata
CI Cam & 64.92556458333333 & 55.999362777777776 & 1e-05 & 8e-06 & HXB & 2020yCat.1350....0G & 2023A\&A...671A.149F \\
IGR J06074+2205 & 91.86088333333332 & 22.096599722222223 & 2e-05 & 1e-05 & HXB & 2020yCat.1350....0G & 2023A\&A...671A.149F \\
HZ Her & 254.4575458333333 & 35.342357222222226 & 1e-05 & 1e-05 & LXB & 2020yCat.1350....0G & 2023A\&A...675A.199A \\
HD 215227 & 340.7387629166666 & 44.721738611111114 & 1e-05 & 1e-05 & HXB & 2020yCat.1350....0G & 2023A\&A...671A.149F \\
SWIFT J174510.8-262411 & 266.2952041666666 & -26.403499999999998 & 0.01 & 0.01 & LXB & 2013ApJS..209...14K & 2023A\&A...675A.199A \\
MAXI J0556-332 & 89.193 & -33.174499999999995 & 2 & 2 & LXB & 2011ATel.3103....1K & 2023A\&A...675A.199A \\
QU TrA & 236.9775 & -62.568333333333335 & nan & nan & LXB & 2003AstL...29..468S & 2023A\&A...675A.199A \\
CRTS J135716.8-093238 & 209.32014874999996 & -9.54411 & 0.001 & 0.001 & LXB & 2020yCat.1350....0G & 2023A\&A...675A.199A \\
XTE J1752-223 & 268.06287499999996 & -22.342322222222222 & 0.001 & 0.001 & LXB & 2016A\&A...587A..61C & 2023A\&A...675A.199A \\
OGLE BLG-ELL-12042 & 268.97191833333335 & -28.276068055555555 & 4e-05 & 3e-05 & LXB & 2020yCat.1350....0G & 2023A\&A...675A.199A \\
$[$JBN2011$]$ 377 & 265.8188333333333 & -27.76038888888889 & nan & nan & LXB & 2011ApJS..194...18J & 2023A\&A...675A.199A \\
MAXI J1807+132 & 272.03145416666666 & 13.2515 & nan & nan & LXB & 2017ApJ...850..155S & 2023A\&A...675A.199A \\
MAXI J0903-531 & 136.27841916666668 & -53.50542305555555 & 1e-05 & 1e-05 & HXB & 2020yCat.1350....0G & 2023A\&A...677A.134N \\
  \multicolumn{8}{c}{...} \\
  \multicolumn{8}{c}{A total of 3964 targets} \\
  \multicolumn{8}{c}{...} \\
SWIFT J1756.9-2508 & 269.2389583333333 & -25.107722222222225 & 4 & 4 & LXB & 2007ApJ...668L.147K & 2023A\&A...675A.199A \\
IGR J17591-2342 & 269.78662083333336 & -23.71297222222222 & nan & nan & LXB & 2018ATel11941....1D & 2023A\&A...675A.199A \\
$[$KRL2007b$]$ 297 & 269.9404166666667 & -22.0275 & nan & nan & LXB & 2007A\&A...469..807L & 2023A\&A...675A.199A \\
4U 1758-25 & 270.29054499999995 & -25.07892222222222 & 2e-05 & 2e-05 & LXB & 2020yCat.1350....0G & 2023A\&A...675A.199A \\
X Sgr X-3 & 270.3845833333333 & -20.528888888888886 & nan & nan & LXB & 2003A\&A...411L..59E & 2023A\&A...675A.199A \\
SAX J1805.5-2031 & 271.3916666666666 & -20.513333333333332 & nan & nan & LXB & 2007A\&A...469..807L & 2023A\&A...675A.199A \\
XMMU J181227.8-181234 & 273.11583333333334 & -18.209444444444443 & nan & nan & LXB & 2006MNRAS.369.1965C & 2023A\&A...675A.199A \\
PSO J274.2181-19.6322 & 274.2208333333333 & -19.511666666666667 & nan & nan & LXB & 2022ATel15418....1N & 2023A\&A...675A.199A \\
SAX J1818.7+1424 & 274.6833333333333 & 14.403333333333332 & nan & nan & LXB & 2007A\&A...469..807L & 2023A\&A...675A.199A \\
AX J1848.8-0129 & 282.19999999999993 & -1.4866666666666666 & nan & nan & LXB & 2001ApJS..134...77S & 2023A\&A...675A.199A \\
SWIFT J185003.2-005627 & 282.51333333333326 & -0.9408333333333333 & nan & nan & LXB & 2011GCN.12101....1G & 2023A\&A...675A.199A \\
SWIFT J2037.2+4151 & 309.24999999999994 & 41.833333333333336 & nan & nan & LXB & 2010A\&A...523A..61K & 2023A\&A...675A.199A \\
SAX J2224.9+5421 & 336.2166666666666 & 54.365 & nan & nan & LXB & 2007A\&A...469..807L & 2023A\&A...675A.199A \\
\enddata
\end{deluxetable*}

The coordinate errors provided by Simbad includes three parameters: the major axis of the error ellipse, the minor axis of the error ellipse, and the position angle. Here, only the major and minor axes are listed. The complete parameters can be obtained from the website \url{https://astrophysics.cc/xray-binary-catalogs} or \url{https://zenodo.org/records/17150017}, including the mass, period, and other parameters provided by \citet{2023A&A...671A.149F, 2023A&A...675A.199A, 2023A&A...677A.134N}.

The X-ray binary type is divided into confirmed but unclassified X-ray binary XB*, confirmed high-mass X-ray Binary HXB, confirmed low-mass X-ray Binary LXB, unclassified X-ray binary candidate XB?, high-mass X-ray Binary candidate HX?, and low-mass X-ray Binary candidate LX?.

The first four columns of Table \ref{tab:Gaia_phy} represent the cross-matching results with \textit{Gaia} DR3, while the subsequent five columns display the absolute parameters we determined based on atmospheric parameters and stellar model \textit{MIST}. The last parameter is the period from \citet{2023A&A...671A.149F, 2023A&A...675A.199A} and \citet{2023A&A...677A.134N}. Same as Table \ref{tab:total}, the full table with all parameters can be downloaded from \url{https://astrophysics.cc/xray-binary-catalogs} or \url{https://zenodo.org/records/17150017}.


\begin{deluxetable*}{@{}p{1.2cm}lllllllll@{}}
\tablecaption{The physical parameters of the donor stars in XRBs} \label{tab:Gaia_phy}
\tabletypesize{\scriptsize}
\tablehead{
\colhead{Name} & \colhead{$T_{\rm eff}$} & \colhead{$\log g$} & \colhead{[M/H]} & \colhead{Mass} & \colhead{Radius} & \colhead{Luminosity} & \colhead{Age} & \colhead{Stage} & \colhead{Period} \\[-8pt]
\colhead{(From Simbad)} & \colhead{(K)} & \colhead{(dex)} & \colhead{(dex)} & \colhead{($M_{\odot}$)} & \colhead{($R_{\odot}$)} & \colhead{($L_{\odot}$)} & \colhead{(Gyr)} & \colhead{(EEP)} & \colhead{(d)} 
}
\startdata  
IGR J06074+2205 & $28984_{-46}^{+70}$ & $4.110_{-0.03}^{+0.032}$ & $+0.007_{-0.011}^{+0.0054}$ & $14.04_{-0.19}^{+0.18}$ & $5.443_{-0.23}^{+0.22}$ & $1.879e+04_{-1626}^{+1503}$ & $0.004621_{-0.00073}^{+0.00084}$ & $328_{-5}^{+6}$ & --- \\
MAXI J0903-531 & $22971_{-112}^{+327}$ & $3.996_{-0.017}^{+0.01}$ & $-0.971_{-0.09}^{+0.024}$ & $7.6_{-0.18}^{+0.18}$ & $4.577_{-0.11}^{+0.11}$ & $5246_{-391.3}^{+355.9}$ & $0.03408_{-0.0018}^{+0.0017}$ & $369_{-2}^{+2}$ & 57.0 \\
MAXI J1820+070 & $5604_{-35}^{+29}$ & $4.694_{-0.022}^{+0.022}$ & $-4.064_{-0.16}^{+0.062}$ & $0.6164_{-0.014}^{+0.0087}$ & $0.5639_{-0.0062}^{+0.0064}$ & $0.286_{-0.0122}^{+0.01224}$ & $11.6_{-1.4}^{+2.5}$ & $349_{-2}^{+5}$ & 0.68549 \\
HD 249179 & $16927_{-683}^{+305}$ & $3.614_{-0.37}^{+0.12}$ & $+0.407_{-0.14}^{+0.42}$ & $6.065_{-0.88}^{+0.65}$ & $4.789_{-1.8}^{+1.2}$ & $1651_{-1535}^{+761.7}$ & $0.04406_{-0.0089}^{+0.01}$ & $354_{-27}^{+19}$ & --- \\
NGC 6649 9 & $35005_{-21}^{+9}$ & $3.835_{-0.022}^{+0.056}$ & $+0.006_{-0.0099}^{+0.004}$ & $25.84_{-0.77}^{+0.7}$ & $9.95_{-0.54}^{+0.51}$ & $1.337e+05_{-1.47e+04}^{+1.355e+04}$ & $0.00451_{-7.5e-05}^{+0.00011}$ & $351_{-1}^{+1}$ & --- \\
V1341 Cyg & $14210_{-54}^{+105}$ & $4.174_{-0.053}^{+0.029}$ & $-1.023_{-0.063}^{+0.078}$ & $2.949_{-0.084}^{+0.082}$ & $2.302_{-0.15}^{+0.13}$ & $194.3_{-25.71}^{+22.03}$ & $0.2216_{-0.012}^{+0.013}$ & $358_{-5}^{+5}$ & 9.841666666666667 \\
PSR J1023+0038 & $5966_{-12}^{+4}$ & $4.654_{-0.015}^{+0.016}$ & $-4.119_{-0.038}^{+0.022}$ & $0.6751_{-0.012}^{+0.0098}$ & $0.6317_{-0.0045}^{+0.0052}$ & $0.4551_{-0.00809}^{+0.008484}$ & $11.01_{-1.5}^{+1.9}$ & $356_{-5}^{+5}$ & 0.198096 \\
  \multicolumn{10}{c}{...} \\
  \multicolumn{10}{c}{A total of 288 targets}  \\
  \multicolumn{10}{c}{...} \\
2MASS J12440380-6322320 & $34793_{-96}^{+88}$ & $3.971_{-0.0061}^{+0.0079}$ & $+0.001_{-0.0023}^{+0.0009}$ & $23.26_{-0.22}^{+0.23}$ & $8.247_{-0.13}^{+0.12}$ & $8.962e+04_{-3034}^{+3087}$ & $0.004022_{-8.6e-05}^{+9e-05}$ & $344_{-0}^{+0}$ & 138.0 \\
BR Cir & $4296_{-17}^{+19}$ & $2.029_{-0.028}^{+0.011}$ & $-0.970_{-0.065}^{+0.053}$ & $0.8773_{-0.016}^{+0.0093}$ & $16.26_{-0.38}^{+0.34}$ & $89.84_{-4.004}^{+3.641}$ & $13.02_{-0.35}^{+0.71}$ & $540_{-0}^{+0}$ & 16.68 \\
$[$MT91$]$ 213 & $29996_{-129}^{+126}$ & $4.037_{-0.01}^{+0.0091}$ & $+0.264_{-0.025}^{+0.034}$ & $16.42_{-0.19}^{+0.2}$ & $6.427_{-0.098}^{+0.095}$ & $3.006e+04_{-1211}^{+1175}$ & $0.003703_{-0.00027}^{+0.00027}$ & $329_{-2}^{+2}$ & 17000.0 \\
XTE J1858+034 & $3392_{-5}^{+15}$ & $4.662_{-0.025}^{+0.1}$ & $-0.418_{-0.043}^{+0.13}$ & $0.2558_{-0.023}^{+0.023}$ & $0.2675_{-0.019}^{+0.021}$ & $0.00856_{-0.00131}^{+0.001317}$ & $10.18_{-2.3}^{+4}$ & $265_{-5}^{+10}$ & --- \\
$[$BPH2004$]$ CX 1 & $5921_{-78}^{+90}$ & $3.686_{-0.054}^{+0.057}$ & $-2.187_{-0.13}^{+0.28}$ & $0.7953_{-0.035}^{+0.02}$ & $2.18_{-0.15}^{+0.13}$ & $5.41_{-0.7936}^{+0.5386}$ & $11.92_{-1.1}^{+1.7}$ & $468_{-1}^{+1}$ & 0.262822 \\
PSR J1723-2837 & $5828_{-46}^{+41}$ & $4.489_{-0.022}^{+0.027}$ & $+0.079_{-0.044}^{+0.05}$ & $1.035_{-0.023}^{+0.022}$ & $0.9668_{-0.025}^{+0.024}$ & $0.9679_{-0.06364}^{+0.06448}$ & $1.319_{-1.1}^{+0.83}$ & $316_{-17}^{+31}$ & 0.6166666666666667 \\
4U 1758-25 & $5816_{-20}^{+11}$ & $4.170_{-0.029}^{+0.009}$ & $-0.294_{-0.017}^{+0.0094}$ & $0.9037_{-0.0094}^{+0.0097}$ & $1.291_{-0.031}^{+0.032}$ & $1.717_{-0.089}^{+0.08999}$ & $11.63_{-0.38}^{+0.36}$ & $437_{-2}^{+2}$ & --- \\
\enddata
\end{deluxetable*}

\section{The description of EEP} \label{eep}

The detailed explanation of EEPs can be found in \citet{2016ApJS..222....8D}. Here, we'll provide a brief introduction. 

EEPs are parameters that reflect a star's different evolutionary stages. To obtain precise EEPs, it is necessary to first establish primary EEPs based on physical definitions. For instance, for a star of one solar mass, primary EEPs includes ten significant evolutionary nodes, ranging from the Pre Main-sequence (PMS), Zero Age Main-Sequence (ZAMS), intermediate age main-sequence (IAMS), terminal age main-sequence (TAMS), RGB tip, to post-AGB and the white dwarf cooling sequence (WDCS).

The physical definition of the same primary EEPs can vary for stars with different initial masses. For instance, in the substellar case, the ZAMS is considered as the maximum of the central temperature along the evolutionary track. Similarly, high-mass stars may not go through a red giant phase, so their RGB tip is defined as the point where the luminosity reaches its maximum or the surface temperature reaches its minimum before a substantial depletion of central helium occurs (when the central helium abundance is greater than the original abundance minus 0.01). For high-mass stars, the last primary EEPs is carbon burning (C-burn) which is equivalent to thermally pulsing AGB (TP-AGB) of low-mass stars. However, for low- and intermediate-mass stars, there are additional stages like post-AGB and WDCS.

After defining the primary EEPs, the space between each pair of adjacent primary EEPs is uniformly divided into a fixed number of secondary EEPs. For example, the interval from PMS to ZAMS is divided into 201 secondary EEPs, and from ZAMS to TAMS into 252 secondary EEPs. The total number of secondary EEPs from PMS to WDCS is 1710.

To ensure secondary EEPs are ``equally spaced" between two adjacent primary EEPs, \textit{MIST} defines a metric function (equation 1 in \citealp{2016ApJS..222....8D}). It derives the metric distance between two points on the evolutionary track. Traditionally, parameters used for this purpose are luminosity and surface temperature (in logarithm), although central temperature and density can also be used, or additional parameters such as age. Any evolutionary parameter can be employed to calculate the metric distance, and each parameter can have an assigned weight. The singular goal is to provide a more detailed division of evolutionary stages. This benefit allows for accurate interpolation between various evolutionary tracks, especially in rapidly evolving post main-sequence stages.

The generation of EEPs is similar to that of parameters such as mass, radius, and age, all obtained by interpolating three atmospheric parameters from a stellar database. Metallicity can also influence the results of EEPs. Unlike other stellar parameters, EEPs cannot be derived directly during stellar evolution. Instead, they are identified after the entire evolutionary track has been computed, by recognizing primary EEPs and then dividing them into a fixed number of secondary EEPs. Therefore, the definition of EEPs is somewhat subjective. Nevertheless, EEPs provide a detailed description of the evolutionary stage, enhancing our understanding of a star's status when combined with other parameters.


As the parameter EEPs is not commonly found in the literature, we cannot cross-validate the EEPs presented in this paper with other studies. We have thoroughly discussed the reliability and uncertainties of parameters like mass and radius, and have demonstrated that the errors provided in Table \ref{tab:Gaia_phy} accurately represent the uncertainties in these parameters. Since EEPs are computed simultaneously with these other parameters using the same interpolation method and atmospheric parameters, their uncertainties should be similarly reliable. 

\section{Statistical validation of the bimodal distribution of donor mass and temperature.} \label{GMM_and_KStest}

To investigate whether the observed distributions of donor mass and temperature can be characterized by two underlying populations, we applied a two-component Gaussian Mixture Model (GMM). The GMM assumes that the observed distribution is a superposition of two normal distributions, each representing a distinct physical population (e.g., low-mass and high-mass systems). The model parameters—including the means, variances, and mixing weights of the two Gaussian components—were estimated using the Expectation-Maximization (EM) algorithm, an iterative maximum likelihood method. The algorithm alternates between estimating the posterior probabilities of component membership for each data point (the E-step) and updating the distribution parameters to maximize the data likelihood (the M-step), continuing until convergence.

To assess the goodness-of-fit of the two-component model, we performed a bootstrap-corrected Kolmogorov-Smirnov (K-S) test, comparing the empirical cumulative distribution function (CDF) of the data with the theoretical CDF derived from the fitted GMM. The use of a standard K-S test is inappropriate here, as the GMM parameters were derived from the data itself, which would lead to an inflated p-value and a higher probability of incorrectly accepting the null hypothesis. The bootstrap procedure accounts for this by correcting the test statistic, thus avoiding inflated p-values and ensuring a robust assessment of the fit.

For donor mass (in log), the fitted parameters for the two Gaussian components are as follows: means of 1.034 and -0.065, variances of 0.244 and 0.265, and mixing weights of 0.657 and 0.343. The maximum vertical distance between the empirical distribution and the fitted Gaussian model was 0.0368. A bootstrap-estimated p-value from 1000 resamples was approximately 0.95, which is significantly greater than 0.05. Therefore, we accept the hypothesis that the stellar mass data originate from a bimodal, two-Gaussian distribution.

Similarly, for donor temperature (in log), the fitted parameters for the two Gaussian components are: means of 4.426 and 3.729, variances of 0.126 and 0.110, and mixing weights of 0.629 and 0.371. The maximum vertical distance was 0.0861. The bootstrap-estimated p-value was approximately 0.68, which is also much greater than 0.05. We therefore accept the hypothesis that the temperature distribution is also bimodal.

\begin{figure*}[h]
\gridline{\fig{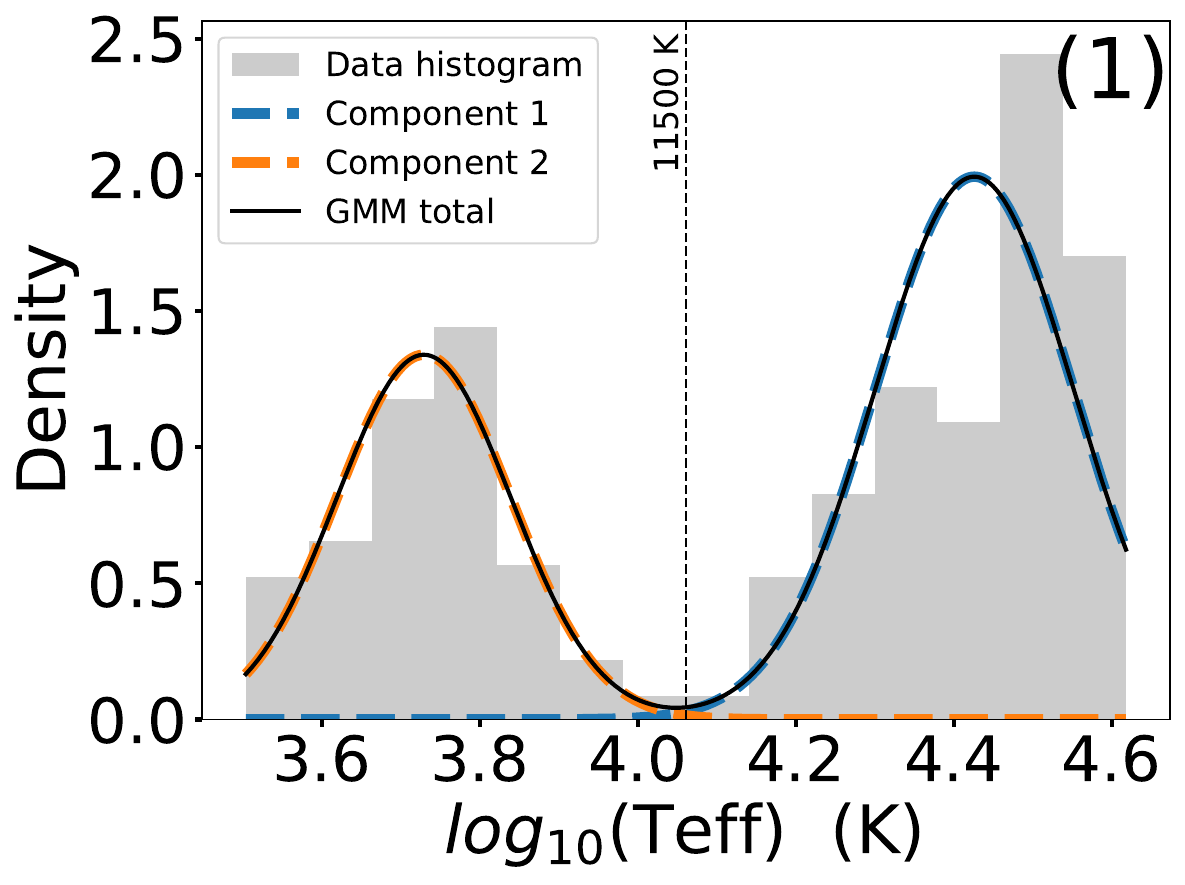}{0.495\textwidth}{}
          \fig{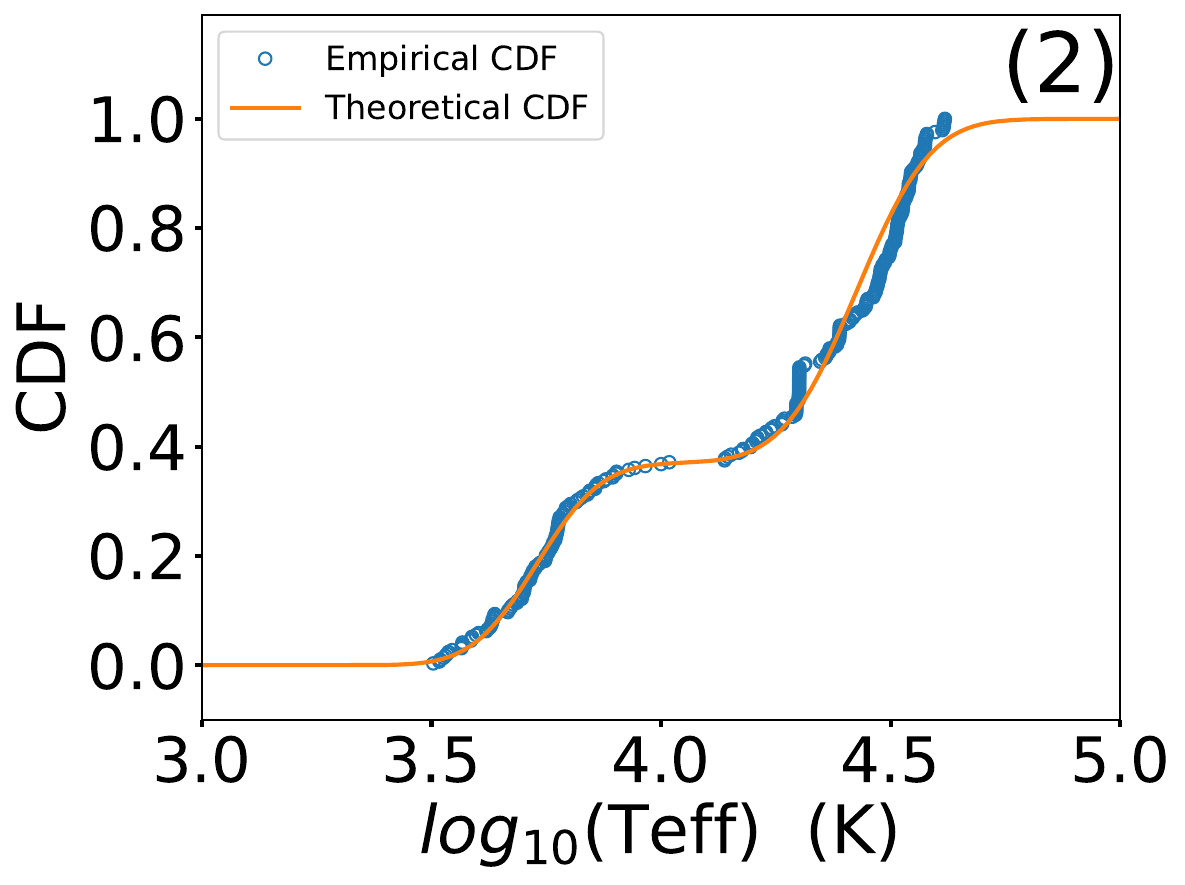}{0.495\textwidth}{}
             }
\vspace{-0.8cm}
\gridline{\fig{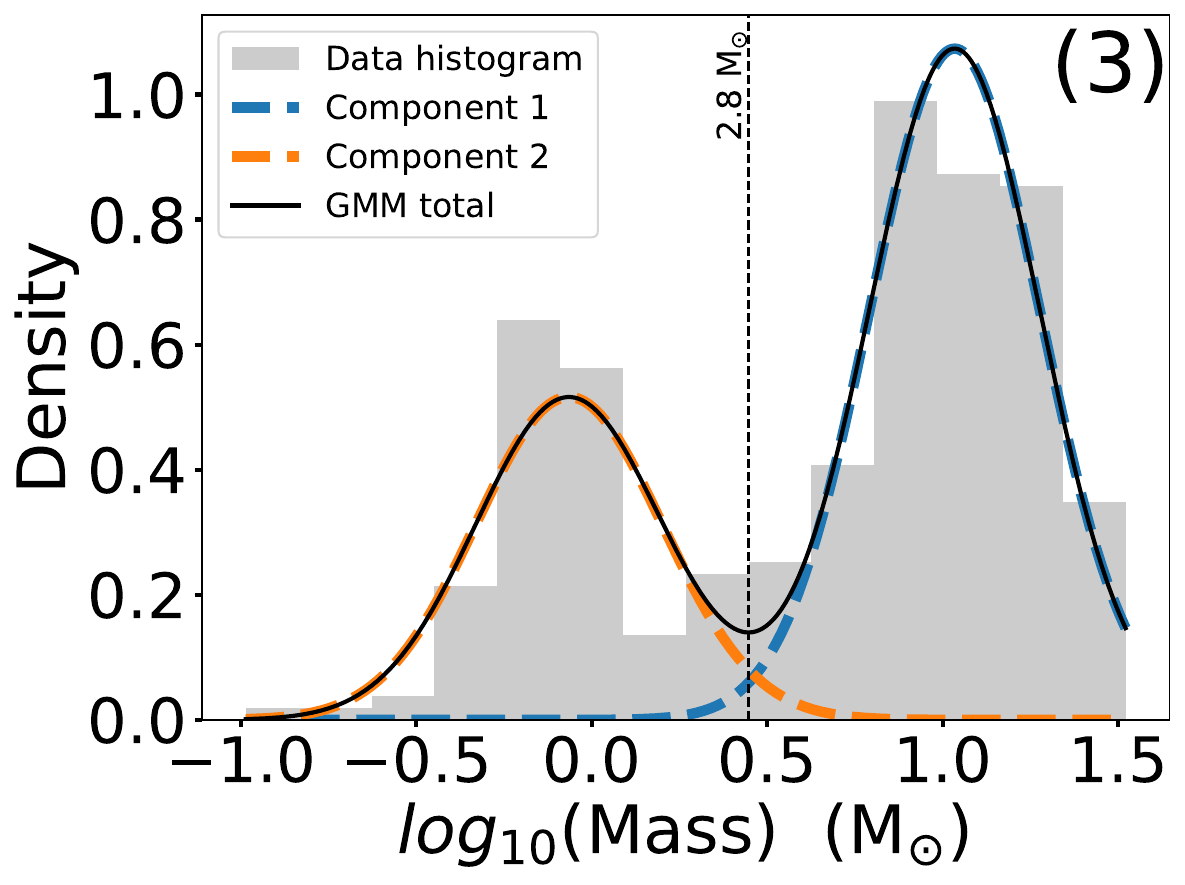}{0.495\textwidth}{}
          \fig{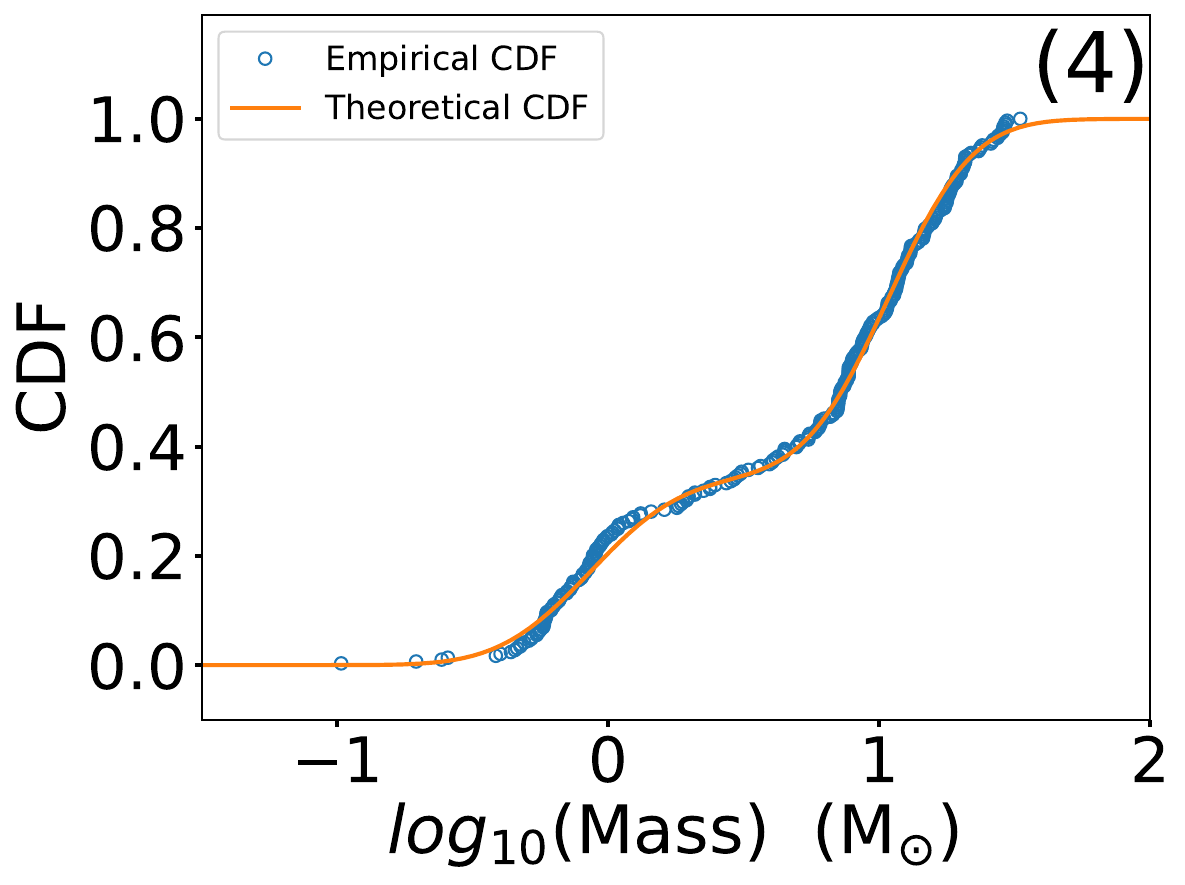}{0.495\textwidth}{}
             }
\vspace{-0.6cm}
\caption{The fitting plot of the bimodal distribution of mass and temperature (upper two panels) and the comparison of the empirical cumulative distribution function (lower two panels). \label{fig:CDF_All_XRBs} }
\end{figure*}

Below Figure \ref{fig:CDF_All_XRBs} are the fitting plots for the two Gaussian components and a comparison of the empirical and theoretical cumulative distribution functions. It is important to note that the GMM fitting process does not require data binning; the fitting is not performed on a distribution curve but rather by calculating and updating the expected probability for each data point. The histogram of the data shown in the left-hand plot must be binned for visualization purposes, whereas the two theoretical Gaussian curves are derived independently of any binning choices.

The above figure \ref{fig:CDF_All_XRBs} shows the analysis for all XRBs from the Milky Way (MW) and the Magellanic Clouds (MCs). The bottom plot, in contrast, shows the bimodal distribution analysis for only the MW XRBs.

For the MW only sample, the bootstrap-estimated p-values from the two-component GMM are approximately 0.8 (for temperature) and 0.9 (for mass), which strongly supports a two-Gaussian distribution. The p-values from the diptest are about 0.00044 (for temperature) and 0.010 (for mass), both of which reject the null hypothesis of a unimodal distribution.

\begin{figure*}[h]
\gridline{\fig{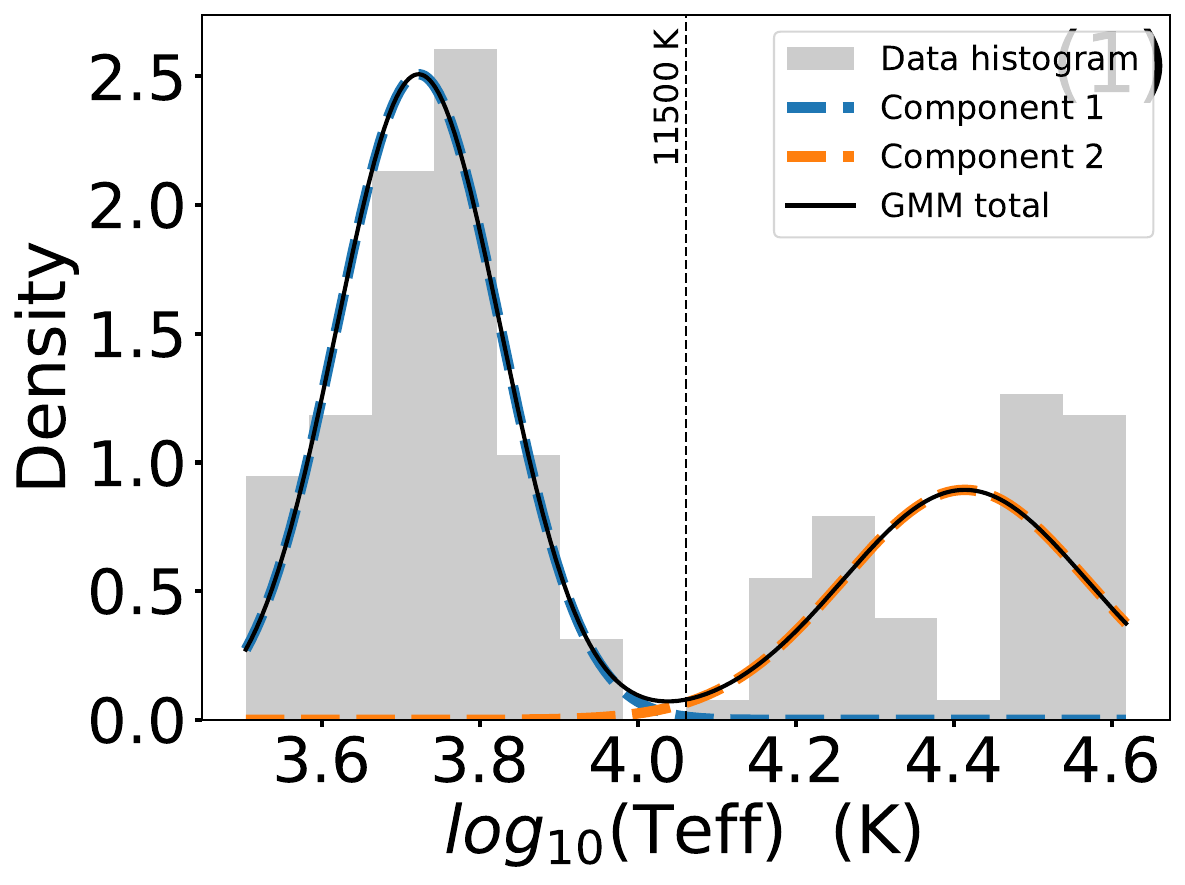}{0.495\textwidth}{}
          \fig{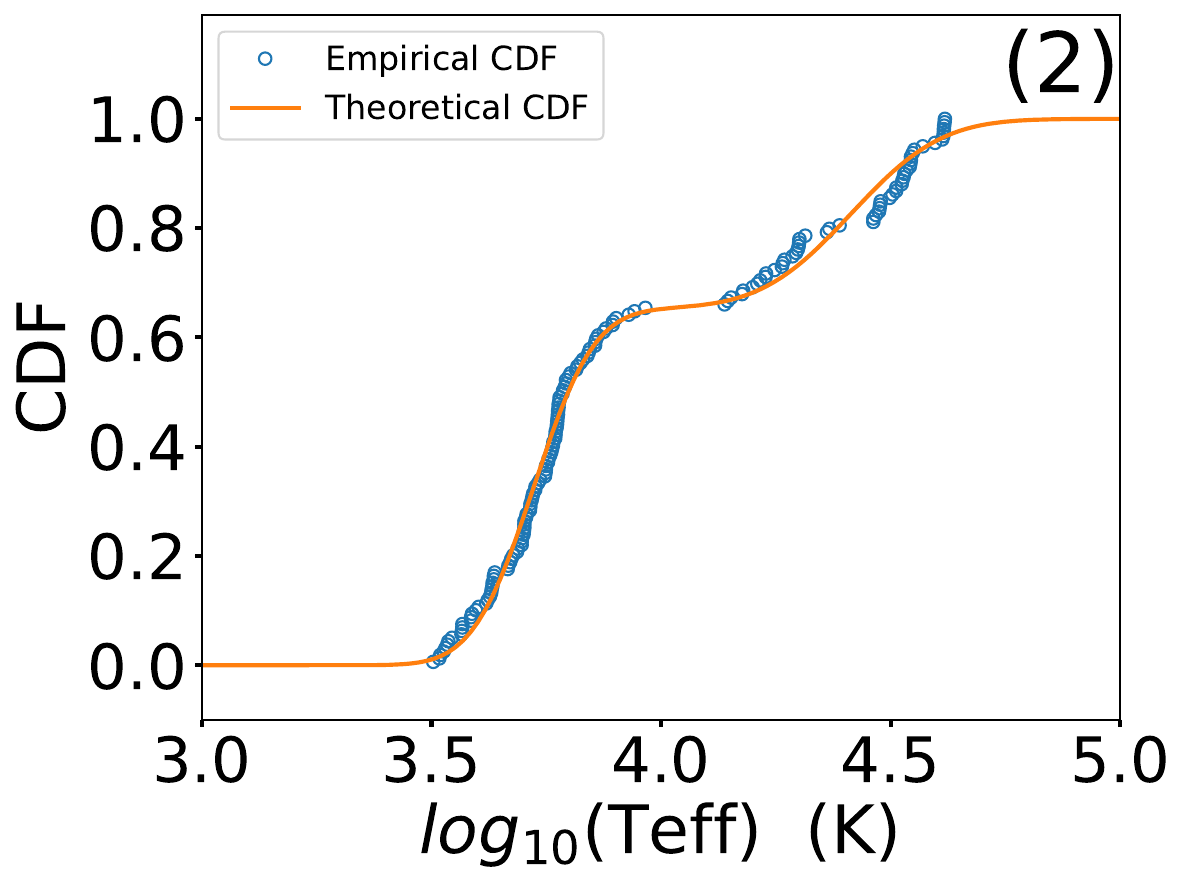}{0.495\textwidth}{}
             }
\vspace{-0.8cm}
\gridline{\fig{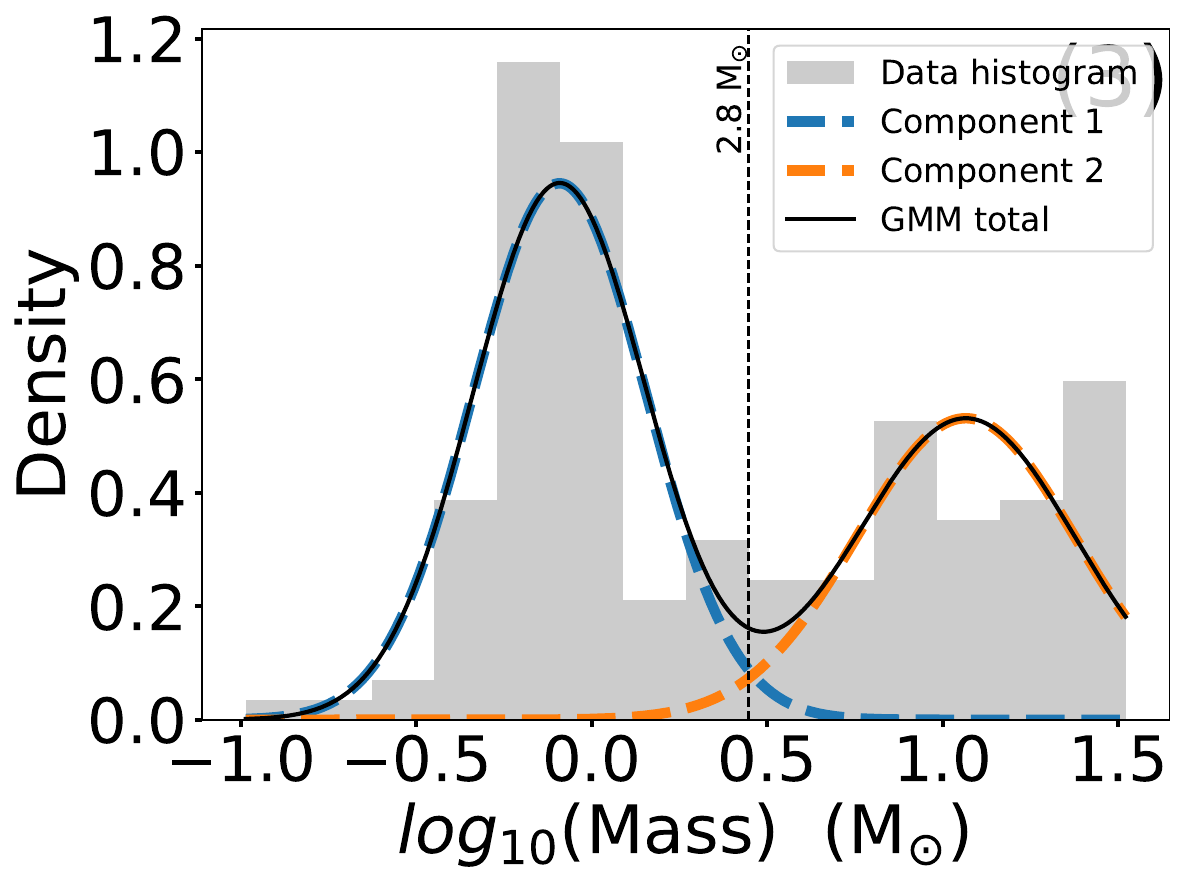}{0.495\textwidth}{}
          \fig{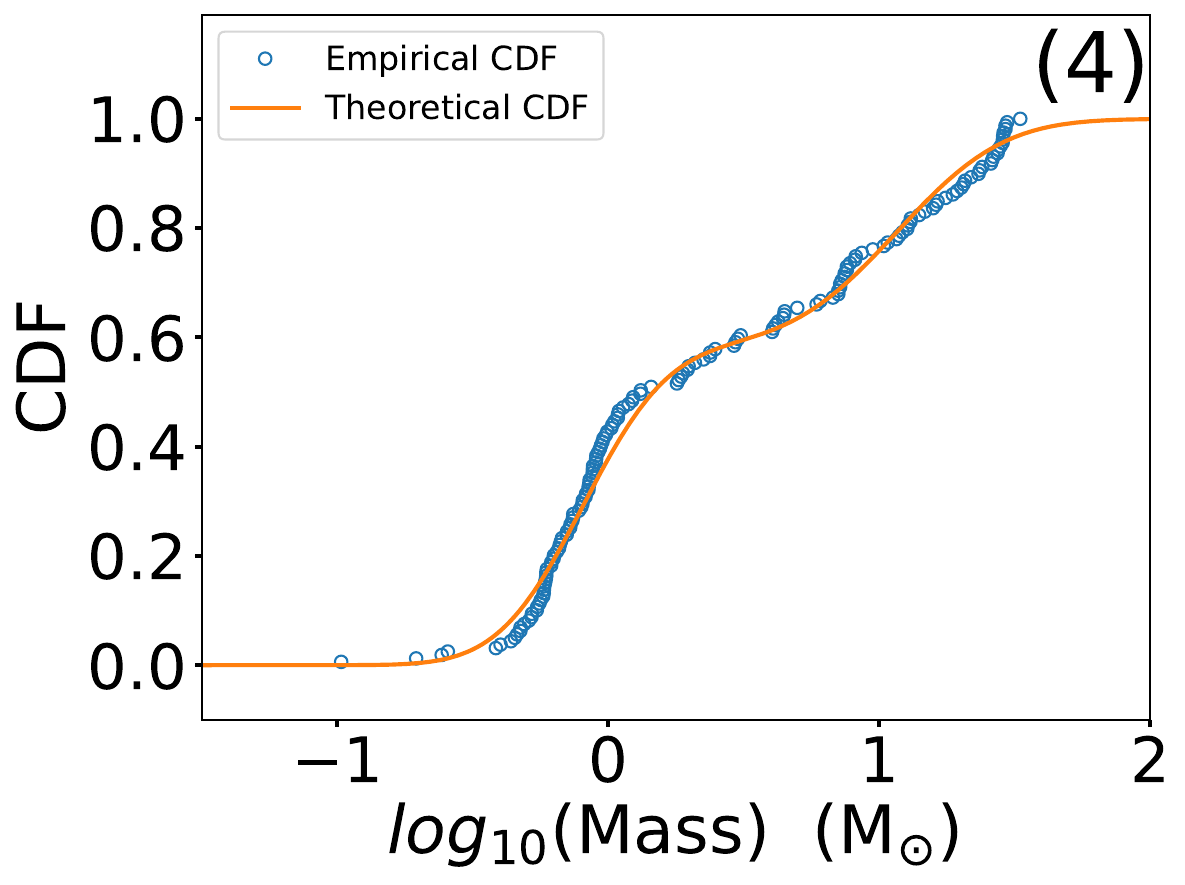}{0.495\textwidth}{}
             }
\vspace{-0.6cm}
\caption{Same as Figure \ref{fig:CDF_All_XRBs} but for XRBs from only MW. \label{fig:CDF_MW}}
\end{figure*}

\section{The Distribution of XRBs and occurrence rate from Monte Carlo testing} \label{distribution_random}

\begin{figure*}[h]
\gridline{\fig{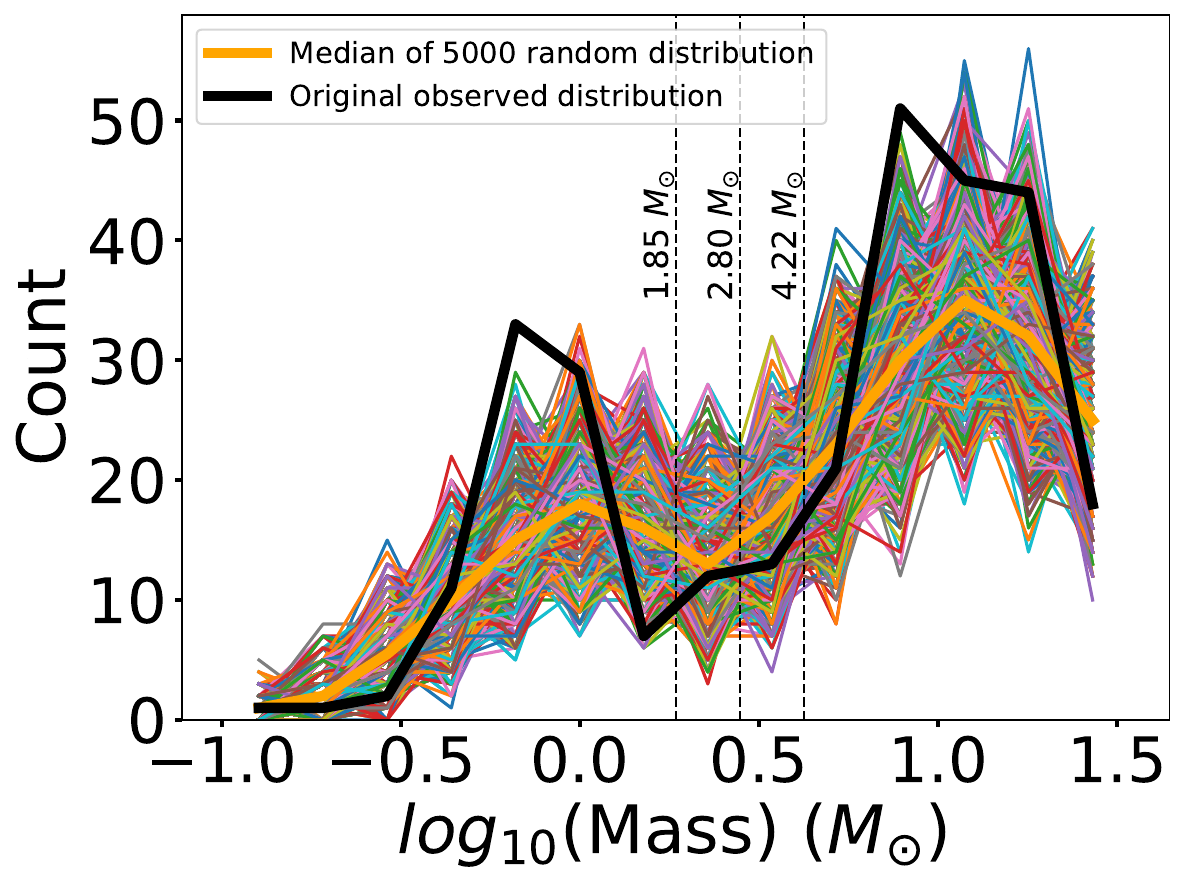}{0.5\textwidth}{(1)}
          \fig{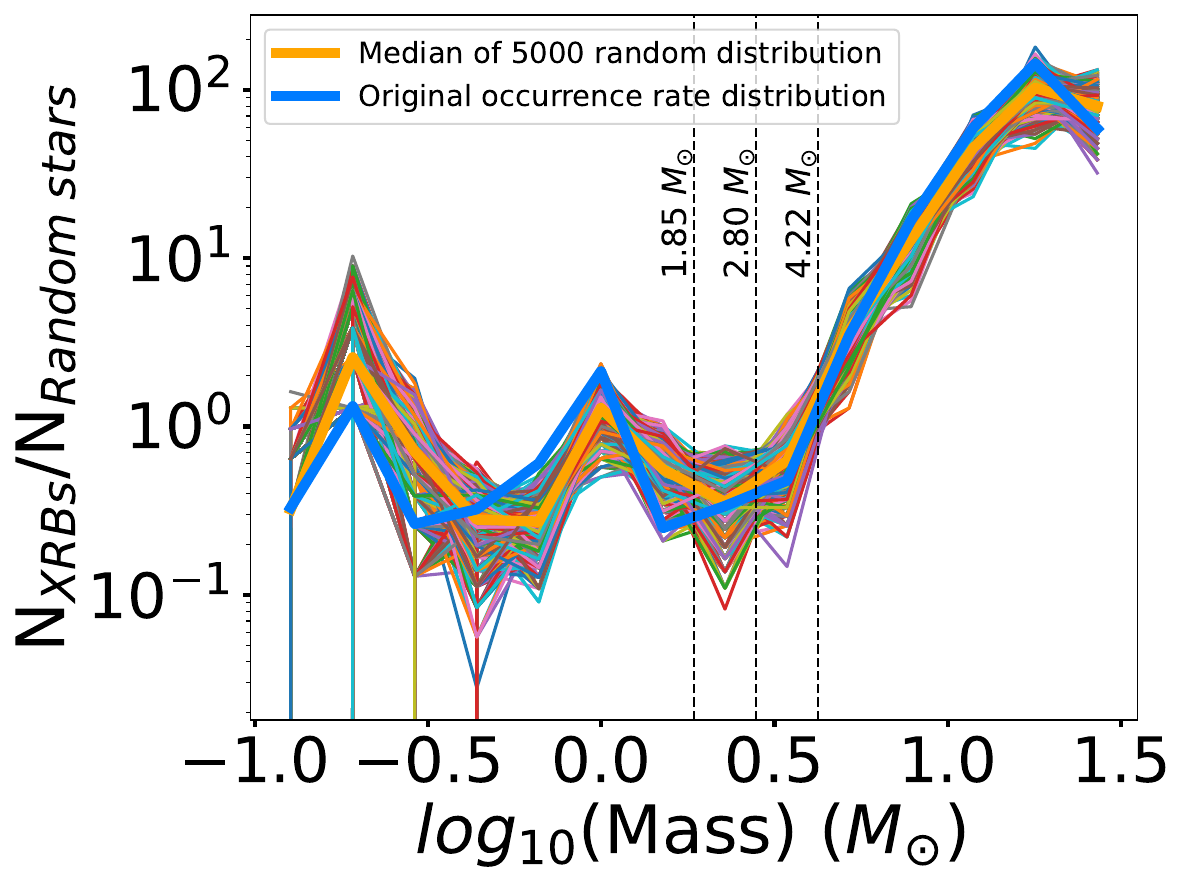}{0.5\textwidth}{(2)}
         }
\caption{The distribution of the donor mass (left) and the occurrence rate (right) from random mass data. Each thin colored line represents one of 5,000 simulated datasets generated using the Monte Carlo method. They are are based on the original data, which correspond to the black line in Panel~2 of Figure~\ref{fig:dist_observed} and the blue line in Panel~2 of Figure~\ref{fig:dist_intrinsic}. The thick orange line indicates the median of the 5,000 simulations. The thick black line (left panel) and thick blue line (right panel) represent the distributions from the original data for comparison.
\label{fig:distribution_random}}
\end{figure*}

This Figure \ref{fig:distribution_random} displays the distribution of donor mass and occurrence rate from simulated data. The simulated datasets were generated by perturbing the parameters of each target with an offset drawn from the distribution of deviations found in the comparison of 64 stars with previous studies.

\section{The reason for the absence of targets for High-Temperature Low-Metallicity area} \label{absence_on_highT_low_metal}

In panel 4 of Figure \ref{fig:relation}, the lower right region is blank. This is not because there are no target in that region, but because the current model cannot measure atmospheric parameters for that region. This is a technical selection effect. The measurement of stellar atmospheric parameters relies on comparing observed spectra with template spectra. For stars with high temperature and low metallicity, their spectra exhibit significant degeneracy, with normalized spectra being highly similar, making it difficult to obtain reliable parameters. Low metallicity results in a reduction and weakening of metal lines, making it impossible to use metal features to measure atmospheric parameters. High temperature also leads to a reduction in metal lines, and both gravity and temperature cause hydrogen lines to broaden, making them difficult to distinguish.

\textit{Gaia} had made significant progress in measuring atmospheric parameters for high-temperature and low-metallicity stars compared to before, allowing us to study high-temperature XRBs in the MCs. However, we still cannot completely eliminate the selection effects caused by high temperature and low metallicity.

If the lower right region of panel 4 is actually filled with XRBs, whether these targets could significantly alter the bimodal distribution shown above and affect the correlations between parameters displayed in other panels? We need to consider whether the blank area might contain a large number of stars.

Our viewpoint is that the blank area, compared to the high-metallicity region above it, is likely to have very few stars, not enough to alter the existing distributions and relationships above. Studies of metallicity in the MW and the MCs \citep{2018MNRAS.475.4279C, 2021MNRAS.507.4752C, 2023A&A...671A.157H, 2024AJ....167..123L} indicate that the metallicity distribution of these galaxies is mostly above -1.5, with very few stars having metallicities below -1.5. If the metallicity of XRBs also follows the overall distribution of the galaxies, then XRBs in the lower right blank area should be rare, and they would not disrupt the distribution characteristics and parameter relationships of XRBs presented in this paper. Massive stars, which are typically hotter, are also younger. Given that young stars tend to have higher metallicities, as shown by \citet{2023NatAs...7..951L}, high-temperature, low-metallicity stars should be uncommon.

\section{The relationship between the radius of a Roche-lobe-filling donor star and the mass ratio}
\label{q0788}


When one star in a binary system fills its Roche lobe, the donor star will begin transferring mass to its companion. This mass transfer process alters the Roche lobe radius $R_{Roche\;Lobe}$ of the donor star and is accompanied by changes in the mass ratio $q$. Here, we define the mass ratio as $q = M_{donor}/M_{accretor}$. According to the expression for the Roche lobe radius $R_{Roche\;Lobe}$ (equal to the donor star radius $R_{donor}$), semi-major axis $a$, and mass ratio $q$ given by \cite{1983ApJ...268..368E},
\[\frac{R_{Roche\;Lobe}}{a} = \frac{0.49q^{2/3}}{0.6q^{2/3}+\ln_{}{(1+q^{1/3})} },\]
and the expression relating the orbital angular momentum $J_{orb}$ to the total mass $M_{tot}$ ($=M_{donor}+M_{accretor}$), mass ratio $q$, and semi-major axis $a$ is
\[J_{orb} = \frac{q\sqrt{GM_{tot}^{3}a} }{(1+q)^2}. \]
From these, we can derive a relation between $R_{Roche\;Lobe}$ and the mass ratio $q$ as follows:
\[\frac{R_{Roche\;Lobe}}{J_{orb}^{2}/(GM_{tot}^{3})} = \frac{0.49q^{-4/3}(1+q)^4}{0.6q^{2/3}+\ln_{}{(1+q^{1/3})}  }. \]
Assuming that during mass transfer, the total mass and orbital angular momentum of the binary remain constant, the denominator on the left-hand side, $J_{orb}^{2}/(GM_{tot}^{3})$, becomes invariant. Therefore, changes in $R_{Roche\;Lobe}$ depend solely on the mass ratio $q$.

\begin{figure*}[h]
\gridline{\fig{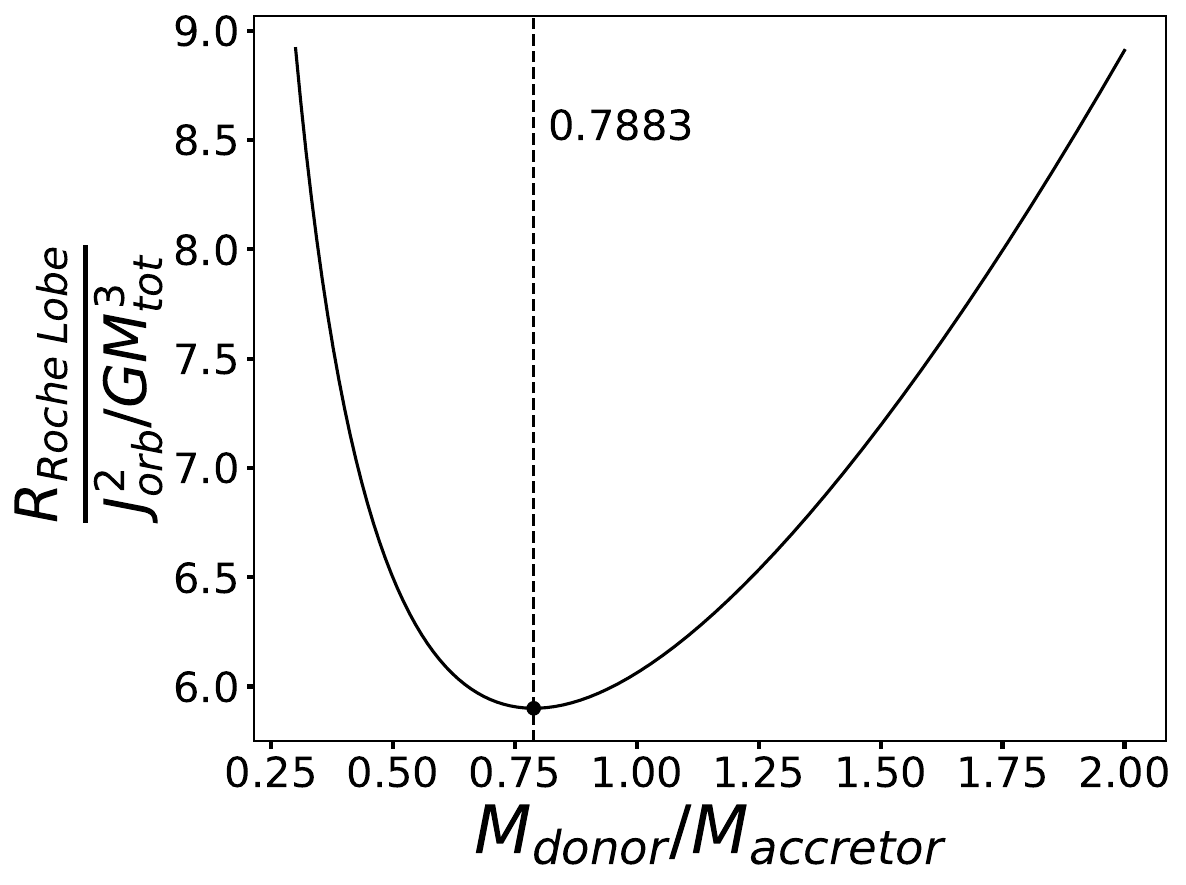}{0.5\textwidth}{}
         }
\caption{The relationship of Roche Lobe radius of donor star $R_{RL}$ with the mass ratio $q$. \label{fig:q0788}}
\end{figure*}

The correlation between $R_{Roche\;Lobe}$ (equal to $R_{donor}$) and $q$ is shown in Figure \ref{fig:q0788}, indicating that $R_{Roche\;Lobe}$ reaches a minimum when $q = 0.788$. Given our definition of the mass ratio $q = M_{donor}/M_{accretor}$, the mass ratio decreases as mass transfer occurs. Therefore, when the mass ratio exceeds 0.788, mass transfer causes $R_{Roche\;Lobe}$ to decrease, leading the donor star, which already fills its Roche lobe, to overflow further and accelerate the mass transfer process. This transfer occurs on a dynamical timescale, rapidly reducing the mass ratio to 0.788 within a short period (months to decades).

Observationally, it is challenging to detect semi-detached binaries undergoing Roche lobe overflow with mass ratios greater than 1. Our statistical analysis shows that for semi-detached binaries with one star filling its Roche lobe, the mass ratio is always less than 1 (excluding near-contact binaries where both stars are nearly filling their Roche lobes).



\begin{acknowledgments}
We sincerely thank the referee for the comprehensive and constructive report, which was instrumental in improving both the scientific rigor and readability of our work. This work is supported by the International Cooperation Projects of the National Key R\&D Program of China (No. 2022YFE0116800), the Chinese Natural Science Foundation (grant Nos. 11933008) and the Science Foundation of Yunnan Province (grant No. 202401AS070046, 202401AW070004), and the International Partership Program of Chinese Academy of Sciences (No. 020GJHZ2023030GC). GZ acknowledges support from the China Manned Space Program with grant No. CMS-CSST-2025-A13.
This research has made use of the SIMBAD database, operated at CDS, Strasbourg, France.
This work has made use of data from the European Space Agency (ESA) mission {\it Gaia} (\url{https://www.cosmos.esa.int/Gaia}), processed by the {\it Gaia} Data Processing and Analysis Consortium (DPAC, \url{https://www.cosmos.esa.int/web/Gaia/dpac/consortium}). Funding for the DPAC has been provided by national institutions, in particular the institutions participating in the {\it Gaia} Multilateral Agreement.
\end{acknowledgments}


\bibliography{sample701}{}
\bibliographystyle{aasjournalv7}



\end{CJK}
\end{document}